\documentclass[aps,prb,twocolumn,amsmath,amssymb,superscriptaddress,floatfix,nofootinbib]{revtex4-1}

\usepackage{amssymb}
\usepackage{amsmath}
\usepackage{graphicx}
\usepackage{textcomp}
\usepackage{color}
\usepackage{comment}
\usepackage{amsfonts}
\usepackage{epsfig}
\usepackage{hyperref}
\usepackage{bm}
\usepackage[caption=false]{subfig} 
\usepackage{booktabs}
\usepackage{tabularx}
\usepackage{array}
\usepackage{soul}
\usepackage{wasysym}
\usepackage{feyn}
\usepackage{esint}
\usepackage{footmisc}

\newcommand{\arctanh}[1]{\operatorname{arctan}}

\newcolumntype{M}[1]{>{\centering\arraybackslash}m{#1}}

\DeclareMathAlphabet\mathbfcal{OMS}{cmsy}{b}{n}

\begin{document}

\title{A theoretical perspective on the modification of the magnetocrystalline anisotropy at molecule-cobalt interfaces}

\author{Anita Halder}
\author{Sumanta Bhandary}
\author{David D. O'Regan}
\author{Stefano Sanvito}
\author{Andrea Droghetti}
\email[]{andrea.droghetti@tcd.ie}
\affiliation{School of Physics and CRANN Institute, Trinity College Dublin, The University of Dublin, Dublin 2, Ireland}

\begin{abstract}
We study the modification of the magnetocrystalline anisotropy (MCA) of Co slabs induced by several different conjugated molecular overlayers, i.e., benzene, cyclooctatetraene, naphthalene, pyrene and coronene. We perform first-principles calculations based on Density Functional Theory and the magnetic force theorem. Our results indicate that molecular adsorption tends to favour a perpendicular MCA at surfaces. A detailed analysis of various atom-resolved quantities, accompanied by an elementary model, demonstrates that the underlying physical mechanism is related to the metal-molecule interfacial hybridization and, in particular, to the chemical bonding between the molecular C $p_z$ and the out-of-plane Co $d_{z^2}$ orbitals. This effect can be estimated from the orbital magnetic moment of the surface Co atoms, a microscopic observable accessible to both theory and experiments. As such, we suggest a way to directly assess the MCA modifications at molecule-decorated surfaces, overcoming the limitations of experimental studies that rely on fits of magnetization hysteresis loops. Finally, we also study the interface between Co and both C$_{60}$ and Alq$_3$, two molecules that find widespread use in organic spintronics. We show that the modification of the surface Co MCA is similar upon adsorption of these two molecules, thereby confirming the results of recent experiments. 
\end{abstract}

\maketitle

\section{Introduction}
Surface magnetic properties dominate in thin films, multilayer structures and nanoparticles of $3d$ transition metals, resulting in a rich physics\cite{Gambardella2020,RevModPhys.89.025006} that is of critical importance for many technological areas, such as high-density magnetic recording\cite{va.ca.16} and spintronics\cite{hi.ke.20}. Most of these properties can be further modified through the adsorption of molecules\cite{Cinchetti2017}. The formation of strong covalent bonds between a molecule and a transition metal surface leads to the hybridization of the molecular orbitals with the $3d$ electronic bands\cite{at.br.10,at.ca.11,at.ra.14}. This affects the surface magnetic moments\cite{tr.la.13,mo.wh.14,Caffery}, spin-polarization\cite{br.at.10,Barraud2010, ja.bo.10,he.ri.17}, and electron correlation\cite{ja.dr.22}.
Furthermore, the hybridization can increase the strength of the magnetic exchange interaction between the surface atoms\cite{blugel_hard,fr.ca.15}, thus contributing to the magnetic hardening observed in recent experiments\cite{Raman2013,bo.ja.18,mo.al.20,be.al.21}. Selected molecules can even lead to a ferromagnetic response in otherwise non-magnetic Cu nanostructures\cite{MaMari2015-jo}. These electronic effects are sometimes further modified through surface reconstruction controlled by temperature annealing\cite{pa.sh.16, fl.le.21}, a feature that offers additional possibilities for materials engineering.\\

Among the surface properties that can be drastically modified through molecular adsorption there is also the magnetic anisotropy. This determines the ``easy'' and ``hard'' directions of magnetization of thin films\cite{stohrmagnetism,Johnson_1996} and, therefore, their magnetization switching dynamics and thermal stability. Bairagi \emph{et al.}\cite{bairagiprl} reported that the deposition of buckyball C$_{60}$ molecules on epitaxial Co hcp(0001) thin-films causes an enhancement of the perpendicular magnetic anisotropy. The same group later extended the study, considering another popular molecule in spintronics, namely Alq$_3$, and comparing ultrathin Co films supported on gold and platinum (Ref. \onlinecite{bairagi_2nd}). They found a considerable enhancement of the perpendicular magnetic anisotropy in all samples and concluded that this effect is rather general in the case of chemisorbed molecules on Co. Along the same lines, Benini \emph{et al.} recently suggested that a similar magnetic anisotropy enhancement could be one of causes for the huge magnetic hardening measured in polycrystalline Co films interfaced with C$_{60}$ and Gaq$_3$ (Ref. \onlinecite{be.al.21}). It must be noted, however, that in their experiment the magnetic field was swept in the films plane rather than out-of-plane. The magnetic anisotropy is, in general, determined by a combination of several extrinsic and intrinsic factors\cite{stohrmagnetism,Johnson_1996}. Nonetheless, in all the mentioned experiments, the magnetocrystalline anisotropy (MCA) is regarded as the main contribution that is modified upon adsorption. The MCA stems from the spin-orbit coupling (SOC) and is related to the atomic structure and bonding of a material. Several works\cite{zh.le.11,ja.na.15,fr.ca.15,st.se.16,at.hi.21} have already shown that the SOC-induced spin-texture of metallic or seminconducting surfaces can be dramatically affected by the chemical bonding with adsorbates of various kinds.
In $3d$ ferromagnetic films with a molecular overlayer, this effect is expected to give rise to MCA changes. \\

The modification of surface magnetic properties in thin films are generally rationalized by means of first-principles calculations based on Density Functional Theory (DFT). The results for Co(0001) show that the contribution of the out-of-plane Co $d_{z^2}$ orbitals to the surface MCA favours in-plane magnetization. This in-plane contribution is completely suppressed upon C$_{60}$ adsorption, when the Co $d_{z^2}$ orbitals hybridize with the C $p_z$ orbitals of C$_{60}$ (Ref. \onlinecite{bairagiprl, bairagi_2nd}). In contrast, the contributions to the MCA of the other $3d$ Co orbitals, which favour out-of-plane magnetization, remain unaffected by C$_{60}$. 
Hence, the perpendicular MCA is enhanced at the C$_{60}$/Co interface.
The DFT results qualitatively support the experimental observations. 
However, a quantitative correlation between interfacial MCA and molecule-metal hybridization remains to be established, and systematic investigations on model systems are needed to address this issue. 
Notably, the comparison between DFT and experimental results appears very problematic, when going beyond phenomenological considerations. On the one hand, the orbital-resolved MCA, computed by DFT and used in previous studies as a descriptor for the effect, is a microscopic quantity that is not directly measurable. On the other hand, in experiments, the MCA is estimated from magnetization hysteresis loops as an average macroscopic quantity. 
In principles, the experiments could be modelled through multi-scale approaches combining DFT and atomistic spin dynamics simulations\cite{Skubic_2008,Evans_2014}, but, in practice, the level of complexity and the number of features (e.g., disorder, temperature, etc.) that such simulations should consider, are out-of-reach. 
To overcome this gap between theory and experiments and to proceed towards a quantitative understanding of the MCA modification at molecule-Co interfaces, 
we then need to identify a MCA-related microscopic observable, which can be directly calculated from first-principles as well as measured in experiments with comparable accuracy.\\

In this work, we explain how the chemical bonding between the Co $d_{z^2}$ and the molecular C $p_z$ orbitals favours perpendicular MCA at molecule-Co interfaces. The physics is firstly described through an elementary model and then analyzed at the quantitative level by means of first-principles DFT calculations for several prototypical systems, namely benzene, cyclooctatetraene, naphthalene, pyrene, and coronene molecules on Co. The results confirm overall the phenomenological arguments used so far to interpret the experimental results in Refs. \onlinecite{bairagiprl,bairagi_2nd,be.al.21}, and provide a strong indication that the modification of the MCA is general for Co with chemisorbed conjugated molecules. Going beyond the results of previous studies, we examine the key electronic parameters determining the magnitude of the interfacial MCA. In particular, we highlight the importance of the energy splitting between the in-plane Co $d_{x^2-y^2}$ orbitals, which are not affected by the bonding with molecules, and the $d_{z^2}$ orbitals, which are instead strongly hybridized with molecular C atoms. When, upon molecular adsorption, the $d_{z^2}$ to $d_{x^2-y^2}$ energy splitting increases, we find an enhanced perpendicular interface MCA. An additional, important observation stemming from our DFT calculations is that the effect of the $p_z$-$d_{z^2}$ hybridization on the MCA can be quantitatively estimated through the orbital magnetic moment, which is a microscopic observable and can be measured in experiments.  As such, we suggest a way to directly assess our first-principles predictions to fill the current gap between theoretical and experimental studies. Finally, we complete our work by comparing the adsorption of the two experimentally studied molecules, namely C$_{60}$ and Alq$_3$, on Co. We show that the MCA modification is quite similar in the two cases, in agreement with experimental observations \cite{bairagi_2nd,be.al.21}.\\

The paper is organized as follows. We begin in Sec. \ref{sec.model} by introducing an elementary model describing the MCA at conjugated molecule/Co interfaces. We then continue by discussing the results of our DFT calculations. After providing the computational details in Sec. \ref{Sec.Details}, we describe in great detail the interface between Co and benzene, which is one the simplest aromatic molecules. Specifically, in Sec. \ref{subsec.ben}, we analyse various atom-resolved quantities, which help us to understand the basic electronic properties and their quantitative correlation with the MCA. We then extend our study to consider the interface with other conjugated molecules and to examine the effect of the chemical reactivity (in Sec. \ref{sec.cot}) and of molecular coverage (in Sec. \ref{sec.coverage}). Finally, we discuss the results for C$_{60}$/Co and Alq$_3$/Co interfaces in Sec. \ref{sec.alq3_and_c60}, and we conclude in Sec. \ref{sec.conclusion}.

\begin{figure}[h]
\centering\includegraphics[width=0.5\textwidth,clip=true]{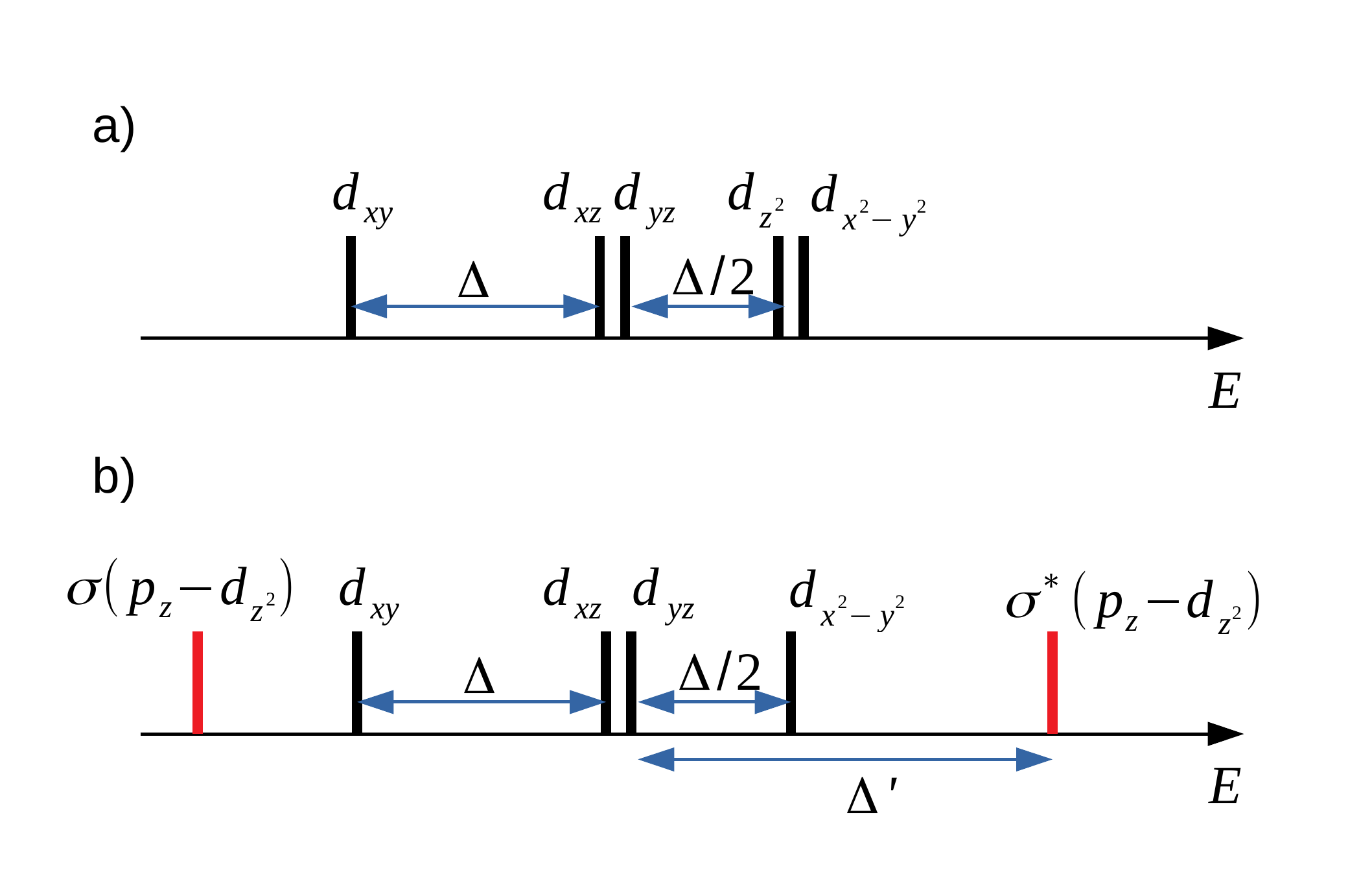}
\caption{(a) Energy diagram for the crystal-field $3d$ orbital splitting at a Co(001) surface. We assume that the $d_{z^2}$ and $d_{x^2-y^2}$ orbitals remain degenerate and that the energy difference, $\Delta$, between $d_{xy}$ and doublet ($d_{xz}$, $d_{yz}$) is twice the energy difference between ($d_{xz}$, $d_{yz}$) and ($d_{z^2}$, $d_{x^2-y^2}$). In panel (b) the same energy diagram is now modified by the adsorption of a conjugated molecule. This displays the bonding, $\sigma$, and anti-boding, $\sigma^*$, states due to the hybridization of the Co $d_{z^2}$ orbital with the $p_z$ orbital, which originally form the $\pi$ molecular system.}
\label{fig.scheme}
\end{figure}

\section{A simple model for surface MCA}\label{sec.model}
The origin of the MCA is the SOC, whose Hamiltonian is written as $\hat H_\textrm{SOC}=\xi \hat{\mathbf{L}}\cdot \hat{\mathbf{S}}/\hbar^2$, where $\hat{\mathbf{L}}$ and $\hat{\mathbf{S}}$ are respectively the orbital and spin angular momentum operator, and $\xi$ is the SOC constant. 
For a ferromagnetic thin film, the MCA energy $E_\mathrm{MCA}$ is defined as the energy difference calculated between the magnetization parallel ($\parallel$) and perpendicular ($\perp$) to the film surface, i.e., $E_\mathrm{MCA}=E_\parallel - E_\perp$. Positive (negative) $E_\mathrm{MCA}$ implies out-of-plane (in-plane) MCA. Throughout this work, we assume a Cartesian frame of reference with the film's perpendicular direction being along the $z$-axis.\\

In $3d$ transition-metals, the SOC constant $\xi$ is relatively small compared to other characteristic energies, such as the band width, the crystal field splitting, and the exchange coupling. Thus, $E_\mathrm{MCA}$ can be evaluated from the SOC-induced ground state energy corrections within second-order perturbation theory as \cite{bruno,stohrmagnetism, sumanta-multi}
\begin{equation}
E_\mathrm{MCA}= \sum_{o,u}\frac{ \vert\langle \psi_o\vert \hat H_\mathrm{SOC}\vert \psi_u\rangle\vert^2_{\parallel}-\vert\langle \psi_o\vert \hat H_\mathrm{SOC}\vert \psi_u\rangle\vert^2_{\perp} }{\epsilon_o-\epsilon_u}. \label{eq.E_MCA}
\end{equation}
Here $\vert\psi_o \rangle$ and $\vert \psi_u \rangle$ are the unperturbed occupied and unoccupied states with energies $\epsilon_o$ and $\epsilon_u$, respectively. These can have either the same or different spin. We set the spin quantization axis along the $z$ direction, which is normal to the film.  
The matrix elements in Eq.~(\ref{eq.E_MCA}) describe virtual transitions, which can promote
in-plane or out-of-plane anisotropy. They strongly contribute to the total $E_\mathrm{MCA}$ when they couple occupied and unoccupied states close to the Fermi level so that the denominator in Eq. (\ref{eq.E_MCA}) is small.
In crystalline films, $\vert\psi_o \rangle$ and $\vert \psi_u \rangle$ are Bloch states. However, to capture some qualitative features of the MCA in a simple way, we consider here a model proposed in Ref. [\onlinecite{zh.lu.17}] using atomic $3d$ orbitals and ignoring electronic band formation. The matrix elements in Eq. (\ref{eq.E_MCA}) can then be calculated analytically\cite{gi.ca.12} and are reported in Sec. S1 of the supplementary material (SM). Based on these, one can then derive some simple rules, which will be used to infer the direction of the MCA \cite{Daalderop,gi.ca.12,Kyuno_1996}. 
Since each $d$ orbital is characterized by its magnetic quantum number $m$ in addition to the spin quantum number, one has that: (i) the transitions between orbitals with $m$ = 0 ($d_{z^2}$) and $\vert m \vert$ = 1 ($d_{yz}$ and $d_{xz}$) in the same spin channel or between orbitals with $m$ = 0 ($d_{z^2}$) and $\vert m \vert$ = 2 ($d_{x^2-y^2}$ and $d_{xy}$) in different spin channels favour in-plane MCA; (ii) the transitions between orbitals with $m$ = 0 and $\vert m \vert$ = 1 in different spin channels and between orbitals with $\vert m \vert$ = 2 in the same spin channel favour out-of-plane MCA.\\

We consider the simplest case of a Co fcc(001) surface (the treatment can be easily generalized to other surfaces). The crystal field splits the five $d$ orbitals into a singlet ($d_{xy}$), a doublet ($d_{xz}$, $d_{yz}$) and two further singlets ($d_{z^2}$ and $d_{x^2-y^2}$). Since generally the separation between $d_{z^2}$ and $d_{x^2-y^2}$ is small, we assume that these orbitals remain degenerate. In addition, to simplify the calculation, we set the energy difference, $\Delta$, between $d_{xy}$ and the doublet ($d_{xz}$, $d_{yz}$) to be twice the energy difference between ($d_{xz}$, $d_{yz}$) and ($d_{z^2}$, $d_{x^2-y^2}$). This is displayed in the energy level diagram of Fig. \ref{fig.scheme}(a). Since (fcc) Co is a strong ferromagnet and the majority-spin band is nearly fully occupied \cite{Li.Pa17}, majority-spin states do not substantially contribute to the MCA energy. Therefore, we only take into account transitions between spin-down occupied states to the spin-down unoccupied ones, and we apply the simple rules mentioned above to infer their contribution. 
Specifically, since Co has two electrons in the spin-down $d$ shell,
the $d_{xy}$ orbital is fully filled, the $d_{xz}$ and $d_{yz}$ are half-filled and the $d_{z^2}$ and $d_{x^2-y^2}$ are empty. Thus the dominant transitions in Eq. (\ref{eq.E_MCA}) are, on the one hand, from $d_{xy}$ to $d_{x^2-y^2}$ favouring out-of-plane MCA, and, on the other hand, from $d_{xy}$ to $d_{xz}$ and from $d_{yz}$ to $d_{z^2}$ and $d_{x^2-y^2}$ (weighted by a factor $1/2$ to account for the half-filling of $d_{xz}$ and $d_{yz}$) favouring in-plane MCA. 
Neglecting degeneracies, the total $E_\mathrm{MCA}$ turns out to be equal to 
\begin{equation}\label{eq.E_MCA_Co}
E_\mathrm{MCA}= -\frac{ 11\xi^2}{12\Delta},  
\end{equation}
and the MCA is therefore in-plane ($E_\mathrm{MCA}<0$) for our choice of parameters. \\

Now, when a conjugated organic molecule is chemisorbed on the Co thin film, the $p_z$ atomic orbitals, originally forming the $\pi$ molecular orbitals, hybridize with the minority Co $d_{z^2}$ states, giving rise to molecule-metal hybrid bonding and antibonding states\cite{at.br.10}. The bonding $\sigma$ state has a predominant $p_{z}$ character and is situated at low energies, while the antibonding $\sigma^*$ state has $d_{z^2}$ character and appears at much higher energies as shown in Fig. \ref{fig.scheme}(b). The modification of the MCA energy induced by molecular adsorption can then be computed via Eq. (\ref{eq.E_MCA}) considering these additional hybrid states. In our simple model, we indicate as $\Delta'$ the energy separation between the doublet ($d_{xz}$, $d_{yz}$) and the antibonding state, and we neglect the low-energy bonding state, which is mostly localized on the molecule rather than at the metal surface. Furthermore, since the anti-bonding state has mostly $d_{z^2}$ character we approximate $\vert\sigma^*\rangle\approx\vert d_{z^2}\rangle$ in the evaluation of the  transition matrix elements. In practice, we assume that the only effect of the absorption is to shift the $d_{z^2}$ orbital up in the energy by an amount $(\Delta' - \Delta/2)$. Hence, the resulting MCA energy is modified and becomes
\begin{equation}
E_\mathrm{MCA}=\frac{7\xi^2}{12\Delta}-\frac{3\xi^2}{4\Delta'}\:.  
\end{equation}
Notably, by analyzing this equation we see that, if $\Delta'>9\Delta/7$, the Co surface MCA will switch from in-plane ($E_\mathrm{MCA}<0$) to out-of-plane ($E_\mathrm{MCA}>0$) in the hybrid Co-molecule system. The reason is that, for a large enough $\Delta'$, the contribution of the transition from $d_{yz}$ to $d_{z^2}$ favouring in-plane MCA in Eq. (\ref{eq.E_MCA}) is drastically suppressed. This remarkably simple result indicates that, in principle, not just the magnitude, but also the direction of the MCA can be modified through molecular adsorption. In spite of the simplicity of our model and the heuristic choice of the crystal-field splitting parameters, we will show in the following that the key qualitative features of the MCA at organic-ferromagnetic metal interfaces are largely captured by such atomic-like picture.

\section{Computational details}\label{Sec.Details}
DFT calculations are performed using the projector augmented wave (PAW)\cite{PAW} method as implemented in the Vienna {\it Ab-initio} Simulation Package (VASP)~\cite{VASP}. The generalized gradient approximation (GGA) in the formulation by Perdew, Burke, and Ernzerhof (PBE)~\cite{PBE} is chosen for the exchange-correlation functional. The tetrahedron integration method~\cite{tetra} with a kinetic-energy cutoff of 600 eV is employed. An energy convergence criterion equal to 10$^{-7}$ eV for total energy calculations and $10^{-3}$ eV for ionic relaxations is adopted. \\

In order to describe the thin-film geometry, we consider finite slabs comprising a few Co fcc(001) or hcp(0001) layers as specified in the following sections and in the SM. Each slab is separated by about 8~\AA~of vacuum along the $z$-direction (orthogonal to the surface) from its periodic image. The lattice constant of fcc Co is fixed to the experimental value of 3.54410~\AA\ as from Ref. [\onlinecite{Owen_1954}].
The molecule-Co systems are modeled with the molecule adsorbed on one side of the slab. During the interface geometry optimization the two bottom layers of the slab are maintained fixed, while all the other atoms are allowed to relax until the ionic forces are lower than 0.001 eV/\AA. \\

The MCA energies are calculated by using the magnetic force theorem \cite{,an.sk.79,we.wa.85,da.ke.90} followings two steps. Firstly, we carry out a scalar relativistic collinear charge self-consistent calculation to obtain the charge density. Then, we use that charge density as input in non-collinear calculations performed non-self-consistently and including SOC, where the magnetization vector is oriented along different directions. In particular, the MCA energy is given by the difference of the total band energies calculated with the magnetization parallel and perpendicular to the slab surface, $E_\mathrm{MCA} = E_{\mathrm{band},\parallel}-E_{\mathrm{band},\perp}$. This method has already been applied in several previous works to estimate the MCA energies of different ferromagnetic multilayer systems (see, for example, Refs. \onlinecite{erikssonbcc,st.kh.16,aa.ha.13,CoFemag,on.ki.15,su.kw.20}), and the results generally agree well with the values obtained from fully self-consistent total-energy calculations\cite{bl.ce.19,su.kw.20}. For the smallest considered systems, we also directly verified that force theorem and total energy calculations give comparable results up to 0.1 meV per supercell.\\

The $\mathbf{k}$-point sampling is performed using a Monkhorst-Pack (MP) grid. All calculations are converged with respect to the number of $\mathbf{k}$-points. We find that a $12 \times 12$ $\mathbf{k}$-point mesh in the two-dimensional Brillouin zone is enough to reach an accuracy below 1 meV on the MCA energy of slabs with  $2 \times 2$ in-plane periodicity. Although the MCA energy is a very small quantity (of the order of one meV or even less), and the DFT results are quite sensitive to the details of the calculations and to the exchange correlation functional \cite{PhysRevLett.75.2871,ya.sa.01,bu.er.04}, we mostly focus on general trends, which are expected to be well captured by DFT calculations. \\

Orbital magnetic moments are calculated without the orbital polarization contribution\cite{re.10} or without any effective orbital polarization corrections\cite{er.br.90}. In case of Co this contribution is negligible \cite{PhysRevB.81.060409}.
The SOC matrix elements are defined as\cite{vaspSOC}
\begin{equation}
E^{I,mm'}_\mathrm{SOC} = \sum_{n\mathbf{k}} \sum_{\sigma\sigma'} P^{lm\sigma}_{n\mathbf{k}} \langle \phi^{I, lm\sigma }\vert \hat H_\mathrm{SOC} \vert \phi^{ I, lm'\sigma'}  \rangle   P^{lm'\sigma' *}_{n\mathbf{k}}.\label{eq.matrixSOC}
\end{equation}
Here, $\phi^{I, lm\sigma }$ is the PAW partial wave of an atom $I$ with orbital quantum number $l$, magnetic quantum number $m$ and spin index $\sigma$. The index of the occupied Kohn-Sham electronic bands is $n$, and $P^{lm\sigma}_{n\mathbf{k}}$ is the projection of the $n$-th Kohn-Sham pseudo-orbital onto a PAW projector sited on the atom $I$. The atomic SOC energy is then obtained as $E^I_\mathrm{SOC}=\sum_{m,m'}E^{I,mm'}_\mathrm{SOC}$ (see Ref. \onlinecite{vaspSOC}).   
The contribution to the MCA energy of the $I$-th atom, $E^I_\mathrm{MCA}$, is estimated from the difference of the atomic SOC energies with the magnetization in-plane and out-of-plane, 
\begin{equation}\label{eq.atomic_MAE}
E^I_\mathrm{MCA}\propto\Delta E^I_\mathrm{SOC}=E^I_\mathrm{SOC,\parallel}-E^I_\mathrm{SOC,\perp},
\end{equation}
where the spin quantization axis is taken perpendicular to the surface (i.e. along the $z$ direction). For most of the considered systems, the sum of $\Delta E^I_\mathrm{SOC}$ over all atoms is found to be twice the MCA energy obtained from the magnetic force theorem, i.e., $E_\mathrm{MCA}\approx 0.5 \sum_I\Delta E^I_\mathrm{SOC}$. Based on the discussion of Ref. [\onlinecite{ANTROPOV201435}], this indicates that the SOC is effectively a second-order perturbative contribution to the system non-relativistic Hamiltonian and, therefore, the interpretation of the results based on Eq. (\ref{eq.E_MCA}) is well justified.

\section{DFT Results}\label{Sec.result}
\begin{figure*}
\includegraphics[width=1.0\linewidth]{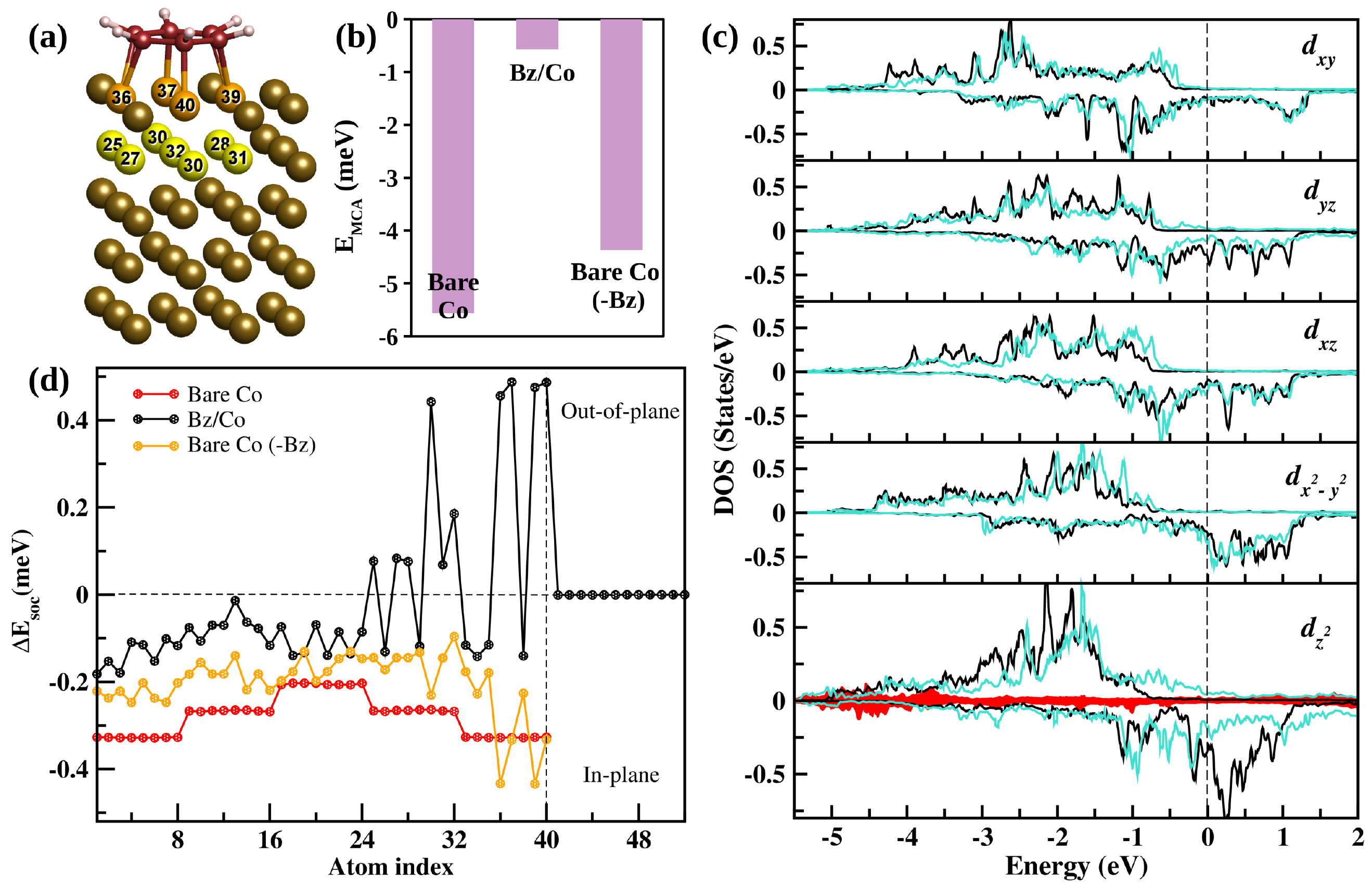}
\caption{DFT results for the Co and Bz/Co slabs. (a) Side view of the Bz/Co slab. Brown, white, and yellow spheres indicate  C, H, and Co atoms, respectively. (b) MCA energy of the bare Co slab, of the Bz/Co slab, and of the distorted slab [bare Co (-Bz)] obtained by removing the molecule from the Bz/Co slab. (c) DOS projected over the $3d$ orbitals of the Co atom that is labeled 36 in (a). The Fermi energy is at 0 eV. The black and cyan lines are for the bare Co slab and the Bz/Co slab, respectively. The red curve in the bottom panel is the DOS projected over the Bz C $p_{z}$ orbital, which hybridizes with the $d_{z^2}$ orbital. (d) $\Delta E^I_\mathrm{SOC}$ evaluated for all atoms. The red, orange and black dots represent the values for the Co slab, the distorted Co slab, and the Bz/Co slab. Atoms 1-8 are at the slab bottom surface, atoms 33-40 are at the top surface, and atoms 41-52 are the Bz atoms. The Co atoms in the first and second layer of the Bz/Co slab, for which $\Delta E^I_\mathrm{SOC}$ is positive, are highlighted with orange and light yellow colors in the slab side view. }\label{fig2.5layers}
\end{figure*}
\subsection{Interface between benzene and Co}\label{subsec.ben}
In order to illustrate how the physics introduced through the elementary model emerges in real systems, we start by performing DFT calculations for an interface between benzene (Bz) and Co.
This can be considered as a prototypical model system, since Bz is the simplest aromatic $\pi$-conjugated molecule that chemisorbs on reactive $3d$ ferromagnetic surfaces \cite{ge.ba.95,ca.zh.18}. We consider fcc Co and select the same
fcc(001) surface as in the atomic model. Note that Co fcc(001) films have been extensively studied for spintronics \cite{ci.he.09,la.al.12,dj.ca.13,dr.st.14,andrea-alq3, la.al.19}. They represent a convenient system to study interface effects, as the bulk contribution to the total MCA vanishes for large films due to the cubic symmetry. Nevertheless, since several recent experiments mentioned in the introduction have focused on hcp(0001) surfaces\cite{bairagiprl}, in the SM (Sec. S2) we also report some additional calculations for Co hcp(0001) slabs. 
The physics driven by molecular adsorption is similar, at the atomic scale, for Co hcp and fcc surfaces and, as such, the discussion and the conclusions presented in the following sections will be qualitatively valid for both cases. \\

 
\subsubsection{Electronic structure}
We study a 5-layer Co slab with a 2$\times$2 in-plane supercell. Results for slabs of different thickness are presented in the SM (Sec. S3). Bz is chemisorbed on the top slab surface in the hollow position [see Fig. \ref{fig2.5layers}(a)],
adopting a non-planar structure, where the H atoms are
situated slightly above the C ones. This is common for conjugated molecules on ferromagnetic surfaces\cite{at.br.10,br.at.10,mi.ha.01,dr.st.14}. The length of the C-Co bonds, which are formed by the two C atoms directly on top of two Co ones, is 2.04~\AA, while the length of the other C-Co bonds is 2.17~\AA. \\

The electronic structure of the Co surface and the modifications induced by molecular adsorption are qualitatively similar to those described by our elementary model. In Fig. \ref{fig2.5layers}(c) we plot the density of states (DOS) projected 
over the $3d$ orbitals of a Co atom forming a bond with the molecule across the interface [specifically, the atom considered is labeled as 36 in Fig. \ref{fig2.5layers}(a)]. The black and the cyan lines are for the bare Co slab and for the Bz/Co slab, respectively. 
For the bare surface, the majority spin (up) states are almost fully occupied, while, the minority spin (down) DOS crosses the Fermi level. The crystal-field splitting is qualitatively comparable to the level splitting assumed in the diagram of Fig. \ref{fig.scheme}(a), although the Co states are now broadened into bands. Consistently, $d_{xy}$ is the orbital with the lowest energy, and most of the corresponding minority DOS is located below the Fermi energy, while $d_{yz}$ and $d_{xz}$ form a doublet. Their minority DOS is identical and centered across the Fermi level. Thus, they are almost half-filled. Finally, $d_{z^2}$ and $d_{x^2-y^2}$ have high energies, and the minority DOS is centered above the Fermi level.\\

Upon Bz adsorption, there is a significant change in the Co electronic structure. By analyzing the DOS projected over the Co $d_{z^2}$ orbital and the C $p_z$ orbital of Bz [red line in Fig.~\ref{fig2.5layers}(c)], we can identify the formation of the Co-Bz $p_z$-$d_{z^2}$ bonding and antibonding states, in particular in the minority spin channel, as assumed in our elementary model.
The minority bonding state has mostly $p_z$ character and is quite localized in energy at about $-4.5$ eV below the Fermi level. In contrast, the minority anti-bonding state with dominating $d_{z^2}$ character, is very broad in energy, and the DOS extends over an energy range as large as 7 eV crossing the Fermi level.
We note that the DFT calculations predict some molecule-metal charge transfer and spin density redistribution leading to a reduction of the DOS spin splitting compared to the bare surface case. As a result, the center of the $d_{z^2}$ minority state is shifted below rather than above the Fermi energy.
These material-specific features were not included in the elementary model. However, we find that the effect of the molecule-Co hybridization on the MCA is similar to that described in Sec. \ref{sec.model}, as explained in the following.

\subsubsection{MCA energy}
The change of the slab MCA due to the adsorption of Bz is summarized in Fig. \ref{fig2.5layers}(b). The MCA energy $E_\mathrm{MCA}$ is equal to $-5.6$ meV for the bare Co slab, indicating an in-plane easy axis, whereas it is drastically reduced to $-0.6$ meV upon molecular adsorption. 
Such a drop in the MCA energy may result from the molecule-Co electronic hybridization, as discussed above, but it may also derive from the displacement of the surface atoms following the Bz adsorption. This displacement was indeed found to induce large changes in the bands spin-texture of some metallic surfaces \cite{st.se.16}. In order to understand which of the two effects is dominant for our specific system, we remove the molecule from the optimized Bz/Co slab and consider the remaining Co layers without re-relaxing their atomic positions [bare Co(-Bz) in Fig. \ref{fig2.5layers}(b)]. In the case of this ``distorted slab'', we find that $E_\mathrm{MCA}$ retains a value of $-4.4$~meV, similar to that of the original bare Co slab. Hence, we conclude that the atomic displacement has only a minor influence on the Co MCA energy, leaving the molecule-metal hybridization as the main effect responsible for the MCA modification.  \\

\begin{figure}[h]
\centering\includegraphics[width=\columnwidth,clip=true]{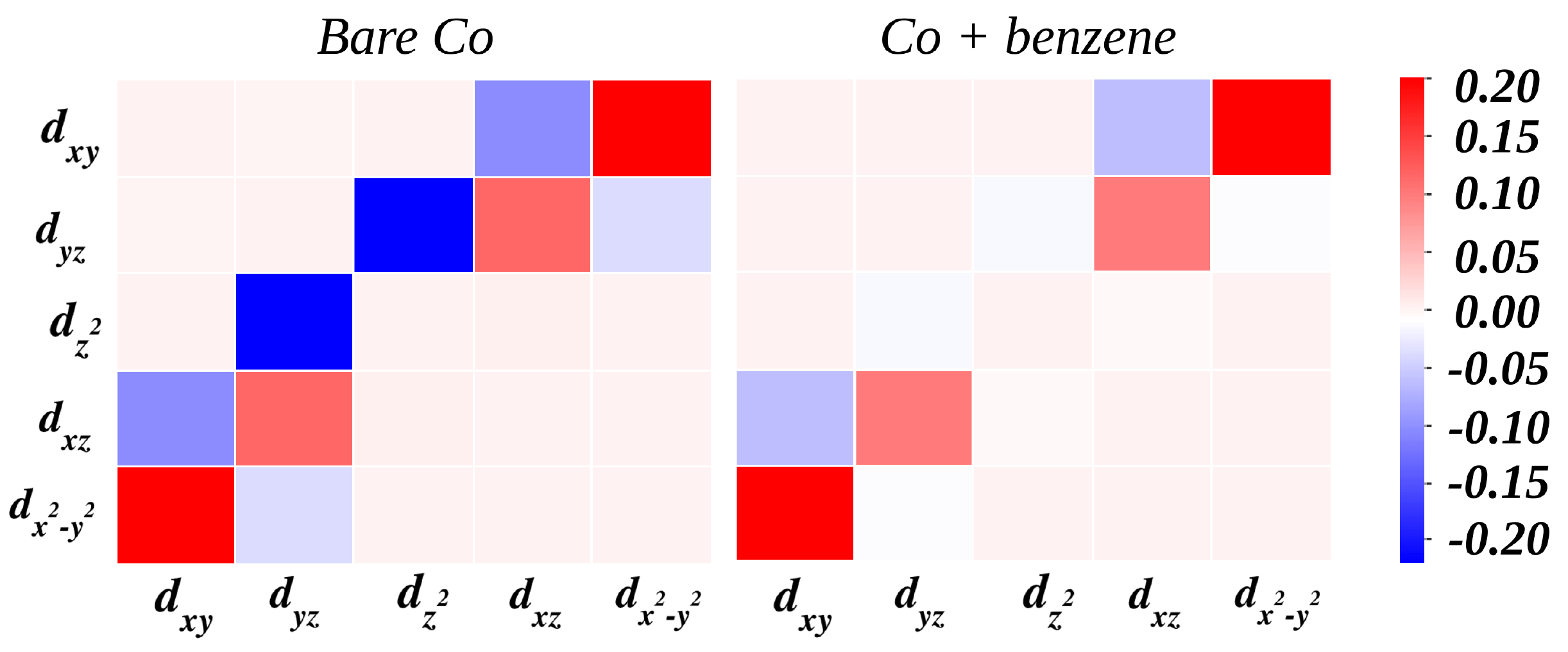}
\caption{In-plane versus out-of-plane difference $\Delta E^{I,mm'}_\mathrm{SOC}= E^{I,mm'}_{\mathrm{SOC},\parallel}-E^{I,mm'}_{\mathrm{SOC},\perp}$ (in meV) for the $3d$ orbitals of the Co atom 36, whose DOS is shown in Fig.~\ref{fig2.5layers}.}
\label{fig.SOC_matrix}
\end{figure}

The drastic MCA reduction induced by Bz adsorption might, at first thought,
be attributed to a complete suppression of the MCA of all interfacial Co layers. This is however not the case. The physics is more complex, as we shall understand through a careful microscopic analysis.
Specifically, we compute $\Delta E^I_\mathrm{SOC}$ defined in Eq. (\ref{eq.atomic_MAE}), which is proportional to the atom-resolved MCA energy, $E^I_\mathrm{MCA}$. The results are plotted in Fig. \ref{fig2.5layers}(d) as a function of the atom index. We label the Co atoms at the bottom (top) slab surface as 1-8 (32-40). 
In the bare Co slab (red dots), $\Delta E^I_\mathrm{SOC}$ is negative for all atoms and the largest values ($\approx 0.33$ meV) are found at the surface, as we expect for a fcc slab. In the distorted slab (orange dots), these values get slightly enhanced (reduced) at the top (bottom) surface but remain negative. 
In contrast, in the Bz/Co slab (black dots), the changes with respect to the bare Co slab become dramatic. Now, $\Delta E^I_\mathrm{SOC}$ is positive and increases in magnitude ($\approx 0.48$ meV) for those Co atoms, which bind to the molecule, and also for some of the atoms in the second layer underneath the interface [cf. Fig \ref{fig2.5layers}(b)]. 
Hence, we understand that the reduction of the slab MCA upon Bz adsorption is due to a balance between positive and negative contributions to $E_\mathrm{MCA}$, which respectively come from the Co atoms hybridized with the molecule and the rest of the Co atoms. Notably, if we consider only the \textit{surface} MCA energy estimated by summing the $\Delta E^I_\mathrm{SOC}$ contributions for the Co atoms at the slab top surface, we will find that this quantity switches from negative to positive after molecular adsorption. The magnetization of the interfacial Co layer would more favorably point along the perpendicular direction, while the magnetization of rest of the slab prefers to lay in-plane, as in the bare Co slab. In practice, however, because of the exchange interaction between the surface and the other layers, which is much stronger than the MCA, the slab will behave as a system with a MCA resulting from the average of the atomic contributions.\\

The effect of the molecule-Co hybridization and, in particular, the positive sign of the MCA energy of those Co atoms that bind to the molecule, are the same, qualitatively, as predicted by the elementary model. This can be seen in detail, by analysing the SOC matrix elements $E^{I,mm'}_\mathrm{SOC}$ of Eq. (\ref{eq.matrixSOC}), which describe transition amplitudes between two Co $3d$ orbitals (labelled through the quantum numbers $m$ and $m'$ as for Ref. [\onlinecite{vaspOrbitals}]). In particular, we here take the difference $\Delta E^{I,mm'}_\mathrm{SOC}= E^{I,mm'}_{\mathrm{SOC},\parallel}-E^{I,mm'}_{\mathrm{SOC},\perp}$ between the matrix elements calculated with in-plane and out-of-plane magnetization. We then proceed along the lines of Sec. \ref{sec.model} investigating what transitions gets suppressed because of the hybridization. Since $\sum_{m,m'}\Delta E^{I,mm'}_\mathrm{SOC}=\Delta E^I_\mathrm{SOC}$ (see Sec. \ref{Sec.Details}), positive (negative) values of $\Delta E^{I,mm'}_\mathrm{SOC}$ for a Co atom $I$ correspond to out-of-plane (in-plane) contributions to the MCA of that atom.\\  

Fig. \ref{fig.SOC_matrix} displays the values of 
$\Delta E^{I,mm'}_\mathrm{SOC}$ for the various combinations of the $3d$ orbitals of the atom 36, which is strongly hybridized with Bz and whose DOS was discussed above. Positive (negative) values are in red (blue). In the case of the bare Co surface, $\Delta E^{I,mm'}_\mathrm{SOC}$ is non-zero and positive for the transition between $d_{xy}$ and $d_{x^2-y^2}$, whereas it is negative for the transitions connecting $d_{xy}$ to $d_{xz}$ and $d_{yz}$ to $d_{z^2}$ or $d_{x^2-y^2}$. These latter transitions overall dominate, favouring in-plane MCA. In contrast, in the case of the Bz-Co interface $\Delta E^{I,mm'}_\mathrm{SOC}$ completely vanishes for the transition connecting $d_{yz}$ to $d_{z^2}$. 
The remaining dominant contribution to $\Delta E^I_\mathrm{SOC}$ is positive, due to the $d_{xy}$ to $ d_{x^2-y^2}$ transition, which remains uncompensated by any negative contributions. Consequently, the out-of-plane MCA is favoured over the in-plane as predicted by the elementary model in the limit of large $\Delta'$.\\

\begin{figure}[h]
\centering\includegraphics[width=\columnwidth,clip=true]{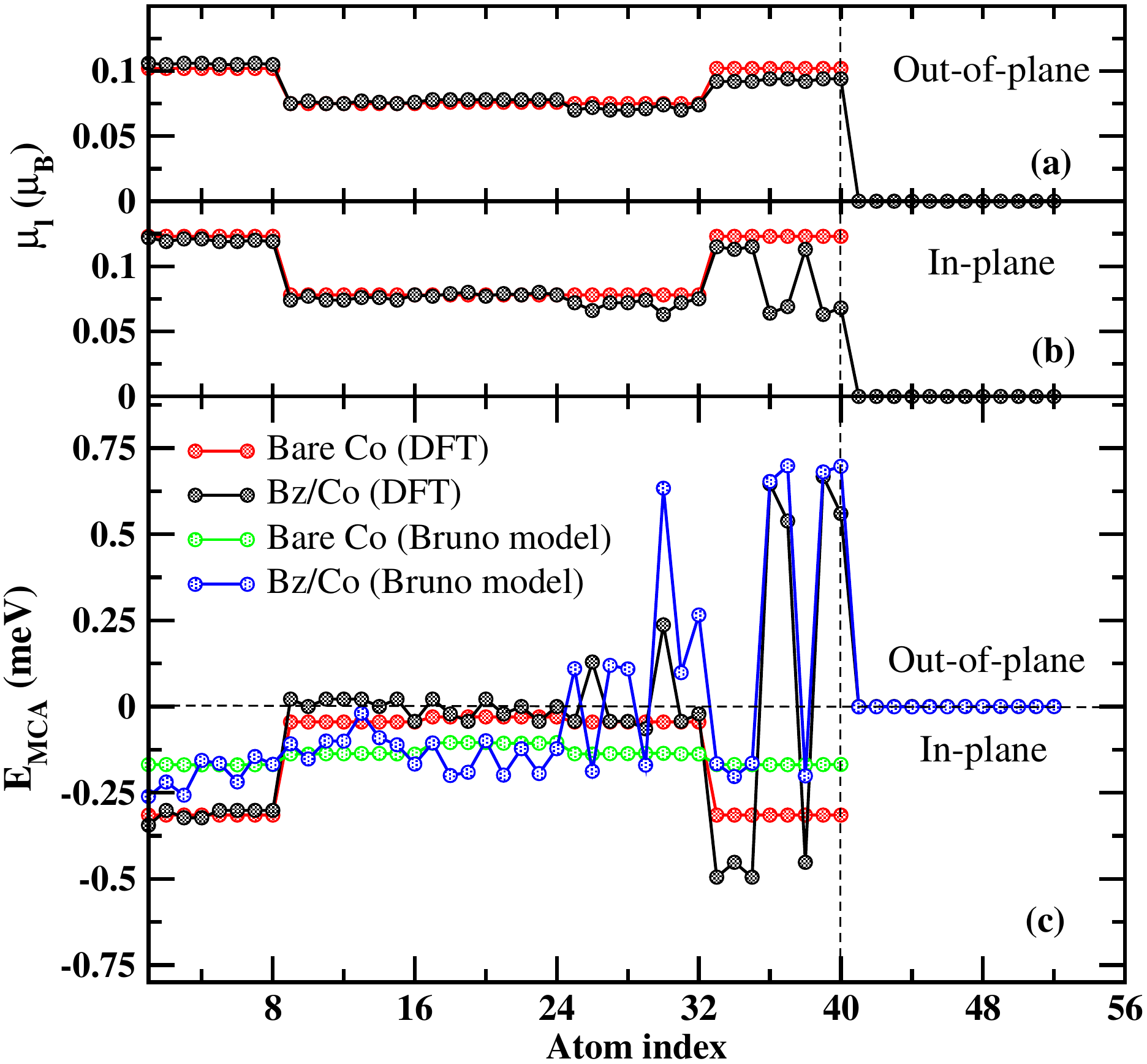}
\caption{Orbital magnetic moments and atomic MCA energy $E^I_\mathrm{MCA}$. (a) $\mu^I_{l,\perp}$ and (b) $\mu^I_{l,\parallel}$ for all atoms. Atoms 1-8 are at the slab bottom surface, atoms 33-40 are at the top surface, and atoms 41-52 belong to Bz. The red (black) dots represent the values for the bare Co (Bz/Co) slab. (c) Atom resolved MCA energy calculated by DFT and compared to the values obtained from Bruno's model, i.e., $E^I_\mathrm{MCA}=- \frac{\xi\hbar}{4 \mu_B}( \mu^I_{l,\parallel}- \mu^I_{l,\perp})$. (Red: DFT for Co. Black: DFT for Bz/Co. Green: the Bruno's model for Co. Blue: the Bruno's model for Bz/Co).}
\label{fig3.bruno}
\end{figure}

\subsubsection{Orbital moments and Bruno's model}
The prediction that the MCA of Co slabs changes upon Bz adsorption has so far relied on an analysis of the SOC energies and of the SOC matrix elements. These quantities can be readily computed in theory, but are not measurable. Therefore, the mechanism for the MCA modification proposed here can not be verified in the lab. In order to overcome this limitation, here we show that the MCA is also related to the orbital magnetic moment of the Co atoms, $\mu^I_{l, \alpha}=\langle \hat L^I_\alpha\rangle\mu_B/\hbar$, where $\mu_B$ is the Bohr magneton, $I$ labels the atom, and $\alpha=x,y,z$. Notably, $\mu^I_{l,alpha}$ is an observable directly accessible by experiments\cite{STOHR1999470}, for example, through X-ray magnetic circular dichroism (XMCD) \cite{PhysRevLett.70.694}. Specifically, under the assumption that the majority spin band is fully occupied (valid for Co), Bruno showed in Ref. [\onlinecite{bruno}] that the perturbative expression of the MCA energy, Eq. (\ref{eq.E_MCA}), can be rewritten as
\begin{equation}
E_\mathrm{MCA} 
= - \frac{\xi\hbar}{4 \mu_B}\sum_I( \mu^I_{l,\parallel}- \mu^I_{l,\perp})\:,
\label{bruno}
\end{equation}
where $\mu_{l,\parallel (\perp)}$ denotes the component of orbital moment of atom $I$ along the in-plane (out-of-plane) spin magnetization direction. 
Effectively, Bruno's model indicates that the easy axis of magnetization coincides with the direction of maximum orbital magnetic moment determined by the bonding and the crystal field\cite{STOHR1999470}.\\

The calculated $\mu^I_{l,\parallel}$ and $ \mu^I_{l,\perp}$ are depicted in Fig. \ref{fig3.bruno}(a) and \ref{fig3.bruno}(b), respectively. In the middle of the slab (atoms 8-32), we see that $\mu^I_{l,\parallel}\approx \mu^I_{l,\perp}$ and
the values are comparable to the atomic orbital magnetic moment of bulk Co, $0.075\mu_B$ (Ref. [\onlinecite{PhysRevLett.75.2871}]), excluding orbital polarization \cite{PhysRevB.81.060409}. In contrast, at the bare Co surface layers, the atomic orbital moments get substantially enhanced as reported in previous studies \cite{PhysRevLett.75.1602}, and, moreover, $\mu^I_{l,\parallel}$ ($\approx 0.125 \mu_B$) becomes larger than $\mu^I_{l,\perp}$ ($\approx 0.1 \mu_B$) and approximately twice the Co bulk value. These results are further drastically modified when we consider the interface with the molecule. In this case, $\mu^I_{l,\perp}$ ($\approx 0.095 \mu_B$) remains somewhat similar to the corresponding value for the bare Co surface, while $\mu^I_{l,\parallel}$ is drastically reduced to about $0.06 \mu_B$ for the Co atoms bonded to the molecule. As a consequence, by using Eq. (\ref{bruno}), we expect that the surface MCA energy will be negative, thus favouring in-plane MCA, for bare Co surfaces, whereas it will be positive, thus favouring out-of-plane MCA, for the Bz-Co interface. This is indeed what we find.\\

Fig. \ref{fig3.bruno}(c) compares the layer-resolved MCA energy discussed in the previous section with the results obtained by using Bruno's model with the Co atomic SOC value $\xi= 86$ meV\cite{gi.ca.12,STOHR1995253}. For the bare Co surface, the result from the model (red curve) and the DFT calculations (green curve) follow the same pattern, despite some quantitative differences. We see that the largest contribution to the MCA energy is due to surface atoms with enhanced orbital moments. The quantitative discrepancies between the DFT results and Bruno's model are consistent with previous reports for metallic slabs\cite{Li-Fe}. In contrast, in the case of the Bz-Co interface, we find a remarkable agreement between the model and the DFT atomic MCA energy (blue) for those atoms bonded to the molecule. This demonstrates that the modulation of the surface layer MCA can indeed be explained in terms of variations of the Co atoms' orbital moments induced by hybridization.\\

The reduction of the in-plane orbital magnetic moments is not a feature specific of the Bz-Co interface but is a rather general effect at interfaces between ferromagnetic transition metals and conjugated molecules, as we will further discuss in the following sections (see Sec. \ref{sec.alq3_and_c60}). It can be understood in a rather simple way by extending a qualitative argument used in Ref. [\onlinecite{STOHR1999470}] to explain orbital moment quenching at surfaces. The orbital moment of a free $d$ electron is perpendicular to the motion of the electron. For a thin-film geometry, the in-plane electron motion will be disrupted due to the Coulomb repulsion from the neighboring atoms. In contrast, the electron motion perpendicular to the surface will be less perturbed due to the loss of neighbors. Hence, the orbital moment normal to the plane gets partially quenched, while that along the plane does not. Upon the adsorption of the molecule on the top of the slab, the orbital motion perpendicular to the surface will be disrupted due to the bonding of the surface atoms with the molecule. As a consequence, the in-plane orbital magnetic moment gets reduced. This is a remarkable effect, which, to our knowledge, has not been explicitly discussed in the literature about molecule-Co interfaces. It can be measured via XMCD experiments\cite{STOHR1995253}, which are common for the interface characterization \cite{Wende2007,zh.ho.10,mi.he.11,he.ta.13,sh.su.14,ja.le.15}, and it does represent a fingerprint of the $p_z$-$d_{z^2}$ hybridization.\\

To date, experiments dedicated to the magnetic anisotropy at hybrid molecule/Co interfaces have relied on measurements of magnetization hysteresis loops of molecule-covered thin films\cite{bairagiprl,bairagi_2nd,be.al.21}. However, hysteresis loops may be affected by many factors besides the MCA. For example, at the electronic level, an important contribution may come from the modification of the Co intra- and inter-layer exchange interaction induced by the bonding with the molecules and by the resulting surface relaxation, as suggested in Refs. \onlinecite{blugel_hard,fr.ca.15,fr.ca.15_2}. Measurements of the orbital moments, as proposed here, will provide a definite way to isolate the MCA contribution from others that do not depend on the SOC. The results will therefore drive a considerable progress towards understanding the magnetic properties of hybrid molecule-metal thin films.

\begin{figure}
\centering\includegraphics[width=\columnwidth,clip=true]{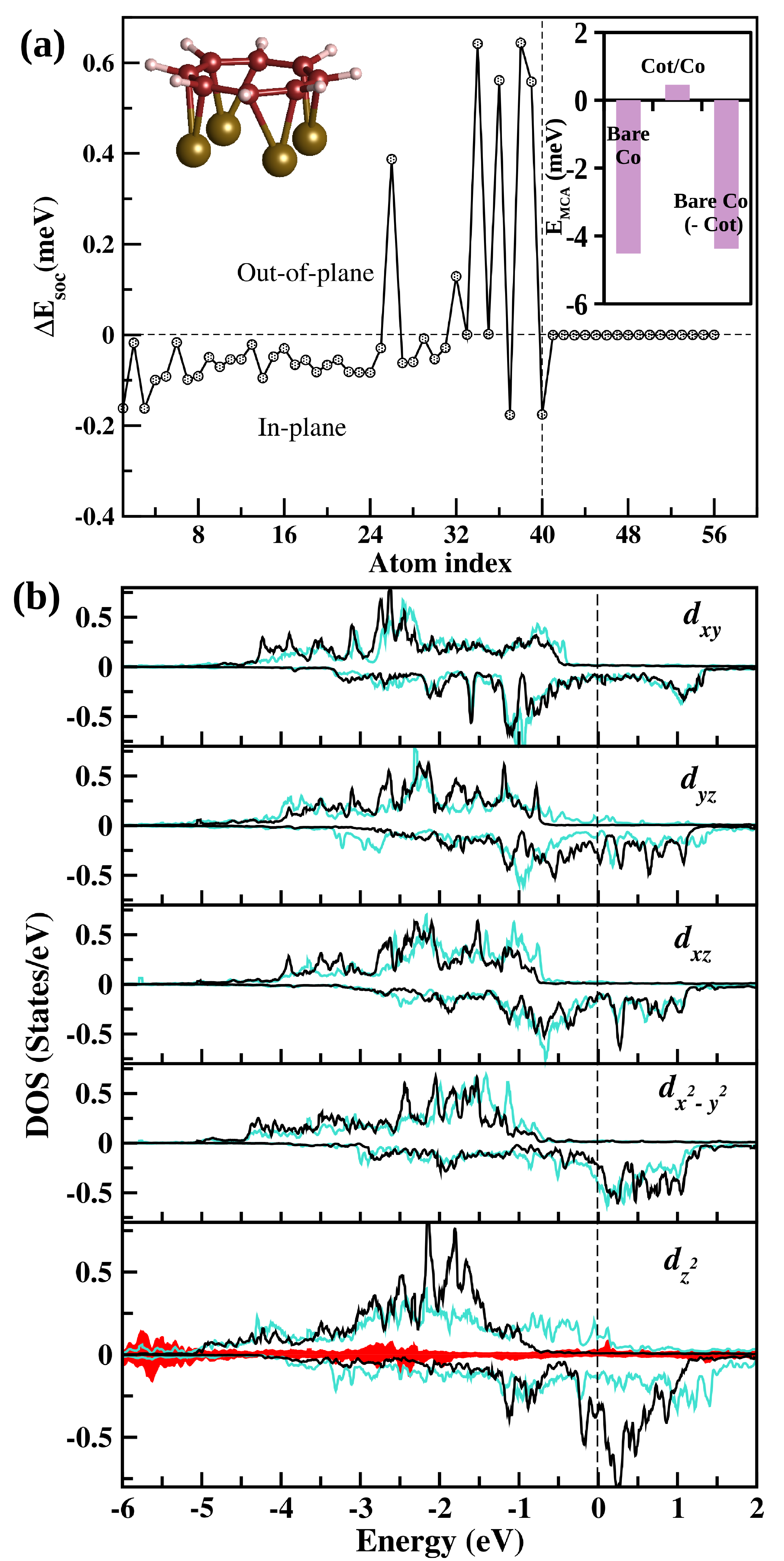}
\caption{MCA at the Cot/Co interface. (a) $\Delta E^I_\mathrm{SOC}$ evaluated for all atoms. Atoms 1-8 are at the slab bottom surface, atoms 33-40 are at the top surface, and atoms 41-56 belong to Cot. The left-hand side inset shows the molecule and the Co atoms to which it binds to. The right-hand side one shows the MCA energy of the bare Co slab (same as in Fig. \ref{fig2.5layers}), of the Cot/Co slab, and of the ``distorted'' slab (calculated by removing the Cot molecules from the optimized  Cot/Co slab). (b) DOS projected over the $3d$ orbitals of a surface Co atom bonded to Cot. The Fermi energy is at 0 eV. The black and cyan lines are for the bare Co slab and the bare Cot/Co slab, respectively. The red curve in the bottom panel is the DOS projected over the C $p_{z}$ orbital of Cot.}
\label{fig.cot}
\end{figure}

\subsection{Interface between cyclooctatetraene and Co}\label{sec.cot}

Having understood the mechanism leading to the MCA modification at conjugated molecule-Co interfaces, we now investigate how the effect can be further controlled by selectively changing the molecular chemical properties. Thus, we consider the same Co slab as in the previous section, but we replace Bz with cyclooctatetraene (Cot), C$_8$H$_8$. This is a more reactive molecule, which binds strongly to $3d$ transition metal surfaces and thus results in a larger hybridization\cite{at.br.10}. The adsorption geometry is shown in the inset of Fig. \ref{fig.cot}(a), where the molecule appears in a hollow position with four C-C bond on top of Co atoms. Like Bz, Cot adopts a nonplanar structure with the H atoms situated slightly above the C ones, and with an average C-Co bond length of 2.07 \AA. \\

Fig. \ref{fig.cot}(b) depicts the DOS projected over the $3d$ orbitals of a Co atom at the bare Co surface (black line) and of a Co atom underneath a Cot C-C bond (cyan line). In the bottom panel, which is dedicated to the $d_{z^2}$ orbital, we also include the DOS projected over the C $p_z$ orbital of Cot for a more complete analysis of the interface electronic structure. We see that the results for the bare surface are the same as in Fig. \ref{fig2.5layers}(c), but the DOS of the Co atoms bonded to the molecule is significantly altered. We clearly recognize the formation of the $p_z$-$d_{z^2}$ bonding and anti-bonding states like in the Bz/Co case. However, since Cot is more reactive than Bz and, therefore, more hybridized with Co, the $p_z$-$d_{z^2}$ antibonding state is greatly broadened and its center is moved towards higher energies compared to the Bz/Co case. On the one hand, the majority-spin DOS extends slightly across the Fermi energy. On the other hand, the minority spin $d_{z^2}$ DOS becomes negligible near the Fermi level, and the most distinct peaks are in the energy region from about 0.8 eV to 1.5 eV above the Fermi level. Thus, the energy splitting between the $d_{z^2}$ orbital and the other Co $3d$ orbitals increases after Cot adsorption. This is in good qualitative agreement with the elementary model of Sec. \ref{sec.model} and, referring to Fig. \ref{fig.scheme}, we estimate that $\Delta'$ is as large as $\sim2$ eV at the Cot/Co interface. As a consequence, the SOC-induced transition from $d_{yz}$ to $d_{z^2}$, which promotes in-plane MCA at the clean Co surface, is entirely suppressed because of the hybridization with the molecule. \\

The atom-resolved MCA estimated through $\Delta E^I_\mathrm{SOC}$ is negative for all atoms in the bare Co slab, whereas it becomes positive for the Co atoms directly bonded to the molecule. As for Bz/Co, these atoms would find it energetically favourable to align their magnetic moments along the perpendicular direction. Yet, this effect is quantitatively larger here than that in Bz/Co. The value of $\Delta E^I_\mathrm{SOC}$ of the hybridized Co atoms is $0.65$ meV for Cot/Co \textit{versus} $0.48$ meV for Bz/Co. Furthermore, although the number of these hybridized atoms at the Cot/Co interface is the same as in Bz/Co, their overall contribution is so important for Cot that the MCA of the entire slab switches from in-plane to out-of-plane. In fact, the total slab MCA energy goes from $-5.57$ meV to $0.71$ meV after molecular adsorption [see inset in Fig. \ref{fig.cot}(a)]. \\

The results for Cot/Co and their comparison with those for Bz/Co demonstrate the dramatic impact that the chemical properties of the adsorbate molecules can potentially have on the Co slab. Our predictions may be readily verified experimentally. More importantly, we hope that they will stimulate additional experimental studies, where the perpendicular MCA of molecular covered Co thin films is varied, and eventually maximized, by chemically tuning the reactivity of the molecules with the film surfaces. 

\begin{figure}[h]
\centering\includegraphics[width=\columnwidth,clip=true]{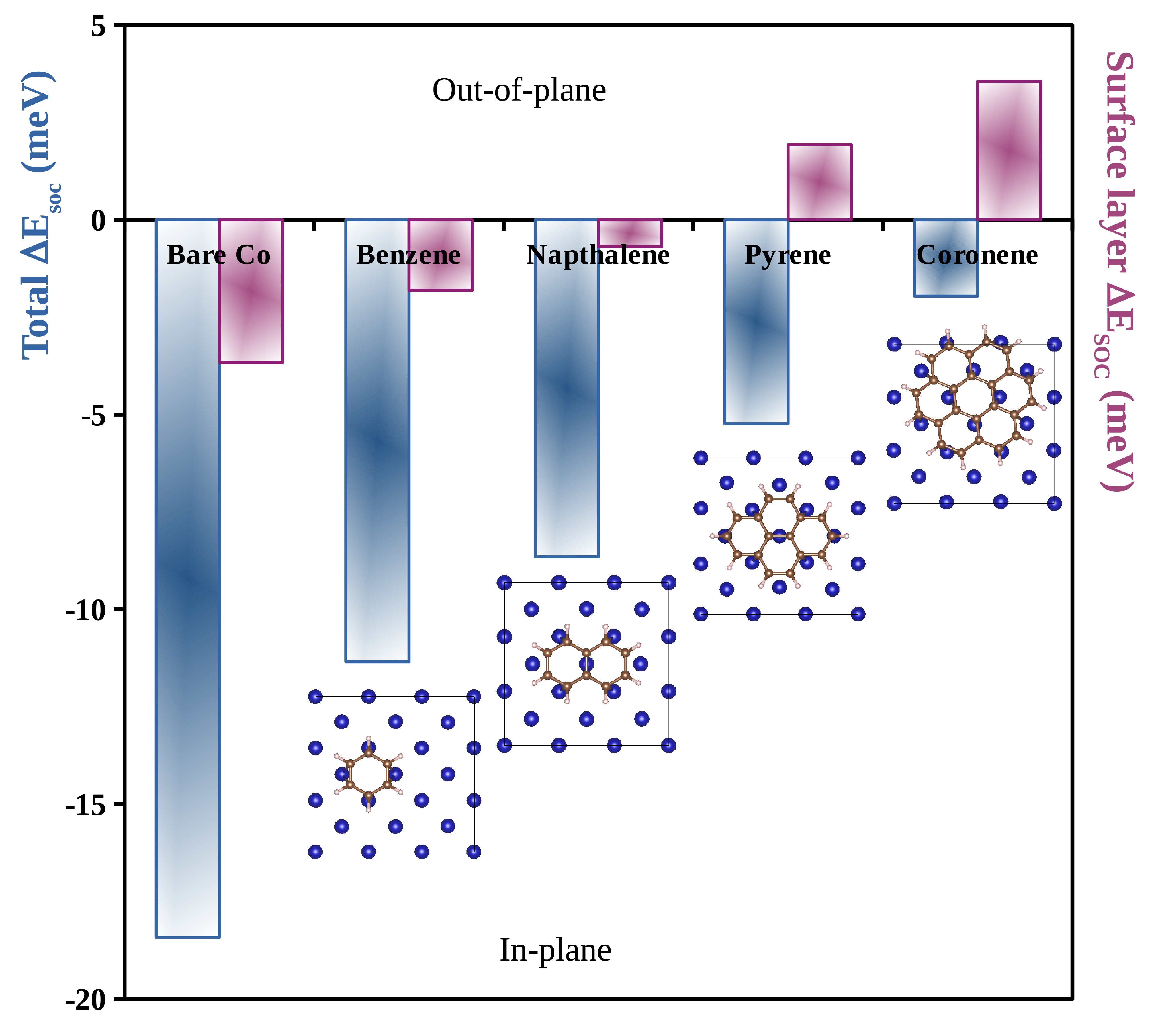}
\caption{MCA modification for a slab covered with a series of aromatic hydrocarbon molecules. The total (surface) MCA energies, represented by the blue (magenta) bars, are estimated by the sum of $\Delta E^I_\mathrm{SOC}$ for all the slab (top surface) atoms.}
\label{fig.coverage}
\end{figure}

\subsection{Dependence of the MCA on the molecular coverage} \label{sec.coverage}

The calculations presented in the previous sections have demonstrated that the MCA modification at hybrid interfaces is mostly a ``local'' phenomenon at the metal atoms forming covalent bonds with the adsorbed conjugated molecules. Therefore, we expect that the effect will be further enhanced by increasing the surface molecular converge, namely the number of C-metal bonds across a molecule/Co interface. This is indeed the case as we now demonstrate. \\

We consider a Co slab, where the top surface is covered with various aromatic hydrocarbon molecules comprising a different number of fused Bz rings: naphthalene (2 Bz rings), pyrene (4 Bz rings) and coronene (6 Bz rings). The slab total MCA energy as well as the top surface MCA energy estimated by the SOC energy $\Delta E_\mathrm{SOC}$, are compared in Fig. \ref{fig.coverage} (note that the slab in-plane unit cell is now larger than the one used in the previous sections). The trend emerging from the figure is clear. The total MCA energy has the largest negative value for the bare Co slab. The absolute magnitude of the MCA energy then decreases as a function of the number of Bz rings in the molecular adsorbate. This effect appears even more dramatic when we inspect only the relative variation of the top surface MCA energy, estimated by summing the $\Delta E^I_\mathrm{SOC}$ contributions over the top surface layer atoms. The sign of such surface MCA switches from negative to positive for pyrene, and the magnitude becomes quite large in the case of the coronone/Co interface, where most of the surface Co atoms form a covalent bond with the molecule. This is a sharp trend, which could be easily verified experimentally, although we are not aware of any study for the hydrocarbon series. We note, however, that a perpendicular surface MCA has been reported for graphene-coated Co films\cite{graphene-Co}. The physics for such system is evidently the same as for coronene/Co, and our phenomenological arguments can be used to interpret those experimental results as well. Overall, our findings demonstrate the possibility to engineer thin films with large out-of-plane MCAs via surface molecular functionalization.   

\begin{figure}[h]
\centering\includegraphics[width=\columnwidth,clip=true]{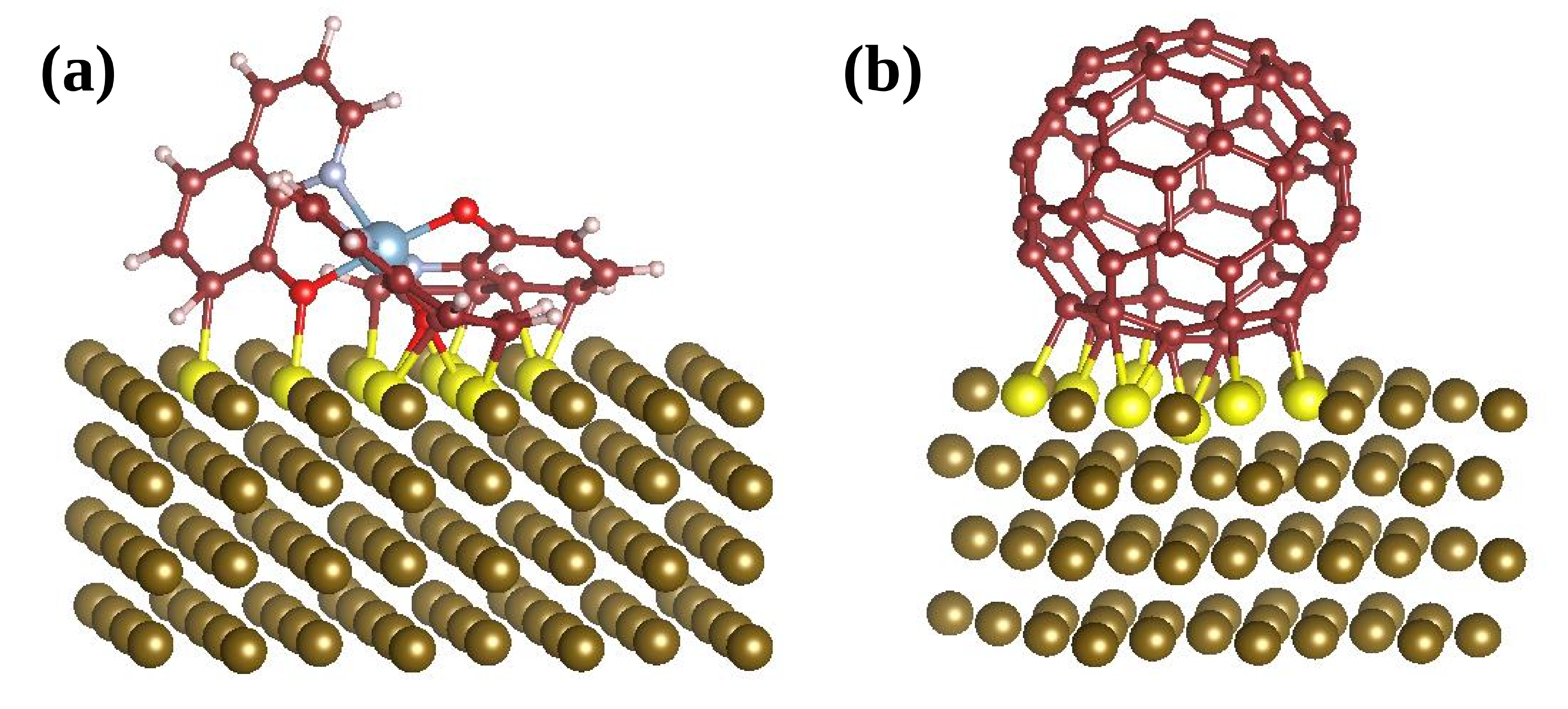}
\caption{Optimized geometry of (a) Alq$_{3}$ and (b) C$_{60}$ deposited on top of a 4-layer Co slab. Yellow, maroon, blue, red, and light pink spheres represent Co, C, Al, O and H atoms, respectively. }
\label{fig7.struc}
\end{figure}

\subsection{Interfaces with C$_{60}$ and Alq$_3$} \label{sec.alq3_and_c60}

The results obtained so far provide a detailed understanding of the mechanism leading to the MCA modifications of Co surfaces upon molecular adsorption. Furthermore, they suggest some potential routes to enhance the effect. Now, we consider two of the most experimentally studied molecule-metal interfaces for spintronics, namely Co/C$_{60}$ and Co/Alq$_{3}$, showing that the same concepts learnt so far, apply also these rather complex systems.   \\

\begin{figure}[h]
\centering\includegraphics[width=\columnwidth,clip=true]{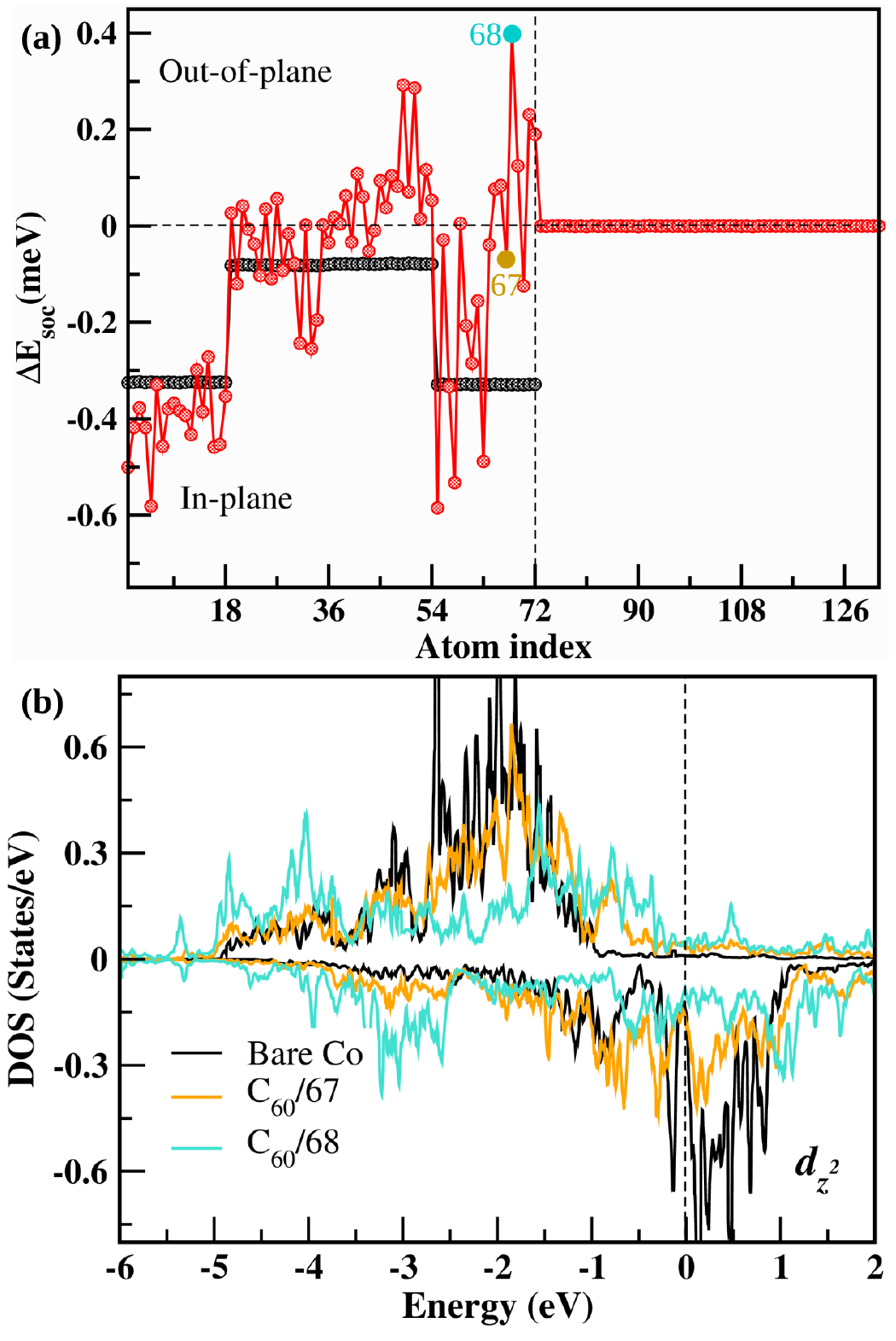}
\caption{MCA at the C$_{60}$/Co interface. (a) $\Delta E^I_\mathrm{SOC}$ evaluated for all atoms. Atoms 1-18 are at the slab bottom surface, atoms 55-72 are at the top surface, and atoms 73-132 belong to C$_{60}$. (b) DOS projected over a surface Co atom in the bare Co slab (black line), and DOS projected over the $3d$ orbitals of the Co atoms $68$ (cyan line) and $67$ (orange line) in the C$_{60}$/Co slab. The Fermi energy is at 0 eV.}
\label{fig8.C60}
\end{figure}

\subsubsection{C$_{60}$/Co interface}

Co thin films with a C$_{60}$ overlayer are the first systems for which an enhanced anisotropy was experimentally reported \cite{bairagiprl,bairagi_2nd}. 
We model the interface through a 3$\times$3 in-plane slab with one molecule in contact with the top surface. We consider only 4 Co layers to reduce the computational cost, since the system is considerably larger than Bz/Co and Cot/Co. Similarly to Ref. [\onlinecite{mo.wh.14}], we compare four different adsorption geometries named hexagon, pentagon, hexagon-hexagon, and hexagon-pentagon. In the first two cases, C$_{60}$ has either a hexagonal or pentagonal face on top of the Co surface. In the hexagon-hexagon case, the molecule bonds to Co through the edge between two hexagonal faces, whereas in the hexagon-pentagon geometry, it bonds through the edge between one hexagonal and one pentagonal face. The lowest energy configuration is the hexagon-pentagon one shown in Fig. \ref{fig7.struc}(b). 
Its energy is 335 meV lower than the energy of the least stable configuration, pentagon, and about 200 meV lower than the energy of the second most stable configuration, hexagon. 
Similar results were also obtained in Ref. [\onlinecite{bairagiprl}] (for hcp Co slabs) and in Ref. [\onlinecite{bairagi_2nd}] (for fcc Co slabs). In the following, we will only discuss the results for the hexagon-pentagon configuration. Although the details of the C$_{60}$/Co electronic structure depend on the adsorption geometry, we find that the overall physics is similar for all cases.\\

The Co surface MCA is strongly affected by C$_{60}$. The $\Delta E^I_\mathrm{SOC}$ is plotted in Fig. \ref{fig8.C60} for all the atoms in the bare Co and in the C$_{60}$/Co slabs (black and red points respectively). The values for the Co atoms, which form a strong covalent bond with the molecule, change sign, switching from negative in the bare Co case to positive in the C$_{60}$/Co case. Specifically, $\Delta E^I_\mathrm{SOC}$
has the largest positive value ($\approx 0.4$ meV) for the Co atom, labelled 68, which is closer to, and therefore more strongly hybridized with, C$_{60}$. For the other Co atoms underneath the molecule, $\Delta E^I_\mathrm{SOC}$ has somewhat smaller values, varying with the Co-C bond length. Finally, $\Delta E^I_\mathrm{SOC}$ remains negative for the surface atoms that are not bonded at all with C$_{60}$. Overall, the order of magnitude of the atomic MCA energy $E^I_\mathrm{MCA}\approx 0.5 \Delta E^I_\mathrm{SOC}$ of the Co atoms hybridized with the molecules is similar to that obtained by Bairagi \emph{et al.} in Fig. 3(c) of Ref. [\onlinecite{bairagiprl}], although our results are for fcc Co(001), whereas Ref. [\onlinecite{bairagiprl}] considers hcp Co(0001). The MCA energy averaged over the interfacial layer is of the order of 0.1 meV$/$atom, which was found consistent with the experimental measurement of Ref. [\onlinecite{bairagiprl}]. \\

The relation between the Co-C hybridization and the MCA energy can be understood by examining the C$_{60}$/Co electronic structure and following the very same reasoning as in the previous sections. For instance, Fig. \ref{fig8.C60}(b) compares the $d_{z^2}$-projected DOS of two Co surface atoms bonded to the molecule, namely atom 68 that, as noted above, has the largest positive atomic MCA energy, and atom 67 that has instead a negative and considerably smaller (in absolute value) MCA energy. 
The DOS of the Co atom 68 is much broader than that of atom 67, indicating a much stronger chemisorption. For atom 68 in the minority spin channel, the C-Co bonding and antibonding states are recognized respectively at about -3 eV below the Fermi level and 1 eV above the Fermi energy. The picture overall resembles the $d_{z^2}$-projected DOS for Cot in Sec. \ref{sec.cot}. Accordingly, we find an enhanced positive MCA energy. Conversely, for atom 67, the formation of the bonding and antibonding states is much less marked and, in the minority channel, the $d_{z^2}$-projected DOS is large near the Fermi level. Hence, the MCA energy remains negative. Even for this complex interface, the overall picture for the MCA modification is fully consistent with that developed starting from our elementary model.\\

\begin{figure}[h]
\centering\includegraphics[width=\columnwidth,clip=true]{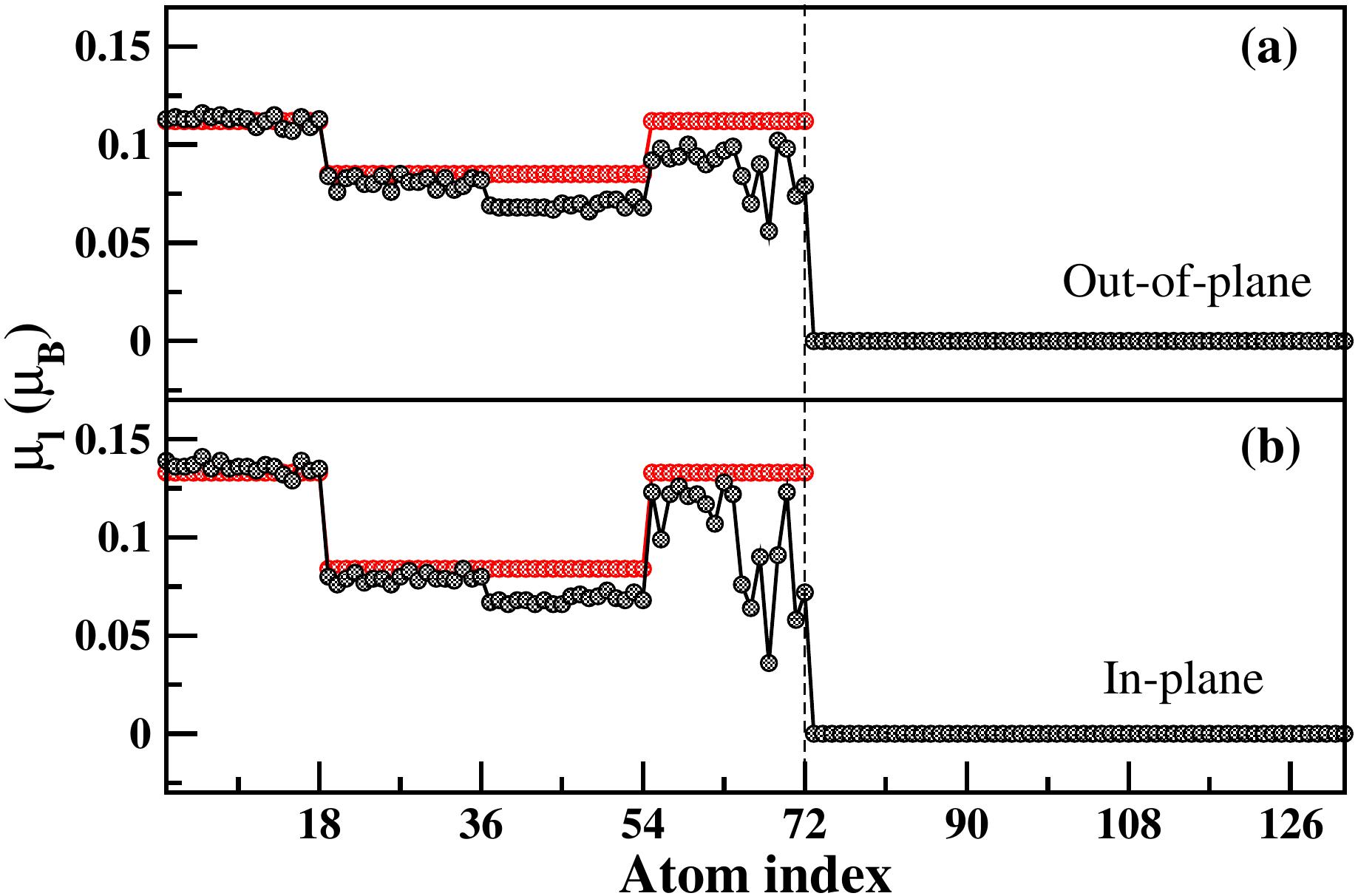}
\caption{Orbital magnetic moments (a) $\mu^I_{l,\perp}$ and (b) $\mu^I_{l,\parallel}$ for all atoms in the Co/C$_{60}$ slab. The red (black) dots represent the values for the bare Co (C$_{60}$/Co) slab. Atoms 1-18 are at the slab bottom surface, atoms 55-72 are at the top surface, and atoms 73-132 belong to C$_{60}$. }
\label{C60_orb}
\end{figure}

The impact of the $p_z$-$d_{z^2}$ hybridization can be further appreciated by noting that it induces a substantial spin redistribution in the DOS. In particular, the spin splitting of the Co $3d$ states and, hence, also the Co spin magnetic moments $\mu^I_s$, are reduced at the interface. The calculated $\mu^I_s$ is equal to $1.895\mu_B$ for the surface atoms in the bare slab. In contrast, $\mu^I_s$ becomes equal to 1.59 $\mu_B$ and 1.25 $\mu_B$, respectively, for atoms 67 and 68 at the C$_{60}$/Co interface. A similar reduction has been obtained in other theoretical studies\cite{bairagiprl} and measured experimentally\cite{mo.wh.14}.  \\  

The atomic MCA or $\Delta E_\mathrm{SOC}^I$ are not quantities accessible in experiments, as already observed in Sec. \ref{subsec.ben}. Yet, like in the case of Bz, they are directly correlated with the orbital magnetic moment, a measurable microscopic observable. In Fig. \ref{C60_orb} we compare the orbital magnetic moment of each atom for the bare Co and the C$_{60}$/Co slab with in-plane and out-of-plane spin magnetization, $\mu^I_{l,\parallel}$ and $ \mu^I_{l,\perp}$. The first and most striking finding is that $\mu^I_{l,\parallel}$ and $ \mu^I_{l,\perp}$ are drastically reduced for those Co atoms that are strongly bonded to C$_{60}$. For example, atom $68$ is found to have the lowest orbital magnetic moment among all Co atoms. The atomic orbital magnetic moment, therefore, shows a similar trend to the spin magnetic moment (although the former is one order of magnitude smaller than the latter). The hybridization tends to quench both of them. Despite that, as anticipated, the difference $\Delta \mu^I_l=\mu^I_{l,\parallel}- \mu^I_{l,\perp}$ is approximately proportional to $\Delta E_\mathrm{SOC}^I$ at the surface. $\Delta \mu_l$ is negative and equal to approximately $-0.02 \mu_B$ for all Co atoms of the bare surface. In contrast, upon molecular absorption, it switches to positive values ranging from about $0.005\mu_B$ to $0.02\mu_B$ for those interface atoms that are more strongly hybridized with the molecule. The measurement of such effect by means, for instance, of XMCD will provide a strong validation for the proposed mechanism for the MCA modification at C$_{60}$/Co interfaces and complement the previous studied based on the fit of the magnetization hysteresis loops. 

\begin{figure}
\includegraphics[width=1.0\linewidth]{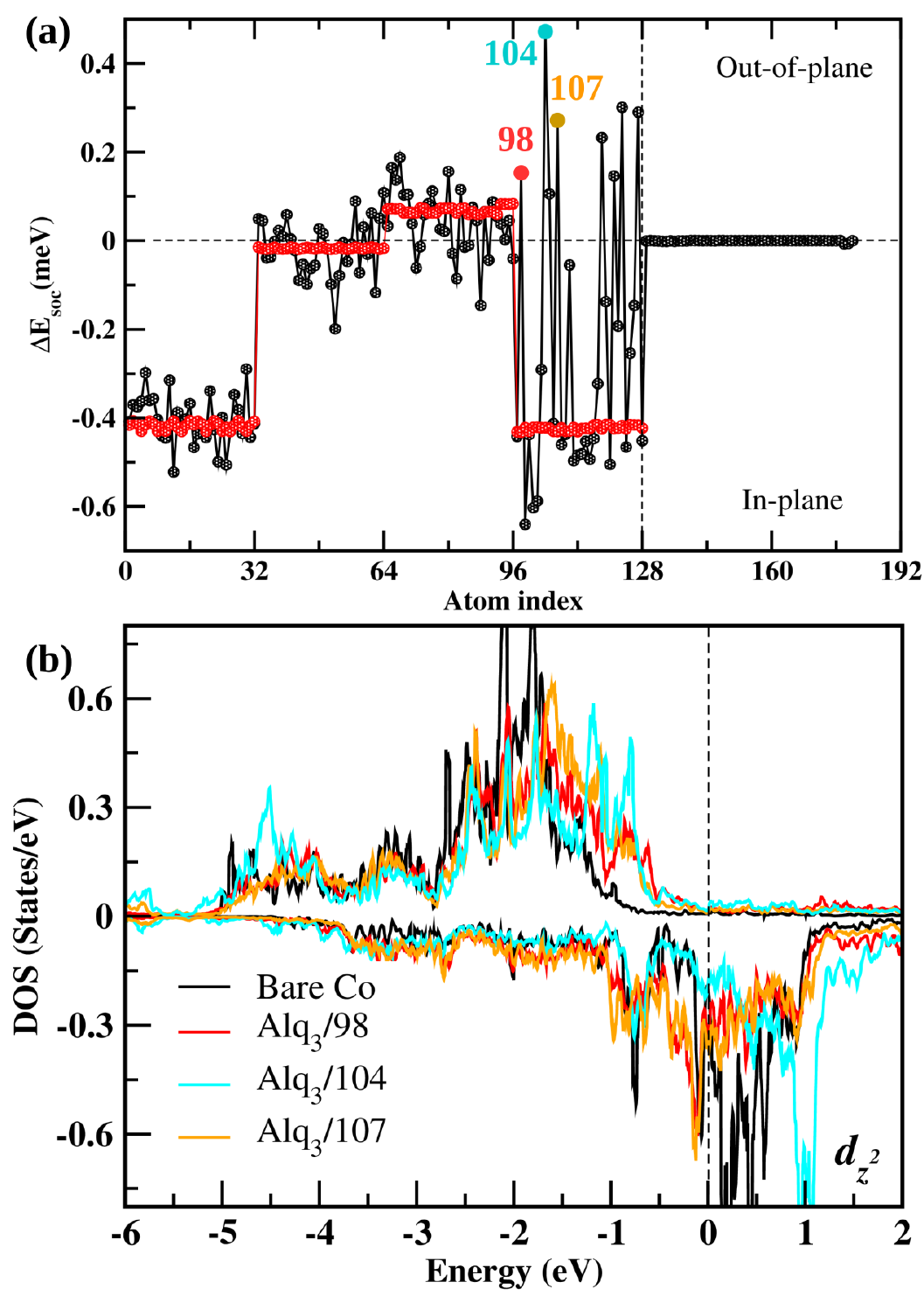}
\caption{MCA at the Alq$_3$/Co interface. (a) $\Delta E^I_\mathrm{SOC}$ evaluated for all atoms. Three Co atoms are shown as representative cases.
Atoms 1-32 are at the bottom surface of the slab, atoms 97-128 are at the top surface in contact with the molecule, and atoms 129-182 belong to the molecule. (b) DOS projected over the out-of-plane $3d$ orbitals for the three Co atoms highlighted in panel (a). }\label{fig.alq3}
\end{figure}

\subsubsection{Alq$_{3}$/Co interface}
 Alq$_3$ is the most popular molecular material investigated for spin transport \cite{xi.wu.04, de.hu.08, Galbiati2016}. Understanding the properties of its interface with Co films, used as electrodes in spin-valve devices, has been an outstanding problem for over a decade in organic spintronics\cite{Barraud2010,Dediu2009}.
 The molecule presents an Al atom octahedrally coordinated with the oxygen and nitrogen atoms of the three 8-hydroxyquinoline ligands. The electronic structure of the molecule\cite{cu.bo.98,ta.sa.10,bi.st.11,dr.ci.14, ad.20} and of its interface with Co\cite{dr.st.14,andrea-alq3} has been already studied in detail by means of DFT calculations in previous works, where results were also compared with spectroscopic measurements. However, to date, there are no first-principles calculations for the MCA of Alq$_3$/Co interfaces. Recent experiments \cite{bairagi_2nd} reported that Alq$_3$, similarly to C$_{60}$, on various Co thin films favours a perpendicular MCA. We confirm these observations here, showing that the underlying physics is indeed rather similar for Alq$_3$/Co and C$_{60}$/Co interfaces.\\
 
In order to model the Alq$_3$/Co interface, we adopt the slab described in Ref. [\onlinecite{andrea-alq3}] and shown in Fig. \ref{fig7.struc}(a) \footnote{The Co slab considered here is slightly different from that used in the other sections. In the case of Alq$_3$/Co, the slab is constrained in-plane to have the same lattice constant of Cu, so to mimic the experiments of Refs. \cite{dr.st.14,andrea-alq3}. Nonetheless this has a minor effect on the results and on the proposed mechanism for the MCA modification.}. 
Two of the molecule's quinoline ligands are chemisorbed almost parallel to the surface, whereas the third one stands vertically. 
Similarly to all the systems studied so far, we find that $\Delta E_\mathrm{SOC}^I$ becomes positive and, on average, equal to about 0.25 meV/atom for the surface Co atoms underneath the molecule, whereas $\Delta E_\mathrm{SOC}^I$ remains negative for the surface Co atoms not bonded with Alq$_3$. These results are displayed  in Fig. \ref{fig.alq3}(a), where the values for the clean Co slab and the Alq$_3$/Co one are respectively represented as black and red dots.\\

The different values of $\Delta E_\mathrm{SOC}^I$ for the various atoms at the interface reflect the strength of their hybridization with the molecule. Similarly to the other interfaces studied, the hybridization leads to the formation of Co-molecule antibonding states with $d_{z^2}$ character, which are sharply defined in particular in the minority spin channel. For instance, in Fig. \ref{fig.alq3}(b), we plot the orbital-projected DOS for three different Co atoms (red, orange, and cyan curve) and we compare it with the DOS of the clean Co surface (black curve). The $d_{z^2}$-projected DOS of the Co atom with largest $\Delta E_\mathrm{SOC}^I(\approx 0.48$ meV) shows a rather sharp resonance at about 1 eV above the Fermi energy in the minority spin channel [cyan curve in Fig. \ref{fig.alq3}(b)]. This feature drives the MCA switch, in qualitative agreement with the model of Sec. \ref{sec.model}. In contrast, the other Co atoms, which are slightly less hybridized with the molecule, present a smaller $\Delta E_\mathrm{SOC}^I$. The center of the $p_z$-$d_{z^2}$ antibonding state in the minority channel is near the Fermi level, and the magnitude of the MCA change is less pronounced. Overall, the results appear rather similar, even at the quantitative level, to those obtained in the case of C$_{60}$/Co, thus confirming the experimental observation of Ref. [\onlinecite{bairagi_2nd}] that the effect of an Alq$_3$ overlayer on Co ultrathin films is rather similar to that of a C$_{60}$ one.\\

Finally we suggest that, instead of comparing Alq$_3$/Co and C$_{60}$/Co, experiments may consider the tris-(9-oxidophenalenone)-aluminum(III) molecule, Al(OP)$_3$, a variation of Alq$_3$ with similar properties, but larger 9-oxidophenalenone ligands \cite{mu.st.13}. These ligands will form a large number of $p_z$-$d_{z^2}$ chemical bonds with Co as already demonstrated in spectroscopic measurements\cite{mu.st.13}, and this will potentially enhance the MCA modifications, as discussed in Sec. \ref{sec.coverage}. Experimental studies along this direction will provide a further confirmation of the phenomenology that we have described in this paper.

\section{Conclusion}\label{sec.conclusion}
We have investigated the modification of the MCA of Co slabs upon molecular adsorption, demonstrating that the hybridization between the out-of-plane Co $d_{z^2}$ orbitals and molecular C $p_z$ orbitals favours perpendicular MCA. \\

Starting from an elementary model, we have employed second-order perturbation theory to identify the virtual electronic transitions between Co occupied and unoccupied states that contribute to the in-plane and out-of-plane MCA. We have then shown that the virtual transitions leading to in-plane MCA are suppressed by the formation of molecule-metal bonding and antibonding states, while those leading to out-of-plane MCA are not. This explains why perpendicular MCA is favoured at molecule-Co interfaces. The key parameter governing the effect is the energy splitting between the in-plane Co $d_{x^2-y^2}$ orbital and the $p_z$-$d_{z^2}$ antibonding hybrid one.\\

The results obtained with the elementary model are confirmed by DFT calculations for several molecular systems. Specifically, we have shown that the electronic structure of the surface layer of Co slabs is significantly modified by the hybridization with C $p_z$ orbitals leading to a suppression of the in-plane contributions to the MCA. This effect is rather general at chemisorbed molecules/Co interfaces, and it can be further enhanced by selecting molecules that are very reactive with Co. Among the various systems studied, there are also the prototypical spintronic interfaces C$_{60}$/Co and Alq$_3$/Co. Our results agree with previous studies for C$_{60}$/Co\cite{bairagiprl} and they support recent experimental observations reporting a similar MCA modification in the two systems\cite{bairagi_2nd}. \\

Finally and importantly, we have demonstrated that the atomic MCA is correlated with the difference between the atomic orbital magnetic moment for in-plane and out-of-plane spin magnetization, in accordance with Bruno's model\cite{bruno}. The orbital magnetic moment is a microscopic observable, which can be measured, for example, via XMCD, and, therefore, experiments could provide a definite validation of the proposed mechanism based on the $p_z$-$d_{z^2}$ hybridization.\\

\section{Acknowledgements}
AD would like to thank V.A. Dediu, I. Bergenti, and M. Benini for stimulating discussions. AH and AD acknowledge funding from Science Foundation Ireland (SFI) and the Royal Society through the University Research Fellowship URF-R1-191769. AD and SS  were  partly  supported  by the H2020  program of the European Commission under the FET-Open project INTERFAST (grant agreement No. 965046). SB, SS, and DO'R acknowledge support from Science Foundation Ireland (19/EPSRC/3605) and the Engineering and Physical Sciences Research Council (EP/S030263/1). The computational resources were provided by Trinity College Dublin Research IT and the Irish Center for High End Computing (ICHEC).


\section{Supplementary Material}

\subsection{Spin-orbit matrix elements}
\begin{table*}
 \begin{tabular*}{\textwidth}{c@{\extracolsep{\fill}}  c  c  c  c  c  c  c  c  c  c}
  \hline\hline
& $\vert d_{xy}, \uparrow \rangle$  &$\vert d_{yz}, \uparrow \rangle$  &$\vert d_{z^2}\uparrow \rangle$   & $\vert d_{xz} \uparrow \rangle$  & $\vert d_{x^2-y^2}\uparrow \rangle$ & $\vert d_{xy}\downarrow \rangle$  &$\vert d_{yz}\downarrow \rangle$  &$\vert d_{z^2}\downarrow \rangle$   & $\vert d_{xz}\downarrow\rangle$  & $\vert d_{x^2-y^2},\downarrow \rangle$\\ 
\hline
 $\vert d_{xy}\uparrow \rangle$ & 0  & 0 & 0 &0  & $i\xi$ & 0 & $\xi/2$ & 0& $i\xi/2$ & 0 \\
$\vert d_{yz}\uparrow \rangle$ & 0 & 0 & 0 & $i\xi/2$ & 0 & -$\xi/2$ & 0 & $i\sqrt{3}\xi/2$ & 0 & $\xi/2$ \\
  $\vert d_{z^2}\uparrow \rangle$ & 0 & 0 & 0 & 0 & 0& 0 & $-i\sqrt{3}\xi/2$ & 0& $-\sqrt{3}\xi/2$ & 0\\ 
   $\vert d_{xz}\uparrow \rangle$ &0 & $-i\xi/2$ &0 &0 & 0& -$i\xi/2$ & 0&$\sqrt{3}\xi/2$ &0 & $-\xi/2$ \\
$\vert d_{x^2-y^2}\uparrow \rangle$ & $-i\xi$ & 0 & 0 & 0 & 0 & 0 &-$\xi/2$ &0 &$\xi/2$ &0\\
$\vert d_{xy} \downarrow \rangle$ & 0 &$-\xi/2$ &0 &$-i\xi/2$ & 0& 0& 0& 0& 0& $-i\xi$\\
$\vert d_{yz}\downarrow \rangle$ & $\xi/2$ &0 &$-i\sqrt{3}\xi/2$ & 0& $-\xi/2$ & 0& 0& 0&$-i\xi/2$ &0\\
  $\vert d_{z^2}\downarrow \rangle$ & 0 & $i\sqrt{3}\xi/2$ & 0 &$\sqrt{3}\xi/2$ &0 &0 & 0& 0& 0&0\\ 
   $\vert d_{xz}\downarrow \rangle$ & $i\xi/2$ & 0& $-\sqrt{3}\xi/2$ &0 &$\xi/2$ &0 &$i\xi/2$ &0 &0 &0 \\
$\vert d_{x^2-y^2}\downarrow \rangle$ &0 & $\xi/2$ & 0& $-\xi/2$ & 0 & $i\xi$ &0 &0 & 0&0\\
 \hline
 \hline
\end{tabular*}\caption{ SOC matrix elements  $\langle d_\alpha, s_z\vert \hat H_\textrm{SOC}\vert d_\beta, s_z' \rangle $. The spin magnetic quantum number $s_z=1/2$ ($-1/2$) is indicated as $\uparrow$ ($\downarrow$).}\label{tab.SOC}
\end{table*}

\begin{table*}
 \begin{tabular*}{\textwidth}{c@{\extracolsep{\fill}}  c  c  c  c  c  c  c  c  c  c}
  \hline\hline
& $\vert d_{xy}\uparrow \rangle$  &$\vert d_{yz}\uparrow \rangle$  &$\vert d_{z^2}\uparrow \rangle$   & $\vert d_{xz}\uparrow \rangle$  & $\vert d_{x^2-y^2}\uparrow \rangle$ & $\vert d_{xy}\downarrow \rangle$  &$\vert d_{yz}\downarrow \rangle$  &$\vert d_{z^2}\downarrow \rangle$   & $\vert d_{xz}\downarrow \rangle$  & $\vert d_{x^2-y^2}\downarrow \rangle$\\ 
\hline
 $\vert d_{xy}\uparrow \rangle$ & 0 & 0 & 0 & $\xi^2/4$  & $\xi^2$ &0 &0 &0 & $-\xi^2/4$ & $\xi^2$ \\
$\vert d_{yz}\uparrow \rangle$ & 0 &0 &$3\xi^2/4$ & $-\xi^2/4$ & $\xi^2/4$ & 0&0 & $-3\xi^2/4$ & $\xi^2/4$ & $-\xi^2/4$ \\
  $\vert d_{z^2}\uparrow \rangle$ & 0 & $3\xi^2/4$ & 0&0 &0 &0 & $-3\xi^2/4$ & 0&0 &0 \\ 
   $\vert d_{xz}\uparrow \rangle$ & $\xi^2/4$ & $-\xi^2/4$ &0 &0 &0  & $-\xi^2/4$ & $\xi^2/4$ & 0 &0 &0\\
$\vert d_{x^2-y^2}\uparrow \rangle$ & $-\xi^2$ & $\xi^2/4$ &0 &0 & 0& $\xi^2$ & $-\xi^2/4$ & 0& 0&0\\
$\vert d_{xy}\downarrow \rangle$ & 0 & 0 & 0 & $-\xi^2/4$ & $\xi^2$ & 0 & 0 & 0 & $\xi^2/4$ & $-\xi^2$\\
$\vert d_{yz}\downarrow \rangle$ & 0 & 0 & $-3\xi^2/4$ & $\xi^2/4$ & $-\xi^2/4$ & 0 & 0 & $3\xi^2/4$ & $-\xi^2/4$ & $\xi^2/4$ \\
  $\vert d_{z^2}\downarrow \rangle$ & 0 & $-3\xi^2/4$ & 0 &0 & 0& 0& $3\xi^2/4$& 0& 0&0\\ 
   $\vert d_{xz}\downarrow \rangle$ & $-\xi^2/4$ & $\xi^2/4$ & 0 & 0 & 0 & $\xi^2/4$ & $-\xi^2/4$ & 0 & 0 & 0 \\
$\vert d_{x^2-y^2}\downarrow \rangle$ & $\xi^2$ & $-\xi^2/4$ & 0 & 0 & 0 & $-\xi^2$ & $\xi^2/4$ & 0 & 0 & 0\\
 \hline
 \hline
\end{tabular*}\caption{ Difference between the SOC matrix elements computed with parallel and perpendicular magnetization $\vert  \langle \hat{H}_\textrm{SOC} \rangle\vert^2_{\parallel}-\vert  \langle \hat{H}_\textrm{SOC} \rangle\vert^2_{\perp}$. The spin magnetic quantum number $s_z=1/2$ ($-1/2$) is indicated as $\uparrow$ ($\downarrow$).}\label{tab.SOC2}
\end{table*}

In this section, we compute the SOC matrix elements used in Eq. (1) of the paper to obtain the MCA energy of the model Co(001) surface via perturbation theory. We assume that the spin quantization axis is along the $z$ Cartesian axis, normal to the surface.   
Following Ref. \onlinecite{gi.ca.12}, we consider the atomic $d_\alpha$ orbitals ($\alpha=$ $xy$, $yz$, $z^2$, $xz$, $x^2-y^2$) with orbital momentum quantum number $l=2$. \\

To begin with, we set the spin magnetic moment to be out-of-plane, i.e. along the $z$ direction. The spin $s$ and the spin magnetic quantum number $s_z=\pm 1/2$ are good quantum numbers for the unperturbed system. The $d_\alpha$ states, $\vert d_\alpha, s_z\rangle $, are then expressed in the basis $\vert m, s_z=\pm 1/2\rangle= \vert l, m,\rangle \otimes \vert s, s_z\rangle$, where $m$ is the orbital magnetic quantum number, $m=0, \pm1, \pm2$. Specifically, we write 
\begin{eqnarray}
&\vert d_{xy}, s_z \rangle=\frac{i}{\sqrt{2}}(\vert -2, s_z\rangle-\vert 2, s_z\rangle),\\ 
&\vert d_{yz}, s_z \rangle=\frac{i}{\sqrt{2}}(\vert -1, s_z\rangle+\vert 1, s_z\rangle), \\
&  \vert d_{z^2}, s_z \rangle=\vert 0, s_z \rangle,\\ 
&  \vert d_{xz}, s_z \rangle=\frac{1}{\sqrt{2}}(\vert -1, s_z\rangle-\vert 1, s_z\rangle), \\
& \vert d_{x^2-y^2}, s_z \rangle=\frac{1}{\sqrt{2}}(\vert -2, s_z\rangle+\vert 2, s_z\rangle). 
  \end{eqnarray}
  We then calculate the matrix elements
  $\langle d_\alpha, s_z \vert \hat H_\textrm{SOC}\vert d_\beta, s_z' \rangle $
  of the SOC Hamiltonian $\hat H_\textrm{SOC}=\xi \hat{\mathbf{L}}\cdot \hat{\mathbf{S}}/\hbar^2= \frac{\xi}{2\hbar^2}(\hat L_{-}\hat S_{+}+\hat L_{+}\hat S_{-})+\frac{\xi}{\hbar^2}\hat L_z \hat S_z$, where $\hat L_{\pm}$ ($\hat S_{\pm}$),  are the ladder orbital (spin) angular momentum operators, and $\hat L_z$ ($\hat S_z$) is the $z$ components of the orbital (spin) angular momentum operator. The results are listed in Table \ref{tab.SOC}.\\
  
  Next, we rotate the spin in such a way that it lays along the in-plane direction, for example, $x$. Thus, we introduce the states 
  \begin{equation}
  \vert d_\alpha, s_x =\pm 1/2 \rangle=\frac{1}{\sqrt{2}}\Big[\vert d_\alpha, s_z=1/2\rangle\pm\vert d_\alpha, s_z=-1/2\rangle\Big],
  \end{equation}
  and we easily calculate the matrix elements $\vert\langle d_\alpha, s_x \vert \hat H_\textrm{SOC}\vert d_\beta, s_x' \rangle\vert^2 $.\\
  
  Finally, we compute the difference between the square modulus of the matrix elements with spin parallel and perpendicular to the surface 
\begin{equation}  
\begin{split}
&\vert \langle \hat{H}_\textrm{SOC} \rangle\vert^2_{\parallel}-\vert \langle \hat{H}_\textrm{SOC} \rangle\vert^2_{\perp}\equiv\\ 
&\vert\langle d_\alpha, s_x \vert \hat H_\textrm{SOC}\vert d_\beta, s_x' \rangle\vert^2 -\vert\langle d_\alpha, s_z \vert \hat H_\textrm{SOC}\vert d_\beta, s_z' \rangle \vert^2,
\end{split}
\end{equation}
where $s_x=s_z=\pm 1/2$ and $s_x'=s_z'=\pm 1/2$.
The results are presented in Table \ref{tab.SOC2} and can be used to derive the simple rules mentioned in Sec. II of the paper and discussed in Refs. \onlinecite{Daalderop,gi.ca.12,Kyuno_1996}.  
  Negative (positive) terms correspond to virtual transitions, which promote in-plane (out-of-plane) MCA. The difference $\vert \langle \hat{H}_\textrm{SOC} \rangle\vert^2_{\parallel}-\vert \langle \hat{H}_\textrm{SOC} \rangle\vert^2_{\perp}$ enters the numerator of Eq. (1) of the paper and is used to obtain the MCA energy in Eqs. (2) and (3).

\subsection{DFT results for the Bz/hcp-Co interface} 
We present here the results of DFT calculations for the Bz/hcp-Co interface showing that the physics driven by molecular adsorption is similar, at the atomic scale, to that described in the paper for Bz/fcc-Co. \\

\begin{figure}[hbt]
\centering\includegraphics[width=\columnwidth,clip=true]{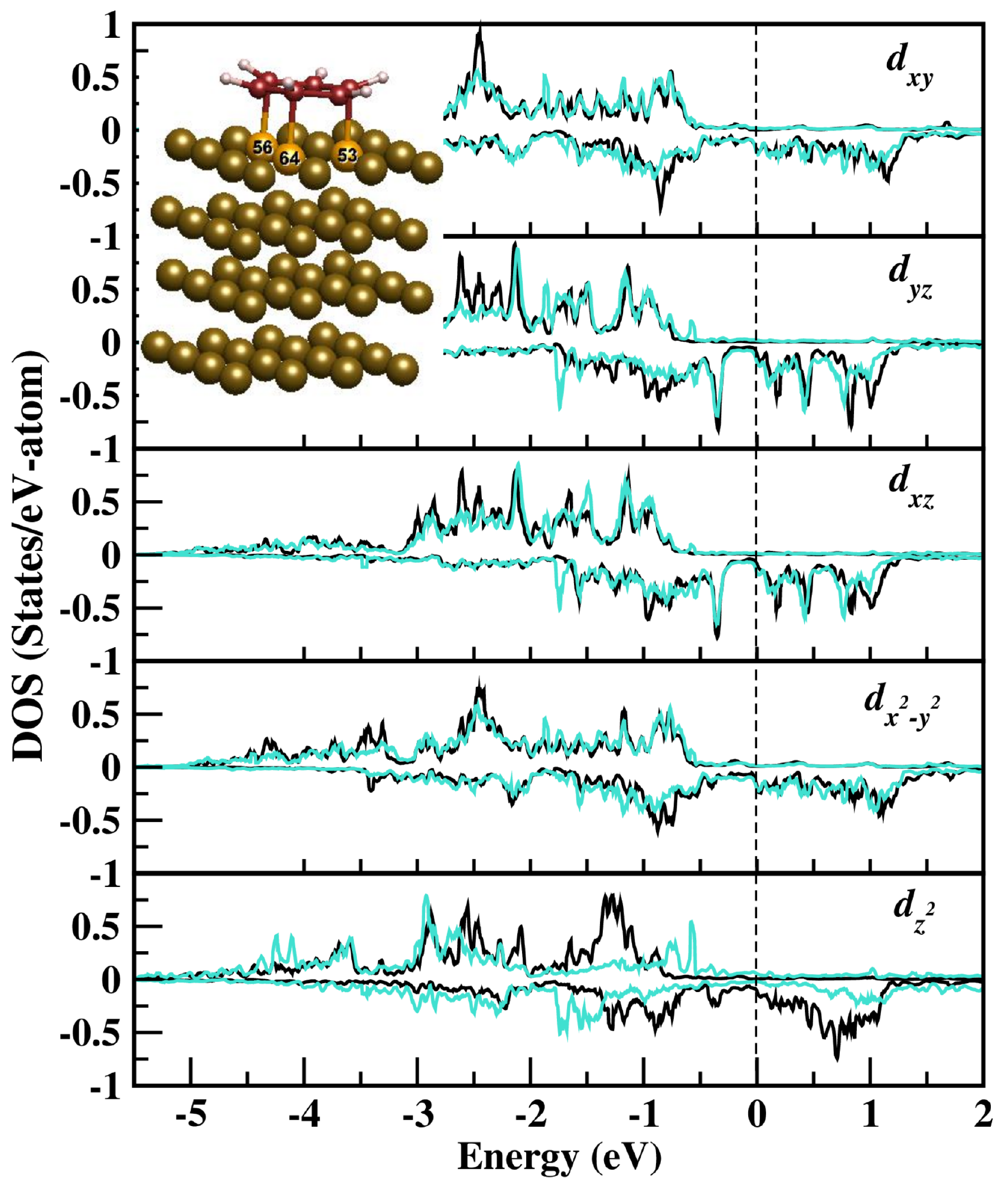}
\caption{DOS projected over the $3d$ orbitals of the Co atom 64 at the Bz/hcp(0001)-Co interface. The Fermi energy is at 0 eV. The black and cyan lines are for the bare Co slab and the Bz/hcp-Co slab, respectively. The inset shows the side view of the Bz/hcp-Co 4-layer slab. Brown, white, and yellow spheres indicate C, H, and Co atoms, respectively.}
\label{hcp_ele}
\end{figure}

The calculations are carried out for a 4-layer hcp(0001) slab with a 4 $\times$ 4 in-plane supercell. The molecule is adsorbed on the top surface as displayed in Fig. \ref{hcp_ele} (inset), where the surface Co atoms bonded with the molecule are highlighted and labelled 53, 56, and 64. To compare the electronic structure of the bare hcp-Co surface and of the Bz/hcp-Co interface, we plot in Fig. \ref{hcp_ele} the DOS projected over the $3d$ orbitals of the atom 64. The black and the cyan lines are for the bare Co and for the Bz/hcp-Co case, respectively.   
The DOS of the bare hcp surface is qualitatively similar to that of the fcc surface (cf. Fig. 2 in the paper). The majority spin states are completely occupied, whereas the minority spin DOS crosses the Fermi energy. However, the crystal field splitting of the hcp surface is different from that of the fcc one. In the hcp case, the $d_{x^2-y^2}$ and the $d_{xy}$ orbitals form a first low energy doublet, while the $d_{xz}$ and the $d_{yz}$ orbitals form a second doublet at higher energies. The $d_{z^2}$ orbital is a singlet with the highest energy. In the minority spin channel, the DOS of the $d_{z^2}$ splits into two distinct peaks appearing on either side of the Fermi level. \\

Despite the difference between hcp and fcc-Co, the change of the electronic structure upon Bz adsorption is qualitatively similar in the two cases. Focusing in particular on the minority spin channel, we see that the $p_z$-$d_{z^2}$ antibonding state is located at about 1 eV above the Fermi energy, whereas the bonding state is broad and centered at about $1.5$ eV below the Fermi energy. \\

\begin{figure}
\centering\includegraphics[width=\columnwidth,clip=true]{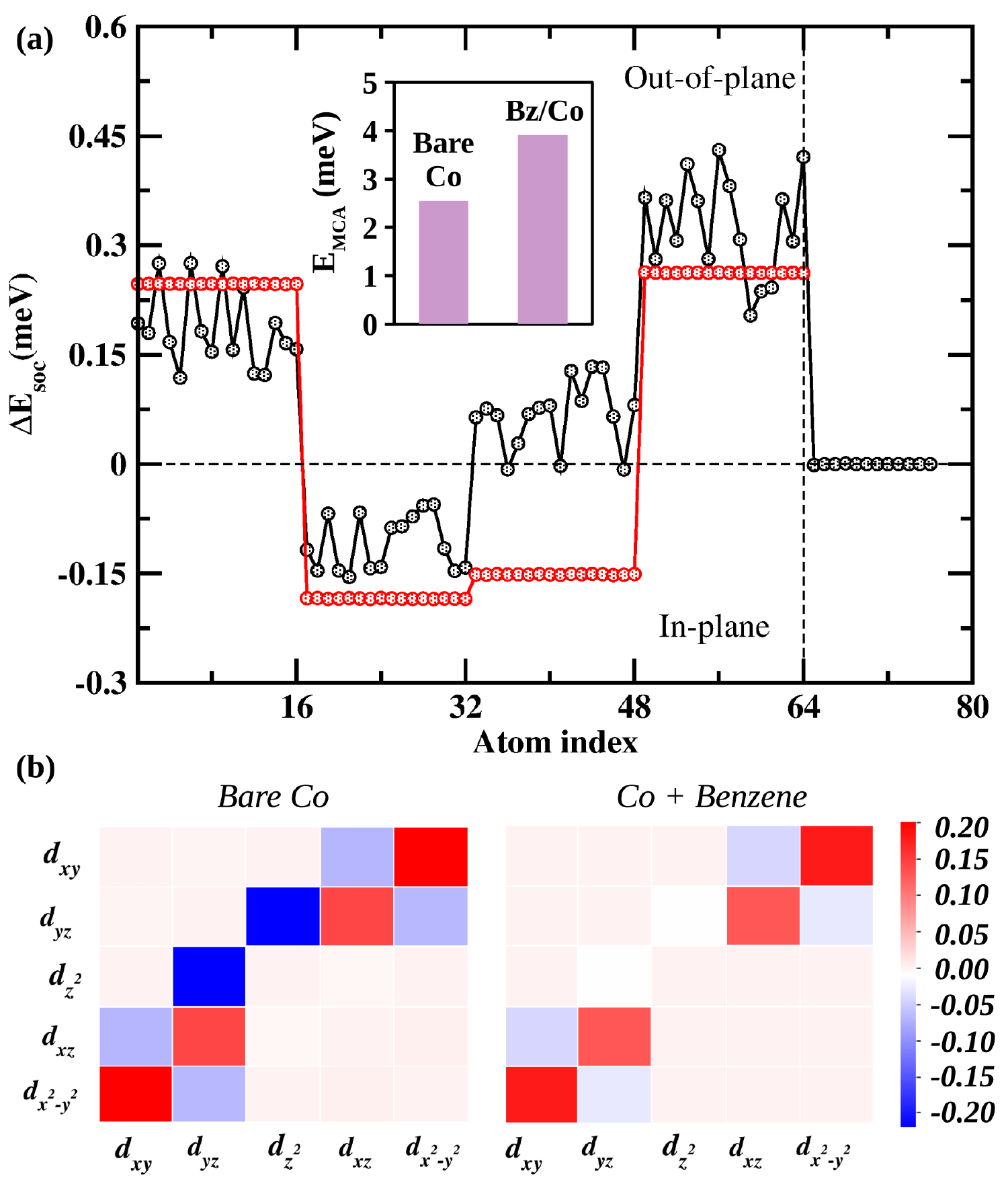}
\caption{DFT results for the MCA of the hcp(0001)-Co and Bz/hcp(0001)-Co slabs. (a) $\Delta E^I_\textrm{SOC}$ evaluated for all atoms. The red, and black dots represent the values for the bare hcp Co slab and the Bz/hcp-Co slab respectively. Atoms 1-16 are at the slab bottom surface, atoms 49-64 are at the top surface, and atoms 65-76 are the Bz atoms. The inset displays the MCA energy of the bare hcp-Co slab and of the Bz/hcp-Co slab. (b) In-plane {\it versus} out-of-plane difference between the SOC matrix elements $\Delta E^{I,mm'}_\textrm{SOC}$ (in meV) for the $3d$ orbitals of the surface Co atom 64, which is bonded with the Bz molecule (see the inset of Fig. \ref{hcp_ele}).}
\label{hcp_mca}
\end{figure}

The modification of the MCA of the hcp slab upon Bz adsorption is summarized in Fig. \ref{hcp_mca}(a), where we plot the total MCA energy $E_\textrm{MCA}$ (inset) and the atomic SOC energy difference $\Delta E^I_\textrm{SOC}$ between the in-plane and out-of-plane magnetization configurations. $E_\textrm{MCA}$ is positive and equal to 2.55 meV for the bare hcp-Co slab, while it is enhanced to 3.91 meV for Bz/hcp-Co. Thus, the easy magnetization direction of hcp-Co is out-of-plane even without Bz, unlike the case of fcc-Co presented in the paper. The effect of the molecule in hcp-Co is to further favour the perpendicular MCA. This is in agreement with the experimental reports by Bairagi {\it et al.} \cite{bairagiprl} for hcp(0001) thin films. \\

The atom resolved $\Delta E^I_\textrm{SOC}$ is displayed in Fig \ref{hcp_mca}(a) as the red (black) dots for the Bz/hcp-Co (bare hcp-Co) slab. For the bare hcp-Co slab, the values are positive for all the atoms in the bottom and top layers (i.e., for the atoms labelled 1-16 and 49-64, respectively). The values are instead negative for the atoms in the central layers (i.e., for the atoms 17-48). In the case of the Bz/hcp-Co slab, $\Delta E^I_\textrm{SOC}$ increases for the atoms at the top surface, which forms now the interface with the molecule, and switches to positive values for the layer just underneath the interface. In particular, the maximum positive $\Delta E^I_\textrm{SOC}$, equal to about 0.43 meV, is found for the Co atoms bonded with the Bz molecule. This behaviour accounts for the overall increase of the out-of-plane MCA induced by the Bz absorption.\\
   
To further understand the effect of the metal-molecule hybridization we present in Fig. \ref{hcp_mca}(b) the SOC matrix elements $\Delta E^{I,mm'}_\textrm{SOC}$ for the various combinations of the $3d$ orbitals of the Co atom 64. Positive (negative) values are in red (blue). The transitions with the largest positive values, promoting out-plane MCA, connect $d_{xy}$ with $d_{x^2-y^2}$ and $d_{yz}$ with $d_{xz}$. They are barely affected by the hybridization with the molecule. 
On the other hand, the main negative transition, which connects $d_{yz}$ with $d_{z^2}$, is fully suppressed once the molecule is present. Besides, we note that there are also other two transitions with negative values, namely those between $d_{xy}$ and $d_{xz}$ and between $d_{yz}$ and $d_{x^2-y^2}$. Although they do not vanish, their magnitude is reduced in the Bz/hcp-Co case compared to the bare hcp-Co case. These results are qualitatively, and even quantitatively, similar those reported for fcc-Co (cf. Fig. 3 in the paper). Therefore, we conclude that the metal-molecule hybridization has the same effect in promoting out-of-plane surface MCA in Bz/fcc-Co as well as in Bz/hcp-Co slabs. 

\begin{figure}
\centering\includegraphics[width=\columnwidth,clip=true]{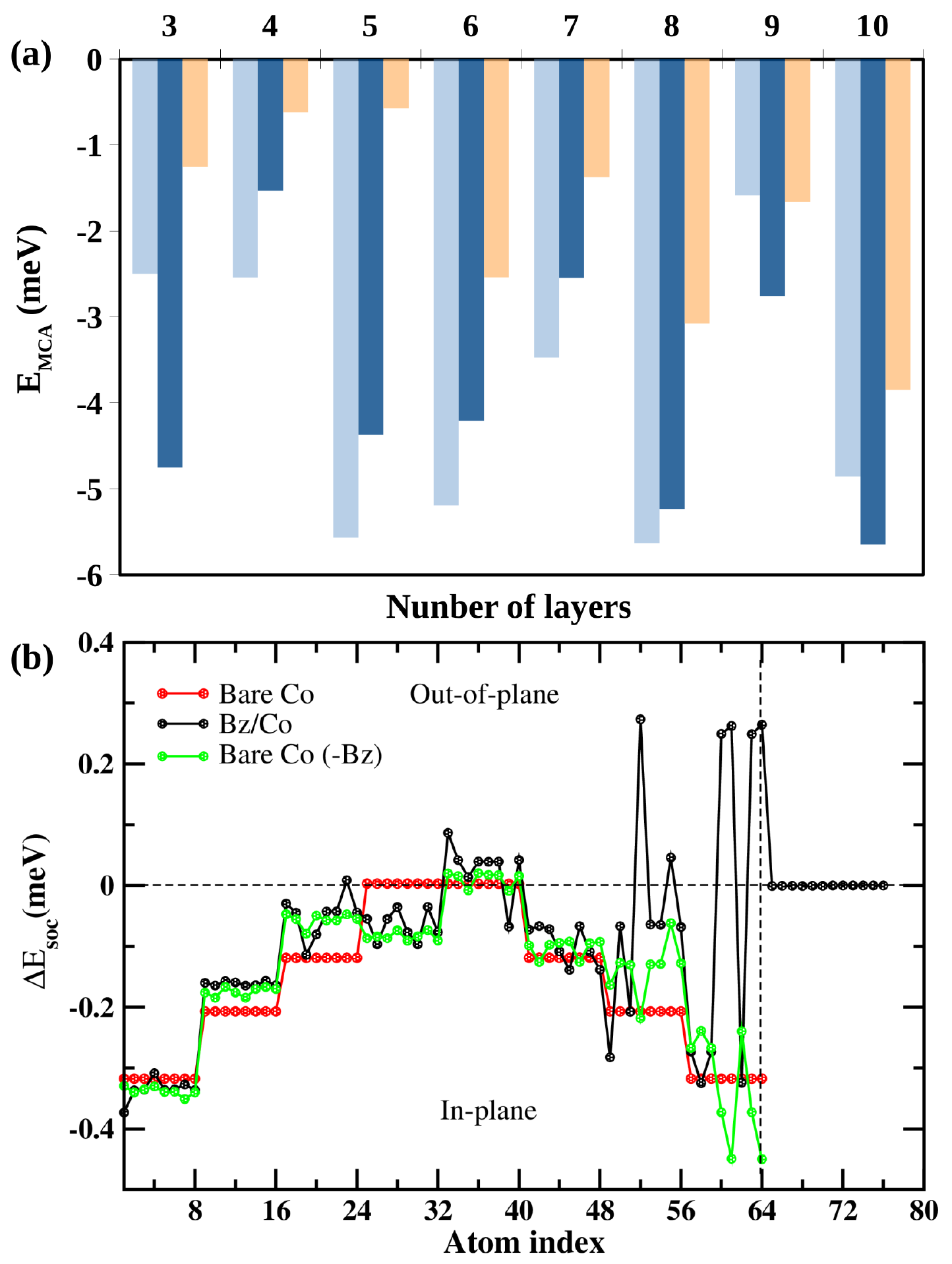}
\caption{ MCA for fcc slabs of different thicknesses. (a) Total MCA energy $E_\textrm{MCA}$ as a function of the number of layers; light blue bars, pink bars, and dark blue bars respectively represent the results for bare fcc-Co slabs, Bz/fcc-Co slabs, and fcc-Co slabs, where the Bz molecule was removed without re-optimization of the atomic coordinates [bare Co (-Bz)]. (b) $\Delta$E$^I_\textrm{SOC}$ for all atoms of a 8-layer slab. The red, black, and green dots represent the results for the bare fcc-Co slab, the Bz/fcc-Co slab, and the bare Co (-Bz) slab. Atoms 57-64 are at the slab's top surface.}
\label{fig.thickness}
\end{figure}

\subsection{Dependence of the MCA on the fcc-Co slab thickness} 
 
The MCA energies $E_\textrm{MCA}$ of ideal fcc-Co slabs and of Bz/fcc-Co slabs with thicknesses varying from 3 to 10 layers are displayed in Fig. \ref{fig.thickness}(a) (the 5-layer slab is the one  extensively discussed in the paper).  
For the bare Co slabs $E_\textrm{MCA}$ is negative (light blue bars), implying an in-plane MCA, regardless of the number of layers, although the results oscillate as a function of the slab thickness. A similar oscillating behaviour has previously been observed in the literature and attributed to quantum well effects\cite{Fe-Co-slab}. The adsorption of Bz on the slabs' top surfaces leads to a considerable reduction of the magnitude of $E_\textrm{MCA}$ in almost all cases (see the pink bars) as we already observed in the paper for the specific case of a 5-layer slab. The removal of the molecule from the slabs' top surfaces without re-optimizing the geometries [bare Co(-Bz) slabs] recovers a negative MCA energy comparable to that of the bare Co slabs thus demonstrating that the displacement of the surface atoms has a minor effect on $E_\textrm{MCA}$ for all cases.  \\

The mechanism leading to the reduction of the MCA energy upon Bz adsorption is the same for thick slabs as for the 5-layer slab discussed in the paper. It is indeed due to a ``local'' effect, which involves only the surface atoms hybridized with the molecules, and which is therefore almost independent on the slab thickness. To see this, we plot in Fig. \ref{fig.thickness}(b) the values of $\Delta E^I_\textrm{SOC}$ for all atoms of a 8-layer slab, which is taken as a representative case of all thick slabs.  
The trend is similar to that reported in the paper [cf. Fig. 2(d)]. At the surfaces of the bare Co slab (atoms 1-8 and 57-64), $\Delta E^I_\textrm{SOC}$ is negative and equal to about $0.31$ eV. In contrast, in the Bz/Co slab, $\Delta E^I_\textrm{SOC}$ switches sign and becomes approximately equal to $0.28$ eV for those Co atoms that are bonded with Bz at the top surface, and for some of the Co atoms in the second layer underneath the interface. Thus, for all the considered slabs, there is a compensation between the positive contributions of the interface and the negative contributions of the rest of the slab to the total MCA energy.


\begin{thebibliography}{99}
\expandafter\ifx\csname natexlab\endcsname\relax\def\natexlab#1{#1}\fi
\expandafter\ifx\csname bibnamefont\endcsname\relax
  \def\bibnamefont#1{#1}\fi
\expandafter\ifx\csname bibfnamefont\endcsname\relax
  \def\bibfnamefont#1{#1}\fi
\expandafter\ifx\csname citenamefont\endcsname\relax
  \def\citenamefont#1{#1}\fi
\expandafter\ifx\csname url\endcsname\relax
  \def\url#1{\texttt{#1}}\fi
\expandafter\ifx\csname urlprefix\endcsname\relax\def\urlprefix{URL }\fi
\providecommand{\bibinfo}[2]{#2}
\providecommand{\eprint}[2][]{\url{#2}}

\bibitem[{\citenamefont{Gambardella and Bl{\"u}gel}(2020)}]{Gambardella2020}
\bibinfo{author}{\bibfnamefont{P.}~\bibnamefont{Gambardella}} \bibnamefont{and}
  \bibinfo{author}{\bibfnamefont{S.}~\bibnamefont{Bl{\"u}gel}},
  \emph{\bibinfo{title}{Magnetic Surfaces, Thin Films and Nanostructures}}
  (\bibinfo{publisher}{Springer International Publishing},
  \bibinfo{address}{Cham}, \bibinfo{year}{2020}), pp.
  \bibinfo{pages}{625--698},
  \urlprefix\url{https://doi.org/10.1007/978-3-030-46906-1_21}.

\bibitem[{\citenamefont{Hellman et~al.}(2017)\citenamefont{Hellman, Hoffmann,
  Tserkovnyak, Beach, Fullerton, Leighton, MacDonald, Ralph, Arena, D\"urr
  et~al.}}]{RevModPhys.89.025006}
\bibinfo{author}{\bibfnamefont{F.}~\bibnamefont{Hellman}},
  \bibinfo{author}{\bibfnamefont{A.}~\bibnamefont{Hoffmann}},
  \bibinfo{author}{\bibfnamefont{Y.}~\bibnamefont{Tserkovnyak}},
  \bibinfo{author}{\bibfnamefont{G.~S.~D.} \bibnamefont{Beach}},
  \bibinfo{author}{\bibfnamefont{E.~E.} \bibnamefont{Fullerton}},
  \bibinfo{author}{\bibfnamefont{C.}~\bibnamefont{Leighton}},
  \bibinfo{author}{\bibfnamefont{A.~H.} \bibnamefont{MacDonald}},
  \bibinfo{author}{\bibfnamefont{D.~C.} \bibnamefont{Ralph}},
  \bibinfo{author}{\bibfnamefont{D.~A.} \bibnamefont{Arena}},
  \bibinfo{author}{\bibfnamefont{H.~A.} \bibnamefont{D\"urr}},
  \bibnamefont{et~al.}, \bibinfo{journal}{Rev. Mod. Phys.}
  \textbf{\bibinfo{volume}{89}}, \bibinfo{pages}{025006}
  (\bibinfo{year}{2017}),
  \urlprefix\url{https://link.aps.org/doi/10.1103/RevModPhys.89.025006}.

\bibitem[{\citenamefont{Varvaro and Casoli}(2016)}]{va.ca.16}
\bibinfo{editor}{\bibfnamefont{G.}~\bibnamefont{Varvaro}} \bibnamefont{and}
  \bibinfo{editor}{\bibfnamefont{F.}~\bibnamefont{Casoli}}, eds.,
  \emph{\bibinfo{title}{Ultra-High-Density Magnetic Recording}}
  (\bibinfo{publisher}{Jenny Stanford Publishing}, \bibinfo{year}{2016}),
  \urlprefix\url{https://doi.org/10.1201/b20044}.

\bibitem[{\citenamefont{Hirohata et~al.}(2020)\citenamefont{Hirohata, Yamada,
  Nakatani, Prejbeanu, Di\'{e}ny, Pirro, and Hillebrands}}]{hi.ke.20}
\bibinfo{author}{\bibfnamefont{A.}~\bibnamefont{Hirohata}},
  \bibinfo{author}{\bibfnamefont{K.}~\bibnamefont{Yamada}},
  \bibinfo{author}{\bibfnamefont{Y.}~\bibnamefont{Nakatani}},
  \bibinfo{author}{\bibfnamefont{I.-L.} \bibnamefont{Prejbeanu}},
  \bibinfo{author}{\bibfnamefont{B.}~\bibnamefont{Di\'{e}ny}},
  \bibinfo{author}{\bibfnamefont{P.}~\bibnamefont{Pirro}}, \bibnamefont{and}
  \bibinfo{author}{\bibfnamefont{B.}~\bibnamefont{Hillebrands}},
  \bibinfo{journal}{Journal of Magnetism and Magnetic Materials}
  \textbf{\bibinfo{volume}{509}}, \bibinfo{pages}{166711}
  (\bibinfo{year}{2020}), ISSN \bibinfo{issn}{0304-8853},
  \urlprefix\url{https://www.sciencedirect.com/science/article/pii/S0304885320302353}.

\bibitem[{\citenamefont{Cinchetti et~al.}(2017)\citenamefont{Cinchetti, Dediu,
  and Hueso}}]{Cinchetti2017}
\bibinfo{author}{\bibfnamefont{M.}~\bibnamefont{Cinchetti}},
  \bibinfo{author}{\bibfnamefont{V.~A.} \bibnamefont{Dediu}}, \bibnamefont{and}
  \bibinfo{author}{\bibfnamefont{L.~E.} \bibnamefont{Hueso}},
  \bibinfo{journal}{Nature Materials} \textbf{\bibinfo{volume}{16}},
  \bibinfo{pages}{507} (\bibinfo{year}{2017}), ISSN \bibinfo{issn}{1476-4660},
  \urlprefix\url{https://doi.org/10.1038/nmat4902}.

\bibitem[{\citenamefont{Atodiresei et~al.}(2010)\citenamefont{Atodiresei,
  Brede, Lazi\ifmmode~\acute{c}\else \'{c}\fi{}, Caciuc, Hoffmann,
  Wiesendanger, and Bl\"ugel}}]{at.br.10}
\bibinfo{author}{\bibfnamefont{N.}~\bibnamefont{Atodiresei}},
  \bibinfo{author}{\bibfnamefont{J.}~\bibnamefont{Brede}},
  \bibinfo{author}{\bibfnamefont{P.}~\bibnamefont{Lazi\ifmmode~\acute{c}\else
  \'{c}\fi{}}}, \bibinfo{author}{\bibfnamefont{V.}~\bibnamefont{Caciuc}},
  \bibinfo{author}{\bibfnamefont{G.}~\bibnamefont{Hoffmann}},
  \bibinfo{author}{\bibfnamefont{R.}~\bibnamefont{Wiesendanger}},
  \bibnamefont{and} \bibinfo{author}{\bibfnamefont{S.}~\bibnamefont{Bl\"ugel}},
  \bibinfo{journal}{Phys. Rev. Lett.} \textbf{\bibinfo{volume}{105}},
  \bibinfo{pages}{066601} (\bibinfo{year}{2010}),
  \urlprefix\url{https://link.aps.org/doi/10.1103/PhysRevLett.105.066601}.

\bibitem[{\citenamefont{Atodiresei et~al.}(2011)\citenamefont{Atodiresei,
  Caciuc, Lazi\ifmmode~\acute{c}\else \'{c}\fi{}, and Bl\"ugel}}]{at.ca.11}
\bibinfo{author}{\bibfnamefont{N.}~\bibnamefont{Atodiresei}},
  \bibinfo{author}{\bibfnamefont{V.}~\bibnamefont{Caciuc}},
  \bibinfo{author}{\bibfnamefont{P.}~\bibnamefont{Lazi\ifmmode~\acute{c}\else
  \'{c}\fi{}}}, \bibnamefont{and}
  \bibinfo{author}{\bibfnamefont{S.}~\bibnamefont{Bl\"ugel}},
  \bibinfo{journal}{Phys. Rev. B} \textbf{\bibinfo{volume}{84}},
  \bibinfo{pages}{172402} (\bibinfo{year}{2011}),
  \urlprefix\url{https://link.aps.org/doi/10.1103/PhysRevB.84.172402}.

\bibitem[{\citenamefont{Atodiresei and Raman}(2014)}]{at.ra.14}
\bibinfo{author}{\bibfnamefont{N.}~\bibnamefont{Atodiresei}} \bibnamefont{and}
  \bibinfo{author}{\bibfnamefont{K.~V.} \bibnamefont{Raman}},
  \bibinfo{journal}{MRS Bulletin} \textbf{\bibinfo{volume}{39}},
  \bibinfo{pages}{596-601} (\bibinfo{year}{2014}).

\bibitem[{\citenamefont{Tran et~al.}(2013)\citenamefont{Tran, Çak?r, Wong,
  Preobrajenski, Brocks, van~der Wiel, and de~Jong}}]{tr.la.13}
\bibinfo{author}{\bibfnamefont{T.~L.~A.} \bibnamefont{Tran}},
  \bibinfo{author}{\bibfnamefont{D.}~\bibnamefont{Cakr}},
  \bibinfo{author}{\bibfnamefont{P.~K.~J.} \bibnamefont{Wong}},
  \bibinfo{author}{\bibfnamefont{A.~B.} \bibnamefont{Preobrajenski}},
  \bibinfo{author}{\bibfnamefont{G.}~\bibnamefont{Brocks}},
  \bibinfo{author}{\bibfnamefont{W.~G.} \bibnamefont{van~der Wiel}},
  \bibnamefont{and} \bibinfo{author}{\bibfnamefont{M.~P.}
  \bibnamefont{de~Jong}}, \bibinfo{journal}{ACS Applied Materials \&
  Interfaces} \textbf{\bibinfo{volume}{5}}, \bibinfo{pages}{837}
  (\bibinfo{year}{2013}), \bibinfo{note}{pMID: 23305202},
  \urlprefix\url{https://doi.org/10.1021/am3024367}.

\bibitem[{\citenamefont{Moorsom et~al.}(2014)\citenamefont{Moorsom, Wheeler,
  Mohd~Khan, Al~Ma'Mari, Kinane, Langridge, Ciudad, Bedoya-Pinto, Hueso,
  Teobaldi et~al.}}]{mo.wh.14}
\bibinfo{author}{\bibfnamefont{T.}~\bibnamefont{Moorsom}},
  \bibinfo{author}{\bibfnamefont{M.}~\bibnamefont{Wheeler}},
  \bibinfo{author}{\bibfnamefont{T.}~\bibnamefont{Mohd~Khan}},
  \bibinfo{author}{\bibfnamefont{F.}~\bibnamefont{Al~Ma'Mari}},
  \bibinfo{author}{\bibfnamefont{C.}~\bibnamefont{Kinane}},
  \bibinfo{author}{\bibfnamefont{S.}~\bibnamefont{Langridge}},
  \bibinfo{author}{\bibfnamefont{D.}~\bibnamefont{Ciudad}},
  \bibinfo{author}{\bibfnamefont{A.}~\bibnamefont{Bedoya-Pinto}},
  \bibinfo{author}{\bibfnamefont{L.}~\bibnamefont{Hueso}},
  \bibinfo{author}{\bibfnamefont{G.}~\bibnamefont{Teobaldi}},
  \bibnamefont{et~al.}, \bibinfo{journal}{Phys. Rev. B}
  \textbf{\bibinfo{volume}{90}}, \bibinfo{pages}{125311}
  (\bibinfo{year}{2014}),
  \urlprefix\url{https://link.aps.org/doi/10.1103/PhysRevB.90.125311}.

\bibitem[{\citenamefont{Caffrey et~al.}(2013)\citenamefont{Caffrey, Ferriani,
  Marocchi, and Heinze}}]{Caffery}
\bibinfo{author}{\bibfnamefont{N.~M.} \bibnamefont{Caffrey}},
  \bibinfo{author}{\bibfnamefont{P.}~\bibnamefont{Ferriani}},
  \bibinfo{author}{\bibfnamefont{S.}~\bibnamefont{Marocchi}}, \bibnamefont{and}
  \bibinfo{author}{\bibfnamefont{S.}~\bibnamefont{Heinze}},
  \bibinfo{journal}{Phys. Rev. B} \textbf{\bibinfo{volume}{88}},
  \bibinfo{pages}{155403} (\bibinfo{year}{2013}),
  \urlprefix\url{https://link.aps.org/doi/10.1103/PhysRevB.88.155403}.

\bibitem[{\citenamefont{Brede et~al.}(2010)\citenamefont{Brede, Atodiresei,
  Kuck, Lazi\ifmmode~\acute{c}\else \'{c}\fi{}, Caciuc, Morikawa, Hoffmann,
  Bl\"ugel, and Wiesendanger}}]{br.at.10}
\bibinfo{author}{\bibfnamefont{J.}~\bibnamefont{Brede}},
  \bibinfo{author}{\bibfnamefont{N.}~\bibnamefont{Atodiresei}},
  \bibinfo{author}{\bibfnamefont{S.}~\bibnamefont{Kuck}},
  \bibinfo{author}{\bibfnamefont{P.}~\bibnamefont{Lazi\ifmmode~\acute{c}\else
  \'{c}\fi{}}}, \bibinfo{author}{\bibfnamefont{V.}~\bibnamefont{Caciuc}},
  \bibinfo{author}{\bibfnamefont{Y.}~\bibnamefont{Morikawa}},
  \bibinfo{author}{\bibfnamefont{G.}~\bibnamefont{Hoffmann}},
  \bibinfo{author}{\bibfnamefont{S.}~\bibnamefont{Bl\"ugel}}, \bibnamefont{and}
  \bibinfo{author}{\bibfnamefont{R.}~\bibnamefont{Wiesendanger}},
  \bibinfo{journal}{Phys. Rev. Lett.} \textbf{\bibinfo{volume}{105}},
  \bibinfo{pages}{047204} (\bibinfo{year}{2010}),
  \urlprefix\url{https://link.aps.org/doi/10.1103/PhysRevLett.105.047204}.

\bibitem[{\citenamefont{Barraud et~al.}(2010)\citenamefont{Barraud, Seneor,
  Mattana, Fusil, Bouzehouane, Deranlot, Graziosi, Hueso, Bergenti, Dediu
  et~al.}}]{Barraud2010}
\bibinfo{author}{\bibfnamefont{C.}~\bibnamefont{Barraud}},
  \bibinfo{author}{\bibfnamefont{P.}~\bibnamefont{Seneor}},
  \bibinfo{author}{\bibfnamefont{R.}~\bibnamefont{Mattana}},
  \bibinfo{author}{\bibfnamefont{S.}~\bibnamefont{Fusil}},
  \bibinfo{author}{\bibfnamefont{K.}~\bibnamefont{Bouzehouane}},
  \bibinfo{author}{\bibfnamefont{C.}~\bibnamefont{Deranlot}},
  \bibinfo{author}{\bibfnamefont{P.}~\bibnamefont{Graziosi}},
  \bibinfo{author}{\bibfnamefont{L.}~\bibnamefont{Hueso}},
  \bibinfo{author}{\bibfnamefont{I.}~\bibnamefont{Bergenti}},
  \bibinfo{author}{\bibfnamefont{V.}~\bibnamefont{Dediu}},
  \bibnamefont{et~al.}, \bibinfo{journal}{Nature Physics}
  \textbf{\bibinfo{volume}{6}}, \bibinfo{pages}{615} (\bibinfo{year}{2010}),
  ISSN \bibinfo{issn}{1745-2481},
  \urlprefix\url{https://doi.org/10.1038/nphys1688}.

\bibitem[{\citenamefont{Javaid et~al.}(2010)\citenamefont{Javaid, Bowen,
  Boukari, Joly, Beaufrand, Chen, Dappe, Scheurer, Kappler, Arabski
  et~al.}}]{ja.bo.10}
\bibinfo{author}{\bibfnamefont{S.}~\bibnamefont{Javaid}},
  \bibinfo{author}{\bibfnamefont{M.}~\bibnamefont{Bowen}},
  \bibinfo{author}{\bibfnamefont{S.}~\bibnamefont{Boukari}},
  \bibinfo{author}{\bibfnamefont{L.}~\bibnamefont{Joly}},
  \bibinfo{author}{\bibfnamefont{J.-B.} \bibnamefont{Beaufrand}},
  \bibinfo{author}{\bibfnamefont{X.}~\bibnamefont{Chen}},
  \bibinfo{author}{\bibfnamefont{Y.~J.} \bibnamefont{Dappe}},
  \bibinfo{author}{\bibfnamefont{F.}~\bibnamefont{Scheurer}},
  \bibinfo{author}{\bibfnamefont{J.-P.} \bibnamefont{Kappler}},
  \bibinfo{author}{\bibfnamefont{J.}~\bibnamefont{Arabski}},
  \bibnamefont{et~al.}, \bibinfo{journal}{Phys. Rev. Lett.}
  \textbf{\bibinfo{volume}{105}}, \bibinfo{pages}{077201}
  (\bibinfo{year}{2010}),
  \urlprefix\url{https://link.aps.org/doi/10.1103/PhysRevLett.105.077201}.

\bibitem[{\citenamefont{Heß et~al.}(2017)\citenamefont{Hess, Friedrich,
  Matthes, Caciuc, Atodiresei, B\"{u}rgler, Bl\"{u}gel, and Schneider}}]{he.ri.17}
\bibinfo{author}{\bibfnamefont{V.}~\bibnamefont{Hess}},
  \bibinfo{author}{\bibfnamefont{R.}~\bibnamefont{Friedrich}},
  \bibinfo{author}{\bibfnamefont{F.}~\bibnamefont{Matthes}},
  \bibinfo{author}{\bibfnamefont{V.}~\bibnamefont{Caciuc}},
  \bibinfo{author}{\bibfnamefont{N.}~\bibnamefont{Atodiresei}},
  \bibinfo{author}{\bibfnamefont{D.~E.} \bibnamefont{B\"{u}rgler}},
  \bibinfo{author}{\bibfnamefont{S.}~\bibnamefont{Bl\"{u}gel}}, \bibnamefont{and}
  \bibinfo{author}{\bibfnamefont{C.~M.} \bibnamefont{Schneider}},
  \bibinfo{journal}{New Journal of Physics} \textbf{\bibinfo{volume}{19}},
  \bibinfo{pages}{053016} (\bibinfo{year}{2017}),
  \urlprefix\url{https://dx.doi.org/10.1088/1367-2630/aa6ece}.

\bibitem[{\citenamefont{Janas et~al.}(2023)\citenamefont{Janas, Droghetti,
  Ponzoni, Cojocariu, Jugovac, Feyer, Radonji, Rungger, Chioncel, Zamborlini
  et~al.}}]{ja.dr.22}
\bibinfo{author}{\bibfnamefont{D.~M.} \bibnamefont{Janas}},
  \bibinfo{author}{\bibfnamefont{A.}~\bibnamefont{Droghetti}},
  \bibinfo{author}{\bibfnamefont{S.}~\bibnamefont{Ponzoni}},
  \bibinfo{author}{\bibfnamefont{I.}~\bibnamefont{Cojocariu}},
  \bibinfo{author}{\bibfnamefont{M.}~\bibnamefont{Jugovac}},
  \bibinfo{author}{\bibfnamefont{V.}~\bibnamefont{Feyer}},
  \bibinfo{author}{\bibfnamefont{M.~M.} \bibnamefont{Radonji}},
  \bibinfo{author}{\bibfnamefont{I.}~\bibnamefont{Rungger}},
  \bibinfo{author}{\bibfnamefont{L.}~\bibnamefont{Chioncel}},
  \bibinfo{author}{\bibfnamefont{G.}~\bibnamefont{Zamborlini}},
  \bibnamefont{et~al.}, \bibinfo{journal}{Advanced Materials}
  \textbf{\bibinfo{volume}{35}}, \bibinfo{pages}{2205698}
  (\bibinfo{year}{2023}),
  \urlprefix\url{https://onlinelibrary.wiley.com/doi/abs/10.1002/adma.202205698}.

\bibitem[{\citenamefont{Callsen et~al.}(2013)\citenamefont{Callsen, Caciuc,
  Kiselev, Atodiresei, and Bl\"ugel}}]{blugel_hard}
\bibinfo{author}{\bibfnamefont{M.}~\bibnamefont{Callsen}},
  \bibinfo{author}{\bibfnamefont{V.}~\bibnamefont{Caciuc}},
  \bibinfo{author}{\bibfnamefont{N.}~\bibnamefont{Kiselev}},
  \bibinfo{author}{\bibfnamefont{N.}~\bibnamefont{Atodiresei}},
  \bibnamefont{and} \bibinfo{author}{\bibfnamefont{S.}~\bibnamefont{Bl\"ugel}},
  \bibinfo{journal}{Phys. Rev. Lett.} \textbf{\bibinfo{volume}{111}},
  \bibinfo{pages}{106805} (\bibinfo{year}{2013}),
  \urlprefix\url{https://link.aps.org/doi/10.1103/PhysRevLett.111.106805}.

\bibitem[{\citenamefont{Friedrich
  et~al.}(2015{\natexlab{a}})\citenamefont{Friedrich, Caciuc, Kiselev,
  Atodiresei, and Bl\"ugel}}]{fr.ca.15}
\bibinfo{author}{\bibfnamefont{R.}~\bibnamefont{Friedrich}},
  \bibinfo{author}{\bibfnamefont{V.}~\bibnamefont{Caciuc}},
  \bibinfo{author}{\bibfnamefont{N.~S.} \bibnamefont{Kiselev}},
  \bibinfo{author}{\bibfnamefont{N.}~\bibnamefont{Atodiresei}},
  \bibnamefont{and} \bibinfo{author}{\bibfnamefont{S.}~\bibnamefont{Bl\"ugel}},
  \bibinfo{journal}{Phys. Rev. B} \textbf{\bibinfo{volume}{91}},
  \bibinfo{pages}{115432} (\bibinfo{year}{2015}{\natexlab{a}}),
  \urlprefix\url{https://link.aps.org/doi/10.1103/PhysRevB.91.115432}.

\bibitem[{\citenamefont{Raman et~al.}(2013)\citenamefont{Raman, Kamerbeek,
  Mukherjee, Atodiresei, Sen, Lazi{\'{c}}, Caciuc, Michel, Stalke, Mandal
  et~al.}}]{Raman2013}
\bibinfo{author}{\bibfnamefont{K.~V.} \bibnamefont{Raman}},
  \bibinfo{author}{\bibfnamefont{A.~M.} \bibnamefont{Kamerbeek}},
  \bibinfo{author}{\bibfnamefont{A.}~\bibnamefont{Mukherjee}},
  \bibinfo{author}{\bibfnamefont{N.}~\bibnamefont{Atodiresei}},
  \bibinfo{author}{\bibfnamefont{T.~K.} \bibnamefont{Sen}},
  \bibinfo{author}{\bibfnamefont{P.}~\bibnamefont{Lazi{\'{c}}}},
  \bibinfo{author}{\bibfnamefont{V.}~\bibnamefont{Caciuc}},
  \bibinfo{author}{\bibfnamefont{R.}~\bibnamefont{Michel}},
  \bibinfo{author}{\bibfnamefont{D.}~\bibnamefont{Stalke}},
  \bibinfo{author}{\bibfnamefont{S.~K.} \bibnamefont{Mandal}},
  \bibnamefont{et~al.}, \bibinfo{journal}{Nature}
  \textbf{\bibinfo{volume}{493}}, \bibinfo{pages}{509} (\bibinfo{year}{2013}),
  ISSN \bibinfo{issn}{1476-4687},
  \urlprefix\url{https://doi.org/10.1038/nature11719}.

\bibitem[{\citenamefont{Boukari et~al.}(2018)\citenamefont{Boukari, Jabbar,
  Schleicher, Gruber, Avedissian, Arabski, Da~Costa, Schmerber, Rengasamy,
  Vileno et~al.}}]{bo.ja.18}
\bibinfo{author}{\bibfnamefont{S.}~\bibnamefont{Boukari}},
  \bibinfo{author}{\bibfnamefont{H.}~\bibnamefont{Jabbar}},
  \bibinfo{author}{\bibfnamefont{F.}~\bibnamefont{Schleicher}},
  \bibinfo{author}{\bibfnamefont{M.}~\bibnamefont{Gruber}},
  \bibinfo{author}{\bibfnamefont{G.}~\bibnamefont{Avedissian}},
  \bibinfo{author}{\bibfnamefont{J.}~\bibnamefont{Arabski}},
  \bibinfo{author}{\bibfnamefont{V.}~\bibnamefont{Da~Costa}},
  \bibinfo{author}{\bibfnamefont{G.}~\bibnamefont{Schmerber}},
  \bibinfo{author}{\bibfnamefont{P.}~\bibnamefont{Rengasamy}},
  \bibinfo{author}{\bibfnamefont{B.}~\bibnamefont{Vileno}},
  \bibnamefont{et~al.}, \bibinfo{journal}{Nano Letters}
  \textbf{\bibinfo{volume}{18}}, \bibinfo{pages}{4659} (\bibinfo{year}{2018}),
  \bibinfo{note}{pMID: 29991266},
  \urlprefix\url{https://doi.org/10.1021/acs.nanolett.8b00570}.

\bibitem[{\citenamefont{Moorsom et~al.}(2020)\citenamefont{Moorsom, Alghamdi,
  Stansill, Poli, Teobaldi, Beg, Fangohr, Rogers, Aslam, Ali
  et~al.}}]{mo.al.20}
\bibinfo{author}{\bibfnamefont{T.}~\bibnamefont{Moorsom}},
  \bibinfo{author}{\bibfnamefont{S.}~\bibnamefont{Alghamdi}},
  \bibinfo{author}{\bibfnamefont{S.}~\bibnamefont{Stansill}},
  \bibinfo{author}{\bibfnamefont{E.}~\bibnamefont{Poli}},
  \bibinfo{author}{\bibfnamefont{G.}~\bibnamefont{Teobaldi}},
  \bibinfo{author}{\bibfnamefont{M.}~\bibnamefont{Beg}},
  \bibinfo{author}{\bibfnamefont{H.}~\bibnamefont{Fangohr}},
  \bibinfo{author}{\bibfnamefont{M.}~\bibnamefont{Rogers}},
  \bibinfo{author}{\bibfnamefont{Z.}~\bibnamefont{Aslam}},
  \bibinfo{author}{\bibfnamefont{M.}~\bibnamefont{Ali}}, \bibnamefont{et~al.},
  \bibinfo{journal}{Phys. Rev. B} \textbf{\bibinfo{volume}{101}},
  \bibinfo{pages}{060408} (\bibinfo{year}{2020}),
  \urlprefix\url{https://link.aps.org/doi/10.1103/PhysRevB.101.060408}.

\bibitem[{\citenamefont{Benini et~al.}(2022)\citenamefont{Benini, Allodi,
  Surpi, Riminucci, Lin, Sanna, Dediu, and Bergenti}}]{be.al.21}
\bibinfo{author}{\bibfnamefont{M.}~\bibnamefont{Benini}},
  \bibinfo{author}{\bibfnamefont{G.}~\bibnamefont{Allodi}},
  \bibinfo{author}{\bibfnamefont{A.}~\bibnamefont{Surpi}},
  \bibinfo{author}{\bibfnamefont{A.}~\bibnamefont{Riminucci}},
  \bibinfo{author}{\bibfnamefont{K.-W.} \bibnamefont{Lin}},
  \bibinfo{author}{\bibfnamefont{S.}~\bibnamefont{Sanna}},
  \bibinfo{author}{\bibfnamefont{V.~A.} \bibnamefont{Dediu}}, \bibnamefont{and}
  \bibinfo{author}{\bibfnamefont{I.}~\bibnamefont{Bergenti}},
  \bibinfo{journal}{Advanced Materials Interfaces}
  \textbf{\bibinfo{volume}{9}}, \bibinfo{pages}{2201394}
  (\bibinfo{year}{2022}),
  \urlprefix\url{https://onlinelibrary.wiley.com/doi/abs/10.1002/admi.202201394}.

\bibitem[{\citenamefont{Ma'Mari et~al.}(2015)\citenamefont{Ma'Mari, Moorsom,
  Teobaldi, Deacon, Prokscha, Luetkens, Lee, Sterbinsky, Arena, MacLaren
  et~al.}}]{MaMari2015-jo}
\bibinfo{author}{\bibfnamefont{F.~A.} \bibnamefont{Ma'Mari}},
  \bibinfo{author}{\bibfnamefont{T.}~\bibnamefont{Moorsom}},
  \bibinfo{author}{\bibfnamefont{G.}~\bibnamefont{Teobaldi}},
  \bibinfo{author}{\bibfnamefont{W.}~\bibnamefont{Deacon}},
  \bibinfo{author}{\bibfnamefont{T.}~\bibnamefont{Prokscha}},
  \bibinfo{author}{\bibfnamefont{H.}~\bibnamefont{Luetkens}},
  \bibinfo{author}{\bibfnamefont{S.}~\bibnamefont{Lee}},
  \bibinfo{author}{\bibfnamefont{G.~E.} \bibnamefont{Sterbinsky}},
  \bibinfo{author}{\bibfnamefont{D.~A.} \bibnamefont{Arena}},
  \bibinfo{author}{\bibfnamefont{D.~A.} \bibnamefont{MacLaren}},
  \bibnamefont{et~al.}, \bibinfo{journal}{Nature}
  \textbf{\bibinfo{volume}{524}}, \bibinfo{pages}{69} (\bibinfo{year}{2015}).

\bibitem[{\citenamefont{Pang et~al.}(2016)\citenamefont{Pang, Shi, and
  Van~Hove}}]{pa.sh.16}
\bibinfo{author}{\bibfnamefont{R.}~\bibnamefont{Pang}},
  \bibinfo{author}{\bibfnamefont{X.}~\bibnamefont{Shi}}, \bibnamefont{and}
  \bibinfo{author}{\bibfnamefont{M.~A.} \bibnamefont{Van~Hove}},
  \bibinfo{journal}{Journal of the American Chemical Society}
  \textbf{\bibinfo{volume}{138}}, \bibinfo{pages}{4029} (\bibinfo{year}{2016}),
  \bibinfo{note}{pMID: 26966934},
  \urlprefix\url{https://doi.org/10.1021/jacs.5b10967}.

\bibitem[{\citenamefont{Fourmental et~al.}(2021)\citenamefont{Fourmental,
  Le~Laurent, Repain, Chacon, Girard, Lagoute, Rousset, Coati, Garreau, Resta
  et~al.}}]{fl.le.21}
\bibinfo{author}{\bibfnamefont{C.}~\bibnamefont{Fourmental}},
  \bibinfo{author}{\bibfnamefont{L.}~\bibnamefont{Le~Laurent}},
  \bibinfo{author}{\bibfnamefont{V.}~\bibnamefont{Repain}},
  \bibinfo{author}{\bibfnamefont{C.}~\bibnamefont{Chacon}},
  \bibinfo{author}{\bibfnamefont{Y.}~\bibnamefont{Girard}},
  \bibinfo{author}{\bibfnamefont{J.}~\bibnamefont{Lagoute}},
  \bibinfo{author}{\bibfnamefont{S.}~\bibnamefont{Rousset}},
  \bibinfo{author}{\bibfnamefont{A.}~\bibnamefont{Coati}},
  \bibinfo{author}{\bibfnamefont{Y.}~\bibnamefont{Garreau}},
  \bibinfo{author}{\bibfnamefont{A.}~\bibnamefont{Resta}},
  \bibnamefont{et~al.}, \bibinfo{journal}{Phys. Rev. B}
  \textbf{\bibinfo{volume}{104}}, \bibinfo{pages}{235413}
  (\bibinfo{year}{2021}),
  \urlprefix\url{https://link.aps.org/doi/10.1103/PhysRevB.104.235413}.

\bibitem[{\citenamefont{St{\"o}hr and Siegmann}()}]{stohrmagnetism}
\bibinfo{author}{\bibfnamefont{J.}~\bibnamefont{St{\"o}hr}} \bibnamefont{and}
  \bibinfo{author}{\bibfnamefont{H.}~\bibnamefont{Siegmann}},
  \emph{\bibinfo{title}{Magnetism: from fundamentals to nanoscale dynamics.
  2006}}.

\bibitem[{\citenamefont{Johnson et~al.}(1996)\citenamefont{Johnson, Bloemen,
  den Broeder, and de~Vries}}]{Johnson_1996}
\bibinfo{author}{\bibfnamefont{M.~T.} \bibnamefont{Johnson}},
  \bibinfo{author}{\bibfnamefont{P.~J.~H.} \bibnamefont{Bloemen}},
  \bibinfo{author}{\bibfnamefont{F.~J.~A.} \bibnamefont{den Broeder}},
  \bibnamefont{and} \bibinfo{author}{\bibfnamefont{J.~J.}
  \bibnamefont{de~Vries}}, \bibinfo{journal}{Reports on Progress in Physics}
  \textbf{\bibinfo{volume}{59}}, \bibinfo{pages}{1409} (\bibinfo{year}{1996}),
  \urlprefix\url{https://doi.org/10.1088/0034-4885/59/11/002}.

\bibitem[{\citenamefont{Bairagi et~al.}(2015)\citenamefont{Bairagi, Bellec,
  Repain, Chacon, Girard, Garreau, Lagoute, Rousset, Breitwieser, Hu
  et~al.}}]{bairagiprl}
\bibinfo{author}{\bibfnamefont{K.}~\bibnamefont{Bairagi}},
  \bibinfo{author}{\bibfnamefont{A.}~\bibnamefont{Bellec}},
  \bibinfo{author}{\bibfnamefont{V.}~\bibnamefont{Repain}},
  \bibinfo{author}{\bibfnamefont{C.}~\bibnamefont{Chacon}},
  \bibinfo{author}{\bibfnamefont{Y.}~\bibnamefont{Girard}},
  \bibinfo{author}{\bibfnamefont{Y.}~\bibnamefont{Garreau}},
  \bibinfo{author}{\bibfnamefont{J.}~\bibnamefont{Lagoute}},
  \bibinfo{author}{\bibfnamefont{S.}~\bibnamefont{Rousset}},
  \bibinfo{author}{\bibfnamefont{R.}~\bibnamefont{Breitwieser}},
  \bibinfo{author}{\bibfnamefont{Y.-C.} \bibnamefont{Hu}},
  \bibnamefont{et~al.}, \bibinfo{journal}{Phys. Rev. Lett.}
  \textbf{\bibinfo{volume}{114}}, \bibinfo{pages}{247203}
  (\bibinfo{year}{2015}),
  \urlprefix\url{https://link.aps.org/doi/10.1103/PhysRevLett.114.247203}.

\bibitem[{\citenamefont{Bairagi et~al.}(2018)\citenamefont{Bairagi, Bellec,
  Repain, Fourmental, Chacon, Girard, Lagoute, Rousset, Le~Laurent, Smogunov
  et~al.}}]{bairagi_2nd}
\bibinfo{author}{\bibfnamefont{K.}~\bibnamefont{Bairagi}},
  \bibinfo{author}{\bibfnamefont{A.}~\bibnamefont{Bellec}},
  \bibinfo{author}{\bibfnamefont{V.}~\bibnamefont{Repain}},
  \bibinfo{author}{\bibfnamefont{C.}~\bibnamefont{Fourmental}},
  \bibinfo{author}{\bibfnamefont{C.}~\bibnamefont{Chacon}},
  \bibinfo{author}{\bibfnamefont{Y.}~\bibnamefont{Girard}},
  \bibinfo{author}{\bibfnamefont{J.}~\bibnamefont{Lagoute}},
  \bibinfo{author}{\bibfnamefont{S.}~\bibnamefont{Rousset}},
  \bibinfo{author}{\bibfnamefont{L.}~\bibnamefont{Le~Laurent}},
  \bibinfo{author}{\bibfnamefont{A.}~\bibnamefont{Smogunov}},
  \bibnamefont{et~al.}, \bibinfo{journal}{Phys. Rev. B}
  \textbf{\bibinfo{volume}{98}}, \bibinfo{pages}{085432}
  (\bibinfo{year}{2018}),
  \urlprefix\url{https://link.aps.org/doi/10.1103/PhysRevB.98.085432}.

\bibitem[{\citenamefont{Zhu et~al.}(2011)\citenamefont{Zhu, Levy, Ludbrook,
  Veenstra, Rosen, Comin, Wong, Dosanjh, Ubaldini, Syers et~al.}}]{zh.le.11}
\bibinfo{author}{\bibfnamefont{Z.-H.} \bibnamefont{Zhu}},
  \bibinfo{author}{\bibfnamefont{G.}~\bibnamefont{Levy}},
  \bibinfo{author}{\bibfnamefont{B.}~\bibnamefont{Ludbrook}},
  \bibinfo{author}{\bibfnamefont{C.~N.} \bibnamefont{Veenstra}},
  \bibinfo{author}{\bibfnamefont{J.~A.} \bibnamefont{Rosen}},
  \bibinfo{author}{\bibfnamefont{R.}~\bibnamefont{Comin}},
  \bibinfo{author}{\bibfnamefont{D.}~\bibnamefont{Wong}},
  \bibinfo{author}{\bibfnamefont{P.}~\bibnamefont{Dosanjh}},
  \bibinfo{author}{\bibfnamefont{A.}~\bibnamefont{Ubaldini}},
  \bibinfo{author}{\bibfnamefont{P.}~\bibnamefont{Syers}},
  \bibnamefont{et~al.}, \bibinfo{journal}{Phys. Rev. Lett.}
  \textbf{\bibinfo{volume}{107}}, \bibinfo{pages}{186405}
  (\bibinfo{year}{2011}),
  \urlprefix\url{https://link.aps.org/doi/10.1103/PhysRevLett.107.186405}.

\bibitem[{\citenamefont{Jakobs et~al.}(2015)\citenamefont{Jakobs, Narayan,
  Stadtm\"{u}ller, Droghetti, Rungger, Hor, Klyatskaya, Jungkenn, St\"{o}ckl, Laux
  et~al.}}]{ja.na.15}
\bibinfo{author}{\bibfnamefont{S.}~\bibnamefont{Jakobs}},
  \bibinfo{author}{\bibfnamefont{A.}~\bibnamefont{Narayan}},
  \bibinfo{author}{\bibfnamefont{B.}~\bibnamefont{Stadtm\"{u}ller}},
  \bibinfo{author}{\bibfnamefont{A.}~\bibnamefont{Droghetti}},
  \bibinfo{author}{\bibfnamefont{I.}~\bibnamefont{Rungger}},
  \bibinfo{author}{\bibfnamefont{Y.~S.} \bibnamefont{Hor}},
  \bibinfo{author}{\bibfnamefont{S.}~\bibnamefont{Klyatskaya}},
  \bibinfo{author}{\bibfnamefont{D.}~\bibnamefont{Jungkenn}},
  \bibinfo{author}{\bibfnamefont{J.}~\bibnamefont{St\"{o]}ckl}},
  \bibinfo{author}{\bibfnamefont{M.}~\bibnamefont{Laux}}, \bibnamefont{et~al.},
  \bibinfo{journal}{Nano Letters} \textbf{\bibinfo{volume}{15}},
  \bibinfo{pages}{6022} (\bibinfo{year}{2015}), \bibinfo{note}{pMID: 26262825},
  \urlprefix\url{https://doi.org/10.1021/acs.nanolett.5b02213}.

\bibitem[{\citenamefont{Stadtm\"uller et~al.}(2016)\citenamefont{Stadtm\"uller,
  Seidel, Haag, Grad, Tusche, van Straaten, Franke, Kirschner, Kumpf, Cinchetti
  et~al.}}]{st.se.16}
\bibinfo{author}{\bibfnamefont{B.}~\bibnamefont{Stadtm\"uller}},
  \bibinfo{author}{\bibfnamefont{J.}~\bibnamefont{Seidel}},
  \bibinfo{author}{\bibfnamefont{N.}~\bibnamefont{Haag}},
  \bibinfo{author}{\bibfnamefont{L.}~\bibnamefont{Grad}},
  \bibinfo{author}{\bibfnamefont{C.}~\bibnamefont{Tusche}},
  \bibinfo{author}{\bibfnamefont{G.}~\bibnamefont{van Straaten}},
  \bibinfo{author}{\bibfnamefont{M.}~\bibnamefont{Franke}},
  \bibinfo{author}{\bibfnamefont{J.}~\bibnamefont{Kirschner}},
  \bibinfo{author}{\bibfnamefont{C.}~\bibnamefont{Kumpf}},
  \bibinfo{author}{\bibfnamefont{M.}~\bibnamefont{Cinchetti}},
  \bibnamefont{et~al.}, \bibinfo{journal}{Phys. Rev. Lett.}
  \textbf{\bibinfo{volume}{117}}, \bibinfo{pages}{096805}
  (\bibinfo{year}{2016}),
  \urlprefix\url{https://link.aps.org/doi/10.1103/PhysRevLett.117.096805}.

\bibitem[{\citenamefont{Alotibi et~al.}(2021)\citenamefont{Alotibi, Hickey,
  Teobaldi, Ali, Barker, Poli, O'Regan, Ramasse, Burnell, Patchett
  et~al.}}]{at.hi.21}
\bibinfo{author}{\bibfnamefont{S.}~\bibnamefont{Alotibi}},
  \bibinfo{author}{\bibfnamefont{B.~J.} \bibnamefont{Hickey}},
  \bibinfo{author}{\bibfnamefont{G.}~\bibnamefont{Teobaldi}},
  \bibinfo{author}{\bibfnamefont{M.}~\bibnamefont{Ali}},
  \bibinfo{author}{\bibfnamefont{J.}~\bibnamefont{Barker}},
  \bibinfo{author}{\bibfnamefont{E.}~\bibnamefont{Poli}},
  \bibinfo{author}{\bibfnamefont{D.~D.} \bibnamefont{O'Regan}},
  \bibinfo{author}{\bibfnamefont{Q.}~\bibnamefont{Ramasse}},
  \bibinfo{author}{\bibfnamefont{G.}~\bibnamefont{Burnell}},
  \bibinfo{author}{\bibfnamefont{J.}~\bibnamefont{Patchett}},
  \bibnamefont{et~al.}, \bibinfo{journal}{ACS Applied Materials \& Interfaces}
  \textbf{\bibinfo{volume}{13}}, \bibinfo{pages}{5228} (\bibinfo{year}{2021}),
  \bibinfo{note}{pMID: 33470108},
  \urlprefix\url{https://doi.org/10.1021/acsami.0c19403}.

\bibitem[{\citenamefont{Skubic et~al.}(2008)\citenamefont{Skubic, Hellsvik,
  Nordstr\"{o}m, and Eriksson}}]{Skubic_2008}
\bibinfo{author}{\bibfnamefont{B.}~\bibnamefont{Skubic}},
  \bibinfo{author}{\bibfnamefont{J.}~\bibnamefont{Hellsvik}},
  \bibinfo{author}{\bibfnamefont{L.}~\bibnamefont{Nordstr\"{o}m}},
  \bibnamefont{and} \bibinfo{author}{\bibfnamefont{O.}~\bibnamefont{Eriksson}},
  \bibinfo{journal}{Journal of Physics: Condensed Matter}
  \textbf{\bibinfo{volume}{20}}, \bibinfo{pages}{315203}
  (\bibinfo{year}{2008}),
  \urlprefix\url{https://dx.doi.org/10.1088/0953-8984/20/31/315203}.

\bibitem[{\citenamefont{Evans et~al.}(2014)\citenamefont{Evans, Fan,
  Chureemart, Ostler, Ellis, and Chantrell}}]{Evans_2014}
\bibinfo{author}{\bibfnamefont{R.~F.~L.} \bibnamefont{Evans}},
  \bibinfo{author}{\bibfnamefont{W.~J.} \bibnamefont{Fan}},
  \bibinfo{author}{\bibfnamefont{P.}~\bibnamefont{Chureemart}},
  \bibinfo{author}{\bibfnamefont{T.~A.} \bibnamefont{Ostler}},
  \bibinfo{author}{\bibfnamefont{M.~O.~A.} \bibnamefont{Ellis}},
  \bibnamefont{and} \bibinfo{author}{\bibfnamefont{R.~W.}
  \bibnamefont{Chantrell}}, \bibinfo{journal}{Journal of Physics: Condensed
  Matter} \textbf{\bibinfo{volume}{26}}, \bibinfo{pages}{103202}
  (\bibinfo{year}{2014}),
  \urlprefix\url{https://dx.doi.org/10.1088/0953-8984/26/10/103202}.

\bibitem[{\citenamefont{Bruno}(1989)}]{bruno}
\bibinfo{author}{\bibfnamefont{P.}~\bibnamefont{Bruno}},
  \bibinfo{journal}{Phys. Rev. B} \textbf{\bibinfo{volume}{39}},
  \bibinfo{pages}{865} (\bibinfo{year}{1989}),
  \urlprefix\url{https://link.aps.org/doi/10.1103/PhysRevB.39.865}.

\bibitem[{\citenamefont{Bhandary et~al.}(2011)\citenamefont{Bhandary,
  Gr\aa{}n\"as, Szunyogh, Sanyal, Nordstr\"om, and Eriksson}}]{sumanta-multi}
\bibinfo{author}{\bibfnamefont{S.}~\bibnamefont{Bhandary}},
  \bibinfo{author}{\bibfnamefont{O.}~\bibnamefont{Gr\aa{}n\"as}},
  \bibinfo{author}{\bibfnamefont{L.}~\bibnamefont{Szunyogh}},
  \bibinfo{author}{\bibfnamefont{B.}~\bibnamefont{Sanyal}},
  \bibinfo{author}{\bibfnamefont{L.}~\bibnamefont{Nordstr\"om}},
  \bibnamefont{and} \bibinfo{author}{\bibfnamefont{O.}~\bibnamefont{Eriksson}},
  \bibinfo{journal}{Phys. Rev. B} \textbf{\bibinfo{volume}{84}},
  \bibinfo{pages}{092401} (\bibinfo{year}{2011}),
  \urlprefix\url{https://link.aps.org/doi/10.1103/PhysRevB.84.092401}.

\bibitem[{\citenamefont{Zhang et~al.}(2017)\citenamefont{Zhang, Lukashev,
  Jaswal, and Tsymbal}}]{zh.lu.17}
\bibinfo{author}{\bibfnamefont{J.}~\bibnamefont{Zhang}},
  \bibinfo{author}{\bibfnamefont{P.~V.} \bibnamefont{Lukashev}},
  \bibinfo{author}{\bibfnamefont{S.~S.} \bibnamefont{Jaswal}},
  \bibnamefont{and} \bibinfo{author}{\bibfnamefont{E.~Y.}
  \bibnamefont{Tsymbal}}, \bibinfo{journal}{Phys. Rev. B}
  \textbf{\bibinfo{volume}{96}}, \bibinfo{pages}{014435}
  (\bibinfo{year}{2017}),
  \urlprefix\url{https://link.aps.org/doi/10.1103/PhysRevB.96.014435}.

\bibitem[{\citenamefont{Gimbert and Calmels}(2012)}]{gi.ca.12}
\bibinfo{author}{\bibfnamefont{F.}~\bibnamefont{Gimbert}} \bibnamefont{and}
  \bibinfo{author}{\bibfnamefont{L.}~\bibnamefont{Calmels}},
  \bibinfo{journal}{Phys. Rev. B} \textbf{\bibinfo{volume}{86}},
  \bibinfo{pages}{184407} (\bibinfo{year}{2012}),
  \urlprefix\url{https://link.aps.org/doi/10.1103/PhysRevB.86.184407}.

\bibitem[{\citenamefont{Daalderop et~al.}(1994)\citenamefont{Daalderop, Kelly,
  and Schuurmans}}]{Daalderop}
\bibinfo{author}{\bibfnamefont{G.~H.~O.} \bibnamefont{Daalderop}},
  \bibinfo{author}{\bibfnamefont{P.~J.} \bibnamefont{Kelly}}, \bibnamefont{and}
  \bibinfo{author}{\bibfnamefont{M.~F.~H.} \bibnamefont{Schuurmans}},
  \bibinfo{journal}{Phys. Rev. B} \textbf{\bibinfo{volume}{50}},
  \bibinfo{pages}{9989} (\bibinfo{year}{1994}),
  \urlprefix\url{https://link.aps.org/doi/10.1103/PhysRevB.50.9989}.

\bibitem[{\citenamefont{Kyuno et~al.}(1996)\citenamefont{Kyuno, Ha, Yamamoto,
  and Asano}}]{Kyuno_1996}
\bibinfo{author}{\bibfnamefont{K.}~\bibnamefont{Kyuno}},
  \bibinfo{author}{\bibfnamefont{J.-G.} \bibnamefont{Ha}},
  \bibinfo{author}{\bibfnamefont{R.}~\bibnamefont{Yamamoto}}, \bibnamefont{and}
  \bibinfo{author}{\bibfnamefont{S.}~\bibnamefont{Asano}},
  \bibinfo{journal}{Japanese Journal of Applied Physics}
  \textbf{\bibinfo{volume}{35}}, \bibinfo{pages}{2774} (\bibinfo{year}{1996}),
  \urlprefix\url{https://doi.org/10.1143/jjap.35.2774}.

\bibitem[{\citenamefont{Liz\'arraga et~al.}(2017)\citenamefont{Liz\'arraga,
  Pan, Bergqvist, Holmstr\"om, Gercsi, and Vitos}}]{Li.Pa17}
\bibinfo{author}{\bibfnamefont{R.}~\bibnamefont{Liz\'arraga}},
  \bibinfo{author}{\bibfnamefont{F.}~\bibnamefont{Pan}},
  \bibinfo{author}{\bibfnamefont{L.}~\bibnamefont{Bergqvist}},
  \bibinfo{author}{\bibfnamefont{E.}~\bibnamefont{Holmstr\"om}},
  \bibinfo{author}{\bibfnamefont{Z.}~\bibnamefont{Gercsi}}, \bibnamefont{and}
  \bibinfo{author}{\bibfnamefont{L.}~\bibnamefont{Vitos}},
  \bibinfo{journal}{Sci. Rep.} \textbf{\bibinfo{volume}{7}},
  \bibinfo{pages}{3778} (\bibinfo{year}{2017}),
  \urlprefix\url{https://doi.org/10.1038/s41598-017-03877-5}.

\bibitem[{\citenamefont{Bl\"ochl}(1994)}]{PAW}
\bibinfo{author}{\bibfnamefont{P.~E.} \bibnamefont{Bl\"ochl}},
  \bibinfo{journal}{Phys. Rev. B} \textbf{\bibinfo{volume}{50}},
  \bibinfo{pages}{17953} (\bibinfo{year}{1994}),
  \urlprefix\url{https://link.aps.org/doi/10.1103/PhysRevB.50.17953}.

\bibitem[{\citenamefont{Kresse and Hafner}(1993)}]{VASP}
\bibinfo{author}{\bibfnamefont{G.}~\bibnamefont{Kresse}} \bibnamefont{and}
  \bibinfo{author}{\bibfnamefont{J.}~\bibnamefont{Hafner}},
  \bibinfo{journal}{Phys. Rev. B} \textbf{\bibinfo{volume}{47}},
  \bibinfo{pages}{558} (\bibinfo{year}{1993}),
  \urlprefix\url{https://link.aps.org/doi/10.1103/PhysRevB.47.558}.

\bibitem[{\citenamefont{Perdew et~al.}(1996)\citenamefont{Perdew, Burke, and
  Ernzerhof}}]{PBE}
\bibinfo{author}{\bibfnamefont{J.~P.} \bibnamefont{Perdew}},
  \bibinfo{author}{\bibfnamefont{K.}~\bibnamefont{Burke}}, \bibnamefont{and}
  \bibinfo{author}{\bibfnamefont{M.}~\bibnamefont{Ernzerhof}},
  \bibinfo{journal}{Phys. Rev. Lett.} \textbf{\bibinfo{volume}{77}},
  \bibinfo{pages}{3865} (\bibinfo{year}{1996}),
  \urlprefix\url{https://link.aps.org/doi/10.1103/PhysRevLett.77.3865}.

\bibitem[{\citenamefont{Bl\"ochl et~al.}(1994)\citenamefont{Bl\"ochl, Jepsen,
  and Andersen}}]{tetra}
\bibinfo{author}{\bibfnamefont{P.~E.} \bibnamefont{Bl\"ochl}},
  \bibinfo{author}{\bibfnamefont{O.}~\bibnamefont{Jepsen}}, \bibnamefont{and}
  \bibinfo{author}{\bibfnamefont{O.~K.} \bibnamefont{Andersen}},
  \bibinfo{journal}{Phys. Rev. B} \textbf{\bibinfo{volume}{49}},
  \bibinfo{pages}{16223} (\bibinfo{year}{1994}),
  \urlprefix\url{https://link.aps.org/doi/10.1103/PhysRevB.49.16223}.

\bibitem[{\citenamefont{Owen and Jones}(1954)}]{Owen_1954}
\bibinfo{author}{\bibfnamefont{E.~A.} \bibnamefont{Owen}} \bibnamefont{and}
  \bibinfo{author}{\bibfnamefont{D.~M.} \bibnamefont{Jones}},
  \bibinfo{journal}{Proceedings of the Physical Society. Section B}
  \textbf{\bibinfo{volume}{67}}, \bibinfo{pages}{456} (\bibinfo{year}{1954}),
  \urlprefix\url{https://doi.org/10.1088/0370-1301/67/6/302}.

\bibitem[{\citenamefont{Andersen et~al.}(1979)\citenamefont{Andersen, Skriver,
  Nohl, and B.}}]{an.sk.79}
\bibinfo{author}{\bibfnamefont{O.~K.} \bibnamefont{Andersen}},
  \bibinfo{author}{\bibfnamefont{H.~L.} \bibnamefont{Skriver}},
  \bibinfo{author}{\bibfnamefont{H.}~\bibnamefont{Nohl}}, \bibnamefont{and}
  \bibinfo{author}{\bibfnamefont{J.}~\bibnamefont{B.}}, \bibinfo{journal}{Pure
  Appl. Chem.} \textbf{\bibinfo{volume}{52}}, \bibinfo{pages}{93}
  (\bibinfo{year}{1979}).

\bibitem[{\citenamefont{Weinert et~al.}(1985)\citenamefont{Weinert, Watson, and
  Davenport}}]{we.wa.85}
\bibinfo{author}{\bibfnamefont{M.}~\bibnamefont{Weinert}},
  \bibinfo{author}{\bibfnamefont{R.~E.} \bibnamefont{Watson}},
  \bibnamefont{and} \bibinfo{author}{\bibfnamefont{J.~W.}
  \bibnamefont{Davenport}}, \bibinfo{journal}{Phys. Rev. B}
  \textbf{\bibinfo{volume}{32}}, \bibinfo{pages}{2115} (\bibinfo{year}{1985}),
  \urlprefix\url{https://link.aps.org/doi/10.1103/PhysRevB.32.2115}.

\bibitem[{\citenamefont{Daalderop et~al.}(1990)\citenamefont{Daalderop, Kelly,
  and Schuurmans}}]{da.ke.90}
\bibinfo{author}{\bibfnamefont{G.~H.~O.} \bibnamefont{Daalderop}},
  \bibinfo{author}{\bibfnamefont{P.~J.} \bibnamefont{Kelly}}, \bibnamefont{and}
  \bibinfo{author}{\bibfnamefont{M.~F.~H.} \bibnamefont{Schuurmans}},
  \bibinfo{journal}{Phys. Rev. B} \textbf{\bibinfo{volume}{41}},
  \bibinfo{pages}{11919} (\bibinfo{year}{1990}),
  \urlprefix\url{https://link.aps.org/doi/10.1103/PhysRevB.41.11919}.

\bibitem[{\citenamefont{Klautau and Eriksson}(2005)}]{erikssonbcc}
\bibinfo{author}{\bibfnamefont{A.~B.} \bibnamefont{Klautau}} \bibnamefont{and}
  \bibinfo{author}{\bibfnamefont{O.}~\bibnamefont{Eriksson}},
  \bibinfo{journal}{Phys. Rev. B} \textbf{\bibinfo{volume}{72}},
  \bibinfo{pages}{014459} (\bibinfo{year}{2005}),
  \urlprefix\url{https://link.aps.org/doi/10.1103/PhysRevB.72.014459}.

\bibitem[{\citenamefont{Steiner et~al.}(2016)\citenamefont{Steiner,
  Khmelevskyi, Marsmann, and Kresse}}]{st.kh.16}
\bibinfo{author}{\bibfnamefont{S.}~\bibnamefont{Steiner}},
  \bibinfo{author}{\bibfnamefont{S.}~\bibnamefont{Khmelevskyi}},
  \bibinfo{author}{\bibfnamefont{M.}~\bibnamefont{Marsmann}}, \bibnamefont{and}
  \bibinfo{author}{\bibfnamefont{G.}~\bibnamefont{Kresse}},
  \bibinfo{journal}{Phys. Rev. B} \textbf{\bibinfo{volume}{93}},
  \bibinfo{pages}{224425} (\bibinfo{year}{2016}),
  \urlprefix\url{https://link.aps.org/doi/10.1103/PhysRevB.93.224425}.

\bibitem[{\citenamefont{Aas et~al.}(2013)\citenamefont{Aas, Hasnip, Cuadrado,
  Plotnikova, Szunyogh, Udvardi, and Chantrell}}]{aa.ha.13}
\bibinfo{author}{\bibfnamefont{C.~J.} \bibnamefont{Aas}},
  \bibinfo{author}{\bibfnamefont{P.~J.} \bibnamefont{Hasnip}},
  \bibinfo{author}{\bibfnamefont{R.}~\bibnamefont{Cuadrado}},
  \bibinfo{author}{\bibfnamefont{E.~M.} \bibnamefont{Plotnikova}},
  \bibinfo{author}{\bibfnamefont{L.}~\bibnamefont{Szunyogh}},
  \bibinfo{author}{\bibfnamefont{L.}~\bibnamefont{Udvardi}}, \bibnamefont{and}
  \bibinfo{author}{\bibfnamefont{R.~W.} \bibnamefont{Chantrell}},
  \bibinfo{journal}{Phys. Rev. B} \textbf{\bibinfo{volume}{88}},
  \bibinfo{pages}{174409} (\bibinfo{year}{2013}),
  \urlprefix\url{https://link.aps.org/doi/10.1103/PhysRevB.88.174409}.

\bibitem[{\citenamefont{Zhang et~al.}(2014)\citenamefont{Zhang, Franz, Czerner,
  and Heiliger}}]{CoFemag}
\bibinfo{author}{\bibfnamefont{J.}~\bibnamefont{Zhang}},
  \bibinfo{author}{\bibfnamefont{C.}~\bibnamefont{Franz}},
  \bibinfo{author}{\bibfnamefont{M.}~\bibnamefont{Czerner}}, \bibnamefont{and}
  \bibinfo{author}{\bibfnamefont{C.}~\bibnamefont{Heiliger}},
  \bibinfo{journal}{Phys. Rev. B} \textbf{\bibinfo{volume}{90}},
  \bibinfo{pages}{184409} (\bibinfo{year}{2014}),
  \urlprefix\url{https://link.aps.org/doi/10.1103/PhysRevB.90.184409}.

\bibitem[{\citenamefont{Ong et~al.}(2015)\citenamefont{Ong, Kioussis, Odkhuu,
  Khalili~Amiri, Wang, and Carman}}]{on.ki.15}
\bibinfo{author}{\bibfnamefont{P.~V.} \bibnamefont{Ong}},
  \bibinfo{author}{\bibfnamefont{N.}~\bibnamefont{Kioussis}},
  \bibinfo{author}{\bibfnamefont{D.}~\bibnamefont{Odkhuu}},
  \bibinfo{author}{\bibfnamefont{P.}~\bibnamefont{Khalili~Amiri}},
  \bibinfo{author}{\bibfnamefont{K.~L.} \bibnamefont{Wang}}, \bibnamefont{and}
  \bibinfo{author}{\bibfnamefont{G.~P.} \bibnamefont{Carman}},
  \bibinfo{journal}{Phys. Rev. B} \textbf{\bibinfo{volume}{92}},
  \bibinfo{pages}{020407} (\bibinfo{year}{2015}),
  \urlprefix\url{https://link.aps.org/doi/10.1103/PhysRevB.92.020407}.

\bibitem[{\citenamefont{Sun et~al.}(2020)\citenamefont{Sun, Kwon, Stamenova,
  Sanvito, and Kioussis}}]{su.kw.20}
\bibinfo{author}{\bibfnamefont{Q.}~\bibnamefont{Sun}},
  \bibinfo{author}{\bibfnamefont{S.}~\bibnamefont{Kwon}},
  \bibinfo{author}{\bibfnamefont{M.}~\bibnamefont{Stamenova}},
  \bibinfo{author}{\bibfnamefont{S.}~\bibnamefont{Sanvito}}, \bibnamefont{and}
  \bibinfo{author}{\bibfnamefont{N.}~\bibnamefont{Kioussis}},
  \bibinfo{journal}{Phys. Rev. B} \textbf{\bibinfo{volume}{101}},
  \bibinfo{pages}{134419} (\bibinfo{year}{2020}),
  \urlprefix\url{https://link.aps.org/doi/10.1103/PhysRevB.101.134419}.

\bibitem[{\citenamefont{Blanco-Rey et~al.}(2019)\citenamefont{Blanco-Rey,
  Cerd{\'{a}}, and Arnau}}]{bl.ce.19}
\bibinfo{author}{\bibfnamefont{M.}~\bibnamefont{Blanco-Rey}},
  \bibinfo{author}{\bibfnamefont{J.~I.} \bibnamefont{Cerd{\'{a}}}},
  \bibnamefont{and} \bibinfo{author}{\bibfnamefont{A.}~\bibnamefont{Arnau}},
  \textbf{\bibinfo{volume}{21}}, \bibinfo{pages}{073054}
  (\bibinfo{year}{2019}),
  \urlprefix\url{https://doi.org/10.1088/1367-2630/ab3060}.

\bibitem[{\citenamefont{Trygg et~al.}(1995)\citenamefont{Trygg, Johansson,
  Eriksson, and Wills}}]{PhysRevLett.75.2871}
\bibinfo{author}{\bibfnamefont{J.}~\bibnamefont{Trygg}},
  \bibinfo{author}{\bibfnamefont{B.}~\bibnamefont{Johansson}},
  \bibinfo{author}{\bibfnamefont{O.}~\bibnamefont{Eriksson}}, \bibnamefont{and}
  \bibinfo{author}{\bibfnamefont{J.~M.} \bibnamefont{Wills}},
  \bibinfo{journal}{Phys. Rev. Lett.} \textbf{\bibinfo{volume}{75}},
  \bibinfo{pages}{2871} (\bibinfo{year}{1995}),
  \urlprefix\url{https://link.aps.org/doi/10.1103/PhysRevLett.75.2871}.

\bibitem[{\citenamefont{Yang et~al.}(2001)\citenamefont{Yang, Savrasov, and
  Kotliar}}]{ya.sa.01}
\bibinfo{author}{\bibfnamefont{I.}~\bibnamefont{Yang}},
  \bibinfo{author}{\bibfnamefont{S.~Y.} \bibnamefont{Savrasov}},
  \bibnamefont{and} \bibinfo{author}{\bibfnamefont{G.}~\bibnamefont{Kotliar}},
  \bibinfo{journal}{Phys. Rev. Lett.} \textbf{\bibinfo{volume}{87}},
  \bibinfo{pages}{216405} (\bibinfo{year}{2001}),
  \urlprefix\url{https://link.aps.org/doi/10.1103/PhysRevLett.87.216405}.

\bibitem[{\citenamefont{Burkert et~al.}(2004)\citenamefont{Burkert, Eriksson,
  James, Simak, Johansson, and Nordstr\"om}}]{bu.er.04}
\bibinfo{author}{\bibfnamefont{T.}~\bibnamefont{Burkert}},
  \bibinfo{author}{\bibfnamefont{O.}~\bibnamefont{Eriksson}},
  \bibinfo{author}{\bibfnamefont{P.}~\bibnamefont{James}},
  \bibinfo{author}{\bibfnamefont{S.~I.} \bibnamefont{Simak}},
  \bibinfo{author}{\bibfnamefont{B.}~\bibnamefont{Johansson}},
  \bibnamefont{and}
  \bibinfo{author}{\bibfnamefont{L.}~\bibnamefont{Nordstr\"om}},
  \bibinfo{journal}{Phys. Rev. B} \textbf{\bibinfo{volume}{69}},
  \bibinfo{pages}{104426} (\bibinfo{year}{2004}),
  \urlprefix\url{https://link.aps.org/doi/10.1103/PhysRevB.69.104426}.

\bibitem[{\citenamefont{Resta}(2010)}]{re.10}
\bibinfo{author}{\bibfnamefont{R.}~\bibnamefont{Resta}},
  \bibinfo{journal}{Journal of Physics: Condensed Matter}
  \textbf{\bibinfo{volume}{22}}, \bibinfo{pages}{123201}
  (\bibinfo{year}{2010}),
  \urlprefix\url{https://dx.doi.org/10.1088/0953-8984/22/12/123201}.

\bibitem[{\citenamefont{Eriksson et~al.}(1990)\citenamefont{Eriksson, Brooks,
  and Johansson}}]{er.br.90}
\bibinfo{author}{\bibfnamefont{O.}~\bibnamefont{Eriksson}},
  \bibinfo{author}{\bibfnamefont{M.~S.~S.} \bibnamefont{Brooks}},
  \bibnamefont{and}
  \bibinfo{author}{\bibfnamefont{B.}~\bibnamefont{Johansson}},
  \bibinfo{journal}{Phys. Rev. B} \textbf{\bibinfo{volume}{41}},
  \bibinfo{pages}{7311} (\bibinfo{year}{1990}),
  \urlprefix\url{https://link.aps.org/doi/10.1103/PhysRevB.41.7311}.

\bibitem[{\citenamefont{Ceresoli et~al.}(2010)\citenamefont{Ceresoli,
  Gerstmann, Seitsonen, and Mauri}}]{PhysRevB.81.060409}
\bibinfo{author}{\bibfnamefont{D.}~\bibnamefont{Ceresoli}},
  \bibinfo{author}{\bibfnamefont{U.}~\bibnamefont{Gerstmann}},
  \bibinfo{author}{\bibfnamefont{A.~P.} \bibnamefont{Seitsonen}},
  \bibnamefont{and} \bibinfo{author}{\bibfnamefont{F.}~\bibnamefont{Mauri}},
  \bibinfo{journal}{Phys. Rev. B} \textbf{\bibinfo{volume}{81}},
  \bibinfo{pages}{060409} (\bibinfo{year}{2010}),
  \urlprefix\url{https://link.aps.org/doi/10.1103/PhysRevB.81.060409}.

\bibitem[{vas({\natexlab{a}})}]{vaspSOC}
\emph{\bibinfo{title}{Vasp wiki}},
  \bibinfo{howpublished}{\url{https://www.vasp.at/wiki/index.php/LSORBIT}}.

\bibitem[{\citenamefont{Antropov et~al.}(2014)\citenamefont{Antropov, Ke, and
  Aberg}}]{ANTROPOV201435}
\bibinfo{author}{\bibfnamefont{V.}~\bibnamefont{Antropov}},
  \bibinfo{author}{\bibfnamefont{L.}~\bibnamefont{Ke}}, \bibnamefont{and}
  \bibinfo{author}{\bibfnamefont{D.}~\bibnamefont{Aberg}},
  \bibinfo{journal}{Solid State Communications} \textbf{\bibinfo{volume}{194}},
  \bibinfo{pages}{35} (\bibinfo{year}{2014}), ISSN \bibinfo{issn}{0038-1098},
  \urlprefix\url{https://www.sciencedirect.com/science/article/pii/S0038109814002476}.

\bibitem[{\citenamefont{Getzlaff et~al.}(1995)\citenamefont{Getzlaff, Bansmann,
  and Sch\"{o}nhense}}]{ge.ba.95}
\bibinfo{author}{\bibfnamefont{M.}~\bibnamefont{Getzlaff}},
  \bibinfo{author}{\bibfnamefont{J.}~\bibnamefont{Bansmann}}, \bibnamefont{and}
  \bibinfo{author}{\bibfnamefont{G.}~\bibnamefont{Sch\"{o}nhense}},
  \bibinfo{journal}{Surface Science} \textbf{\bibinfo{volume}{323}},
  \bibinfo{pages}{118} (\bibinfo{year}{1995}), ISSN \bibinfo{issn}{0039-6028},
  \urlprefix\url{https://www.sciencedirect.com/science/article/pii/0039602894006415}.

\bibitem[{\citenamefont{Carey et~al.}(2018)\citenamefont{Carey, Zhao, and
  Campbell}}]{ca.zh.18}
\bibinfo{author}{\bibfnamefont{S.~J.} \bibnamefont{Carey}},
  \bibinfo{author}{\bibfnamefont{W.}~\bibnamefont{Zhao}}, \bibnamefont{and}
  \bibinfo{author}{\bibfnamefont{C.~T.} \bibnamefont{Campbell}},
  \bibinfo{journal}{Surface Science} \textbf{\bibinfo{volume}{676}},
  \bibinfo{pages}{9} (\bibinfo{year}{2018}), ISSN \bibinfo{issn}{0039-6028},
  \bibinfo{note}{special Issue Dedicated to the 75 birthday of Peter R Norton},
  \urlprefix\url{https://www.sciencedirect.com/science/article/pii/S0039602817309779}.

\bibitem[{\citenamefont{Cinchetti et~al.}(2009)\citenamefont{Cinchetti, Heimer,
  W{\"u}stenberg, Andreyev, Bauer, Lach, Ziegler, Gao, and
  Aeschlimann}}]{ci.he.09}
\bibinfo{author}{\bibfnamefont{M.}~\bibnamefont{Cinchetti}},
  \bibinfo{author}{\bibfnamefont{K.}~\bibnamefont{Heimer}},
  \bibinfo{author}{\bibfnamefont{J.-P.} \bibnamefont{W{\"u}stenberg}},
  \bibinfo{author}{\bibfnamefont{O.}~\bibnamefont{Andreyev}},
  \bibinfo{author}{\bibfnamefont{M.}~\bibnamefont{Bauer}},
  \bibinfo{author}{\bibfnamefont{S.}~\bibnamefont{Lach}},
  \bibinfo{author}{\bibfnamefont{C.}~\bibnamefont{Ziegler}},
  \bibinfo{author}{\bibfnamefont{Y.}~\bibnamefont{Gao}}, \bibnamefont{and}
  \bibinfo{author}{\bibfnamefont{M.}~\bibnamefont{Aeschlimann}},
  \bibinfo{journal}{Nature Materials} \textbf{\bibinfo{volume}{8}},
  \bibinfo{pages}{115} (\bibinfo{year}{2009}), ISSN \bibinfo{issn}{1476-4660},
  \urlprefix\url{https://doi.org/10.1038/nmat2334}.

\bibitem[{\citenamefont{Lach et~al.}(2012)\citenamefont{Lach, Altenhof,
  Tarafder, Schmitt, Ali, Vogel, Sauther, Oppeneer, and Ziegler}}]{la.al.12}
\bibinfo{author}{\bibfnamefont{S.}~\bibnamefont{Lach}},
  \bibinfo{author}{\bibfnamefont{A.}~\bibnamefont{Altenhof}},
  \bibinfo{author}{\bibfnamefont{K.}~\bibnamefont{Tarafder}},
  \bibinfo{author}{\bibfnamefont{F.}~\bibnamefont{Schmitt}},
  \bibinfo{author}{\bibfnamefont{M.~E.} \bibnamefont{Ali}},
  \bibinfo{author}{\bibfnamefont{M.}~\bibnamefont{Vogel}},
  \bibinfo{author}{\bibfnamefont{J.}~\bibnamefont{Sauther}},
  \bibinfo{author}{\bibfnamefont{P.~M.} \bibnamefont{Oppeneer}},
  \bibnamefont{and} \bibinfo{author}{\bibfnamefont{C.}~\bibnamefont{Ziegler}},
  \bibinfo{journal}{Advanced Functional Materials}
  \textbf{\bibinfo{volume}{22}}, \bibinfo{pages}{989} (\bibinfo{year}{2012}),
  \urlprefix\url{https://onlinelibrary.wiley.com/doi/abs/10.1002/adfm.201102297}.

\bibitem[{\citenamefont{Djeghloul et~al.}(2013)\citenamefont{Djeghloul,
  Ibrahim, Cantoni, Bowen, Joly, Boukari, Ohresser, Bertran, Le~F{\`e}vre,
  Thakur et~al.}}]{dj.ca.13}
\bibinfo{author}{\bibfnamefont{F.}~\bibnamefont{Djeghloul}},
  \bibinfo{author}{\bibfnamefont{F.}~\bibnamefont{Ibrahim}},
  \bibinfo{author}{\bibfnamefont{M.}~\bibnamefont{Cantoni}},
  \bibinfo{author}{\bibfnamefont{M.}~\bibnamefont{Bowen}},
  \bibinfo{author}{\bibfnamefont{L.}~\bibnamefont{Joly}},
  \bibinfo{author}{\bibfnamefont{S.}~\bibnamefont{Boukari}},
  \bibinfo{author}{\bibfnamefont{P.}~\bibnamefont{Ohresser}},
  \bibinfo{author}{\bibfnamefont{F.}~\bibnamefont{Bertran}},
  \bibinfo{author}{\bibfnamefont{P.}~\bibnamefont{Le~F{\`e}vre}},
  \bibinfo{author}{\bibfnamefont{P.}~\bibnamefont{Thakur}},
  \bibnamefont{et~al.}, \bibinfo{journal}{Scientific Reports}
  \textbf{\bibinfo{volume}{3}}, \bibinfo{pages}{1272} (\bibinfo{year}{2013}),
  ISSN \bibinfo{issn}{2045-2322},
  \urlprefix\url{https://doi.org/10.1038/srep01272}.

\bibitem[{\citenamefont{Droghetti
  et~al.}(2014{\natexlab{a}})\citenamefont{Droghetti, Steil, Gro\ss{}mann,
  Haag, Zhang, Willis, Gillin, Drew, Aeschlimann, Sanvito et~al.}}]{dr.st.14}
\bibinfo{author}{\bibfnamefont{A.}~\bibnamefont{Droghetti}},
  \bibinfo{author}{\bibfnamefont{S.}~\bibnamefont{Steil}},
  \bibinfo{author}{\bibfnamefont{N.}~\bibnamefont{Gro\ss{}mann}},
  \bibinfo{author}{\bibfnamefont{N.}~\bibnamefont{Haag}},
  \bibinfo{author}{\bibfnamefont{H.}~\bibnamefont{Zhang}},
  \bibinfo{author}{\bibfnamefont{M.}~\bibnamefont{Willis}},
  \bibinfo{author}{\bibfnamefont{W.~P.} \bibnamefont{Gillin}},
  \bibinfo{author}{\bibfnamefont{A.~J.} \bibnamefont{Drew}},
  \bibinfo{author}{\bibfnamefont{M.}~\bibnamefont{Aeschlimann}},
  \bibinfo{author}{\bibfnamefont{S.}~\bibnamefont{Sanvito}},
  \bibnamefont{et~al.}, \bibinfo{journal}{Phys. Rev. B}
  \textbf{\bibinfo{volume}{89}}, \bibinfo{pages}{094412}
  (\bibinfo{year}{2014}{\natexlab{a}}),
  \urlprefix\url{https://link.aps.org/doi/10.1103/PhysRevB.89.094412}.

\bibitem[{\citenamefont{Droghetti et~al.}(2016)\citenamefont{Droghetti,
  Thielen, Rungger, Haag, Gro{\ss}mann, St{\"o}ckl, Stadtm{\"u}ller,
  Aeschlimann, Sanvito, and Cinchetti}}]{andrea-alq3}
\bibinfo{author}{\bibfnamefont{A.}~\bibnamefont{Droghetti}},
  \bibinfo{author}{\bibfnamefont{P.}~\bibnamefont{Thielen}},
  \bibinfo{author}{\bibfnamefont{I.}~\bibnamefont{Rungger}},
  \bibinfo{author}{\bibfnamefont{N.}~\bibnamefont{Haag}},
  \bibinfo{author}{\bibfnamefont{N.}~\bibnamefont{Gro{\ss}mann}},
  \bibinfo{author}{\bibfnamefont{J.}~\bibnamefont{St{\"o}ckl}},
  \bibinfo{author}{\bibfnamefont{B.}~\bibnamefont{Stadtm{\"u}ller}},
  \bibinfo{author}{\bibfnamefont{M.}~\bibnamefont{Aeschlimann}},
  \bibinfo{author}{\bibfnamefont{S.}~\bibnamefont{Sanvito}}, \bibnamefont{and}
  \bibinfo{author}{\bibfnamefont{M.}~\bibnamefont{Cinchetti}},
  \bibinfo{journal}{Nature Communications} \textbf{\bibinfo{volume}{7}},
  \bibinfo{pages}{12668} (\bibinfo{year}{2016}), ISSN
  \bibinfo{issn}{2041-1723},
  \urlprefix\url{https://doi.org/10.1038/ncomms12668}.

\bibitem[{\citenamefont{Lach et~al.}(2019)\citenamefont{Lach, Altenhof, Shi,
  Fahlman, and Ziegler}}]{la.al.19}
\bibinfo{author}{\bibfnamefont{S.}~\bibnamefont{Lach}},
  \bibinfo{author}{\bibfnamefont{A.}~\bibnamefont{Altenhof}},
  \bibinfo{author}{\bibfnamefont{S.}~\bibnamefont{Shi}},
  \bibinfo{author}{\bibfnamefont{M.}~\bibnamefont{Fahlman}}, \bibnamefont{and}
  \bibinfo{author}{\bibfnamefont{C.}~\bibnamefont{Ziegler}},
  \bibinfo{journal}{Phys. Chem. Chem. Phys.} \textbf{\bibinfo{volume}{21}},
  \bibinfo{pages}{15833} (\bibinfo{year}{2019}),
  \urlprefix\url{http://dx.doi.org/10.1039/C9CP02205H}.

\bibitem[{\citenamefont{Mittendorfer and Hafner}(2001)}]{mi.ha.01}
\bibinfo{author}{\bibfnamefont{F.}~\bibnamefont{Mittendorfer}}
  \bibnamefont{and} \bibinfo{author}{\bibfnamefont{J.}~\bibnamefont{Hafner}},
  \bibinfo{journal}{Surface Science} \textbf{\bibinfo{volume}{472}},
  \bibinfo{pages}{133} (\bibinfo{year}{2001}), ISSN \bibinfo{issn}{0039-6028},
  \urlprefix\url{https://www.sciencedirect.com/science/article/pii/S0039602800009298}.

\bibitem[{vas({\natexlab{b}})}]{vaspOrbitals}
\emph{\bibinfo{title}{Vasp wiki}},
  \bibinfo{howpublished}{\url{https://www.vasp.at/wiki/index.php/Angular_functions}}.

\bibitem[{\citenamefont{St\"{o}hr}(1999)}]{STOHR1999470}
\bibinfo{author}{\bibfnamefont{J.}~\bibnamefont{St\"{o}hr}},
  \bibinfo{journal}{Journal of Magnetism and Magnetic Materials}
  \textbf{\bibinfo{volume}{200}}, \bibinfo{pages}{470} (\bibinfo{year}{1999}),
  ISSN \bibinfo{issn}{0304-8853},
  \urlprefix\url{https://www.sciencedirect.com/science/article/pii/S0304885399004072}.

\bibitem[{\citenamefont{Carra et~al.}(1993)\citenamefont{Carra, Thole,
  Altarelli, and Wang}}]{PhysRevLett.70.694}
\bibinfo{author}{\bibfnamefont{P.}~\bibnamefont{Carra}},
  \bibinfo{author}{\bibfnamefont{B.~T.} \bibnamefont{Thole}},
  \bibinfo{author}{\bibfnamefont{M.}~\bibnamefont{Altarelli}},
  \bibnamefont{and} \bibinfo{author}{\bibfnamefont{X.}~\bibnamefont{Wang}},
  \bibinfo{journal}{Phys. Rev. Lett.} \textbf{\bibinfo{volume}{70}},
  \bibinfo{pages}{694} (\bibinfo{year}{1993}),
  \urlprefix\url{https://link.aps.org/doi/10.1103/PhysRevLett.70.694}.

\bibitem[{\citenamefont{Tischer et~al.}(1995)\citenamefont{Tischer, Hjortstam,
  Arvanitis, Hunter~Dunn, May, Baberschke, Trygg, Wills, Johansson, and
  Eriksson}}]{PhysRevLett.75.1602}
\bibinfo{author}{\bibfnamefont{M.}~\bibnamefont{Tischer}},
  \bibinfo{author}{\bibfnamefont{O.}~\bibnamefont{Hjortstam}},
  \bibinfo{author}{\bibfnamefont{D.}~\bibnamefont{Arvanitis}},
  \bibinfo{author}{\bibfnamefont{J.}~\bibnamefont{Hunter~Dunn}},
  \bibinfo{author}{\bibfnamefont{F.}~\bibnamefont{May}},
  \bibinfo{author}{\bibfnamefont{K.}~\bibnamefont{Baberschke}},
  \bibinfo{author}{\bibfnamefont{J.}~\bibnamefont{Trygg}},
  \bibinfo{author}{\bibfnamefont{J.~M.} \bibnamefont{Wills}},
  \bibinfo{author}{\bibfnamefont{B.}~\bibnamefont{Johansson}},
  \bibnamefont{and} \bibinfo{author}{\bibfnamefont{O.}~\bibnamefont{Eriksson}},
  \bibinfo{journal}{Phys. Rev. Lett.} \textbf{\bibinfo{volume}{75}},
  \bibinfo{pages}{1602} (\bibinfo{year}{1995}),
  \urlprefix\url{https://link.aps.org/doi/10.1103/PhysRevLett.75.1602}.

\bibitem[{\citenamefont{St\"{o}hr}(1995)}]{STOHR1995253}
\bibinfo{author}{\bibfnamefont{J.}~\bibnamefont{St\"{o}hr}},
  \bibinfo{journal}{Journal of Electron Spectroscopy and Related Phenomena}
  \textbf{\bibinfo{volume}{75}}, \bibinfo{pages}{253} (\bibinfo{year}{1995}),
  ISSN \bibinfo{issn}{0368-2048}, \bibinfo{note}{future Perspectives for
  Electron Spectroscopy with Synchrotron Radiation},
  \urlprefix\url{https://www.sciencedirect.com/science/article/pii/0368204895025375}.

\bibitem[{\citenamefont{Li et~al.}(2013)\citenamefont{Li, Smogunov, Barreteau,
  Ducastelle, and Spanjaard}}]{Li-Fe}
\bibinfo{author}{\bibfnamefont{D.}~\bibnamefont{Li}},
  \bibinfo{author}{\bibfnamefont{A.}~\bibnamefont{Smogunov}},
  \bibinfo{author}{\bibfnamefont{C.}~\bibnamefont{Barreteau}},
  \bibinfo{author}{\bibfnamefont{F.~m.~c.} \bibnamefont{Ducastelle}},
  \bibnamefont{and}
  \bibinfo{author}{\bibfnamefont{D.}~\bibnamefont{Spanjaard}},
  \bibinfo{journal}{Phys. Rev. B} \textbf{\bibinfo{volume}{88}},
  \bibinfo{pages}{214413} (\bibinfo{year}{2013}),
  \urlprefix\url{https://link.aps.org/doi/10.1103/PhysRevB.88.214413}.

\bibitem[{\citenamefont{Wende et~al.}(2007)\citenamefont{Wende, Bernien, Luo,
  Sorg, Ponpandian, Kurde, Miguel, Piantek, Xu, Eckhold et~al.}}]{Wende2007}
\bibinfo{author}{\bibfnamefont{H.}~\bibnamefont{Wende}},
  \bibinfo{author}{\bibfnamefont{M.}~\bibnamefont{Bernien}},
  \bibinfo{author}{\bibfnamefont{J.}~\bibnamefont{Luo}},
  \bibinfo{author}{\bibfnamefont{C.}~\bibnamefont{Sorg}},
  \bibinfo{author}{\bibfnamefont{N.}~\bibnamefont{Ponpandian}},
  \bibinfo{author}{\bibfnamefont{J.}~\bibnamefont{Kurde}},
  \bibinfo{author}{\bibfnamefont{J.}~\bibnamefont{Miguel}},
  \bibinfo{author}{\bibfnamefont{M.}~\bibnamefont{Piantek}},
  \bibinfo{author}{\bibfnamefont{X.}~\bibnamefont{Xu}},
  \bibinfo{author}{\bibfnamefont{P.}~\bibnamefont{Eckhold}},
  \bibnamefont{et~al.}, \bibinfo{journal}{Nature Materials}
  \textbf{\bibinfo{volume}{6}}, \bibinfo{pages}{516} (\bibinfo{year}{2007}),
  ISSN \bibinfo{issn}{1476-4660},
  \urlprefix\url{https://doi.org/10.1038/nmat1932}.

\bibitem[{\citenamefont{Zhan et~al.}(2010)\citenamefont{Zhan, Holmstr\"{o}m,
  Liz\'{a}rraga, Eriksson, Liu, Li, Carlegrim, Stafstr\"{o}m, and
  Fahlman}}]{zh.ho.10}
\bibinfo{author}{\bibfnamefont{Y.}~\bibnamefont{Zhan}},
  \bibinfo{author}{\bibfnamefont{E.}~\bibnamefont{Holmstr\"{o}m}},
  \bibinfo{author}{\bibfnamefont{R.}~\bibnamefont{Liz\'{a}rraga}},
  \bibinfo{author}{\bibfnamefont{O.}~\bibnamefont{Eriksson}},
  \bibinfo{author}{\bibfnamefont{X.}~\bibnamefont{Liu}},
  \bibinfo{author}{\bibfnamefont{F.}~\bibnamefont{Li}},
  \bibinfo{author}{\bibfnamefont{E.}~\bibnamefont{Carlegrim}},
  \bibinfo{author}{\bibfnamefont{S.}~\bibnamefont{Stafstr\"{o}m}},
  \bibnamefont{and} \bibinfo{author}{\bibfnamefont{M.}~\bibnamefont{Fahlman}},
  \bibinfo{journal}{Advanced Materials} \textbf{\bibinfo{volume}{22}},
  \bibinfo{pages}{1626} (\bibinfo{year}{2010}),
  \urlprefix\url{https://onlinelibrary.wiley.com/doi/abs/10.1002/adma.200903556}.

\bibitem[{\citenamefont{Miguel et~al.}(2011)\citenamefont{Miguel, Hermanns,
  Bernien, Kr\"{u}ger, and Kuch}}]{mi.he.11}
\bibinfo{author}{\bibfnamefont{J.}~\bibnamefont{Miguel}},
  \bibinfo{author}{\bibfnamefont{C.~F.} \bibnamefont{Hermanns}},
  \bibinfo{author}{\bibfnamefont{M.}~\bibnamefont{Bernien}},
  \bibinfo{author}{\bibfnamefont{A.}~\bibnamefont{Kr\"{u}ger}}, \bibnamefont{and}
  \bibinfo{author}{\bibfnamefont{W.}~\bibnamefont{Kuch}}, \bibinfo{journal}{The
  Journal of Physical Chemistry Letters} \textbf{\bibinfo{volume}{2}},
  \bibinfo{pages}{1455} (\bibinfo{year}{2011}),
  \urlprefix\url{https://doi.org/10.1021/jz200489y}.

\bibitem[{\citenamefont{Hermanns et~al.}(2013)\citenamefont{Hermanns, Tarafder,
  Bernien, Kr\"{u}ger, Chang, Oppeneer, and Kuch}}]{he.ta.13}
\bibinfo{author}{\bibfnamefont{C.~F.} \bibnamefont{Hermanns}},
  \bibinfo{author}{\bibfnamefont{K.}~\bibnamefont{Tarafder}},
  \bibinfo{author}{\bibfnamefont{M.}~\bibnamefont{Bernien}},
  \bibinfo{author}{\bibfnamefont{A.}~\bibnamefont{Kr\"{u}ger}},
  \bibinfo{author}{\bibfnamefont{Y.-M.} \bibnamefont{Chang}},
  \bibinfo{author}{\bibfnamefont{P.~M.} \bibnamefont{Oppeneer}},
  \bibnamefont{and} \bibinfo{author}{\bibfnamefont{W.}~\bibnamefont{Kuch}},
  \bibinfo{journal}{Advanced Materials} \textbf{\bibinfo{volume}{25}},
  \bibinfo{pages}{3473} (\bibinfo{year}{2013}),
  \urlprefix\url{https://onlinelibrary.wiley.com/doi/abs/10.1002/adma.201205275}.

\bibitem[{\citenamefont{Shi et~al.}(2014)\citenamefont{Shi, Sun, Bedoya-Pinto,
  Graziosi, Li, Liu, Hueso, Dediu, Luo, and Fahlman}}]{sh.su.14}
\bibinfo{author}{\bibfnamefont{S.}~\bibnamefont{Shi}},
  \bibinfo{author}{\bibfnamefont{Z.}~\bibnamefont{Sun}},
  \bibinfo{author}{\bibfnamefont{A.}~\bibnamefont{Bedoya-Pinto}},
  \bibinfo{author}{\bibfnamefont{P.}~\bibnamefont{Graziosi}},
  \bibinfo{author}{\bibfnamefont{X.}~\bibnamefont{Li}},
  \bibinfo{author}{\bibfnamefont{X.}~\bibnamefont{Liu}},
  \bibinfo{author}{\bibfnamefont{L.}~\bibnamefont{Hueso}},
  \bibinfo{author}{\bibfnamefont{V.~A.} \bibnamefont{Dediu}},
  \bibinfo{author}{\bibfnamefont{Y.}~\bibnamefont{Luo}}, \bibnamefont{and}
  \bibinfo{author}{\bibfnamefont{M.}~\bibnamefont{Fahlman}},
  \bibinfo{journal}{Advanced Functional Materials}
  \textbf{\bibinfo{volume}{24}}, \bibinfo{pages}{4812} (\bibinfo{year}{2014}),
  \urlprefix\url{https://onlinelibrary.wiley.com/doi/abs/10.1002/adfm.201400125}.

\bibitem[{\citenamefont{Jang et~al.}(2015)\citenamefont{Jang, Lee,
  Pookpanratana, Hacker, Tran, and Richter}}]{ja.le.15}
\bibinfo{author}{\bibfnamefont{H.-J.} \bibnamefont{Jang}},
  \bibinfo{author}{\bibfnamefont{J.-S.} \bibnamefont{Lee}},
  \bibinfo{author}{\bibfnamefont{S.~J.} \bibnamefont{Pookpanratana}},
  \bibinfo{author}{\bibfnamefont{C.~A.} \bibnamefont{Hacker}},
  \bibinfo{author}{\bibfnamefont{I.~C.} \bibnamefont{Tran}}, \bibnamefont{and}
  \bibinfo{author}{\bibfnamefont{C.~A.} \bibnamefont{Richter}},
  \bibinfo{journal}{The Journal of Physical Chemistry C}
  \textbf{\bibinfo{volume}{119}}, \bibinfo{pages}{12949}
  (\bibinfo{year}{2015}),
  \urlprefix\url{https://doi.org/10.1021/acs.jpcc.5b01222}.

\bibitem[{\citenamefont{Friedrich
  et~al.}(2015{\natexlab{b}})\citenamefont{Friedrich, Caciuc, Atodiresei, and
  Bl\"ugel}}]{fr.ca.15_2}
\bibinfo{author}{\bibfnamefont{R.}~\bibnamefont{Friedrich}},
  \bibinfo{author}{\bibfnamefont{V.}~\bibnamefont{Caciuc}},
  \bibinfo{author}{\bibfnamefont{N.}~\bibnamefont{Atodiresei}},
  \bibnamefont{and} \bibinfo{author}{\bibfnamefont{S.}~\bibnamefont{Bl\"ugel}},
  \bibinfo{journal}{Phys. Rev. B} \textbf{\bibinfo{volume}{92}},
  \bibinfo{pages}{195407} (\bibinfo{year}{2015}{\natexlab{b}}),
  \urlprefix\url{https://link.aps.org/doi/10.1103/PhysRevB.92.195407}.

\bibitem[{\citenamefont{Yang et~al.}(2016)\citenamefont{Yang, Vu, Hallal,
  Rougemaille, Coraux, Chen, Schmid, and Chshiev}}]{graphene-Co}
\bibinfo{author}{\bibfnamefont{H.}~\bibnamefont{Yang}},
  \bibinfo{author}{\bibfnamefont{A.~D.} \bibnamefont{Vu}},
  \bibinfo{author}{\bibfnamefont{A.}~\bibnamefont{Hallal}},
  \bibinfo{author}{\bibfnamefont{N.}~\bibnamefont{Rougemaille}},
  \bibinfo{author}{\bibfnamefont{J.}~\bibnamefont{Coraux}},
  \bibinfo{author}{\bibfnamefont{G.}~\bibnamefont{Chen}},
  \bibinfo{author}{\bibfnamefont{A.~K.} \bibnamefont{Schmid}},
  \bibnamefont{and} \bibinfo{author}{\bibfnamefont{M.}~\bibnamefont{Chshiev}},
  \bibinfo{journal}{Nano Letters} \textbf{\bibinfo{volume}{16}},
  \bibinfo{pages}{145} (\bibinfo{year}{2016}), \bibinfo{note}{pMID: 26641927},
  \urlprefix\url{https://doi.org/10.1021/acs.nanolett.5b03392}.

\bibitem[{\citenamefont{Xiong et~al.}(2004)\citenamefont{Xiong, Wu,
  Valy~Vardeny, and Shi}}]{xi.wu.04}
\bibinfo{author}{\bibfnamefont{Z.~H.} \bibnamefont{Xiong}},
  \bibinfo{author}{\bibfnamefont{D.}~\bibnamefont{Wu}},
  \bibinfo{author}{\bibfnamefont{Z.}~\bibnamefont{Valy~Vardeny}},
  \bibnamefont{and} \bibinfo{author}{\bibfnamefont{J.}~\bibnamefont{Shi}},
  \bibinfo{journal}{Nature} \textbf{\bibinfo{volume}{427}},
  \bibinfo{pages}{821} (\bibinfo{year}{2004}), ISSN \bibinfo{issn}{1476-4687},
  \urlprefix\url{https://doi.org/10.1038/nature02325}.

\bibitem[{\citenamefont{Dediu et~al.}(2008)\citenamefont{Dediu, Hueso,
  Bergenti, Riminucci, Borgatti, Graziosi, Newby, Casoli, De~Jong, Taliani
  et~al.}}]{de.hu.08}
\bibinfo{author}{\bibfnamefont{V.}~\bibnamefont{Dediu}},
  \bibinfo{author}{\bibfnamefont{L.~E.} \bibnamefont{Hueso}},
  \bibinfo{author}{\bibfnamefont{I.}~\bibnamefont{Bergenti}},
  \bibinfo{author}{\bibfnamefont{A.}~\bibnamefont{Riminucci}},
  \bibinfo{author}{\bibfnamefont{F.}~\bibnamefont{Borgatti}},
  \bibinfo{author}{\bibfnamefont{P.}~\bibnamefont{Graziosi}},
  \bibinfo{author}{\bibfnamefont{C.}~\bibnamefont{Newby}},
  \bibinfo{author}{\bibfnamefont{F.}~\bibnamefont{Casoli}},
  \bibinfo{author}{\bibfnamefont{M.~P.} \bibnamefont{De~Jong}},
  \bibinfo{author}{\bibfnamefont{C.}~\bibnamefont{Taliani}},
  \bibnamefont{et~al.}, \bibinfo{journal}{Phys. Rev. B}
  \textbf{\bibinfo{volume}{78}}, \bibinfo{pages}{115203}
  (\bibinfo{year}{2008}),
  \urlprefix\url{https://link.aps.org/doi/10.1103/PhysRevB.78.115203}.

\bibitem[{\citenamefont{Galbiati}(2016)}]{Galbiati2016}
\bibinfo{author}{\bibfnamefont{M.}~\bibnamefont{Galbiati}},
  \emph{\bibinfo{title}{State of the Art in Alq3-Based Spintronic Devices}}
  \bibinfo{publisher}{Springer International Publishing},
  \bibinfo{address}{Cham}, \bibinfo{year}{2016}), pp.
  \bibinfo{pages}{139--151},
  \urlprefix\url{https://doi.org/10.1007/978-3-319-22611-8_7}.

\bibitem[{\citenamefont{Dediu et~al.}(2009)\citenamefont{Dediu, Hueso,
  Bergenti, and Taliani}}]{Dediu2009}
\bibinfo{author}{\bibfnamefont{V.~A.} \bibnamefont{Dediu}},
  \bibinfo{author}{\bibfnamefont{L.~E.} \bibnamefont{Hueso}},
  \bibinfo{author}{\bibfnamefont{I.}~\bibnamefont{Bergenti}}, \bibnamefont{and}
  \bibinfo{author}{\bibfnamefont{C.}~\bibnamefont{Taliani}},
  \bibinfo{journal}{Nature Materials} \textbf{\bibinfo{volume}{8}},
  \bibinfo{pages}{707} (\bibinfo{year}{2009}),
  \urlprefix\url{https://doi.org/10.1038/nmat2510}.

\bibitem[{\citenamefont{Curioni et~al.}(1998)\citenamefont{Curioni, Boero, and
  Andreoni}}]{cu.bo.98}
\bibinfo{author}{\bibfnamefont{A.}~\bibnamefont{Curioni}},
  \bibinfo{author}{\bibfnamefont{M.}~\bibnamefont{Boero}}, \bibnamefont{and}
  \bibinfo{author}{\bibfnamefont{W.}~\bibnamefont{Andreoni}},
  \bibinfo{journal}{Chemical Physics Letters} \textbf{\bibinfo{volume}{294}},
  \bibinfo{pages}{263} (\bibinfo{year}{1998}), ISSN \bibinfo{issn}{0009-2614},
  \urlprefix\url{https://www.sciencedirect.com/science/article/pii/S000926149800829X}.

\bibitem[{\citenamefont{Tarafder et~al.}(2010)\citenamefont{Tarafder, Sanyal,
  and Oppeneer}}]{ta.sa.10}
\bibinfo{author}{\bibfnamefont{K.}~\bibnamefont{Tarafder}},
  \bibinfo{author}{\bibfnamefont{B.}~\bibnamefont{Sanyal}}, \bibnamefont{and}
  \bibinfo{author}{\bibfnamefont{P.~M.} \bibnamefont{Oppeneer}},
  \bibinfo{journal}{Phys. Rev. B} \textbf{\bibinfo{volume}{82}},
  \bibinfo{pages}{060413} (\bibinfo{year}{2010}),
  \urlprefix\url{https://link.aps.org/doi/10.1103/PhysRevB.82.060413}.

\bibitem[{\citenamefont{Bisti et~al.}(2011)\citenamefont{Bisti, Stroppa,
  Donarelli, Picozzi, and Ottaviano}}]{bi.st.11}
\bibinfo{author}{\bibfnamefont{F.}~\bibnamefont{Bisti}},
  \bibinfo{author}{\bibfnamefont{A.}~\bibnamefont{Stroppa}},
  \bibinfo{author}{\bibfnamefont{M.}~\bibnamefont{Donarelli}},
  \bibinfo{author}{\bibfnamefont{S.}~\bibnamefont{Picozzi}}, \bibnamefont{and}
  \bibinfo{author}{\bibfnamefont{L.}~\bibnamefont{Ottaviano}},
  \bibinfo{journal}{Phys. Rev. B} \textbf{\bibinfo{volume}{84}},
  \bibinfo{pages}{195112} (\bibinfo{year}{2011}),
  \urlprefix\url{https://link.aps.org/doi/10.1103/PhysRevB.84.195112}.

\bibitem[{\citenamefont{Droghetti
  et~al.}(2014{\natexlab{b}})\citenamefont{Droghetti, Cinchetti, and
  Sanvito}}]{dr.ci.14}
\bibinfo{author}{\bibfnamefont{A.}~\bibnamefont{Droghetti}},
  \bibinfo{author}{\bibfnamefont{M.}~\bibnamefont{Cinchetti}},
  \bibnamefont{and} \bibinfo{author}{\bibfnamefont{S.}~\bibnamefont{Sanvito}},
  \bibinfo{journal}{Phys. Rev. B} \textbf{\bibinfo{volume}{89}},
  \bibinfo{pages}{245137} (\bibinfo{year}{2014}{\natexlab{b}}),
  \urlprefix\url{https://link.aps.org/doi/10.1103/PhysRevB.89.245137}.

\bibitem[{\citenamefont{Droghetti}(2020)}]{ad.20}
\bibinfo{author}{\bibfnamefont{A.}~\bibnamefont{Droghetti}},
  \bibinfo{journal}{Journal of Magnetism and Magnetic Materials}
  \textbf{\bibinfo{volume}{502}}, \bibinfo{pages}{166578}
  (\bibinfo{year}{2020}), ISSN \bibinfo{issn}{0304-8853},
  \urlprefix\url{https://www.sciencedirect.com/science/article/pii/S0304885319340156}.

\bibitem[{Note1()}]{Note1}
Note1, \bibinfo{note}{the Co slab considered here is slightly different from
  that used in the other sections. In the case of Alq$_3$/Co, the slab is
  constrained in-plane to have the same lattice constant of Cu, so to mimic the
  experiments of Refs. \cite {dr.st.14,andrea-alq3}. Nonetheless this has a
  minor effect on the results and on the proposed mechanism for the MCA
  modification.}

\bibitem[{\citenamefont{M\"{u}ller et~al.}(2013)\citenamefont{M\"{u}ller, Steil,
  Droghetti, Großmann, Meded, Magri, Sch\"{a}fer, Fuhr, Sanvito, Ruben
  et~al.}}]{mu.st.13}
\bibinfo{author}{\bibfnamefont{S.}~\bibnamefont{M\"{u}ller}},
  \bibinfo{author}{\bibfnamefont{S.}~\bibnamefont{Steil}},
  \bibinfo{author}{\bibfnamefont{A.}~\bibnamefont{Droghetti}},
  \bibinfo{author}{\bibfnamefont{N.}~\bibnamefont{Grossmann}},
  \bibinfo{author}{\bibfnamefont{V.}~\bibnamefont{Meded}},
  \bibinfo{author}{\bibfnamefont{A.}~\bibnamefont{Magri}},
  \bibinfo{author}{\bibfnamefont{B.}~\bibnamefont{Sch\"{a}fer}},
  \bibinfo{author}{\bibfnamefont{O.}~\bibnamefont{Fuhr}},
  \bibinfo{author}{\bibfnamefont{S.}~\bibnamefont{Sanvito}},
  \bibinfo{author}{\bibfnamefont{M.}~\bibnamefont{Ruben}},
  \bibnamefont{et~al.}, \bibinfo{journal}{New Journal of Physics}
  \textbf{\bibinfo{volume}{15}}, \bibinfo{pages}{113054}
  (\bibinfo{year}{2013}),
  \urlprefix\url{https://dx.doi.org/10.1088/1367-2630/15/11/113054}.

\bibitem[{\citenamefont{Laurent et~al.}(2021)\citenamefont{
  Le~Laurent, Barreteau, Markussen}}]{Fe-Co-slab}
  \bibinfo{author}{\bibfnamefont{L.}~\bibnamefont{Le~Laurent}},
  \bibinfo{author}{\bibfnamefont{C.}~\bibnamefont{Barreteau}},
  \bibinfo{author}{\bibfnamefont{T.}~\bibnamefont{Markussen}},
  , \bibinfo{journal}{Phys. Rev. B}
  \textbf{\bibinfo{volume}{100}}, \bibinfo{pages}{174426}
  (\bibinfo{year}{2019}),
  \urlprefix\url{https://link.aps.org/doi/10.1103/PhysRevB.100.174426}.

\end{thebibliography}

\end{document}